\documentclass[12pt]{article}

\usepackage[a4paper,text={16.8cm,22.4cm}]{geometry}
\usepackage{amsmath,amsfonts,slashed,amssymb,tikz,bm,psfrag,graphicx,feynarts,color,textcomp,multirow}

\RequirePackage[sort&compress,square,comma,numbers]{natbib}
\allowdisplaybreaks 
\newcommand{\ord}{{\cal O}}

\newcommand{\im}{\operatorname{Im}}
\newcommand{\re}{\operatorname{Re}}

\newcommand{\e}{\epsilon}
\newcommand{\s}{\slashed}

\newcommand{\tx}{\text}
\newcommand{\da}{\dagger}
\newcommand{\pa}{\partial}
\newcommand{\La}{\mathcal{L}}
\newcommand{\bs}{\boldsymbol}
 
\newcommand{\sgn}{\tx{sgn}}

\newcommand{\one}{\mathbf{1}}

\newcommand{\TeV}{\tx{TeV}}
\newcommand{\GeV}{\tx{GeV}}
\newcommand{\MeV}{\tx{MeV}}
\newcommand{\SM}{\tx{SM}}
\newcommand{\RS}{\tx{RS}}

\newcommand{\NP}{\tx{NP}}

\newcommand{\Sb}{\boldsymbol{S}} 
\newcommand{\Vb}{\boldsymbol{V}} 
 
\newcommand{\Xb}{\boldsymbol{X}}

\newcommand{\MKK}{M_{\text{KK}}} 
\newcommand{\muKK}{\mu_{\text{KK}}}

\newcommand{\al}{\alpha}

\newcommand{\ga}{\gamma}

\newcommand{\hc}{\tx{h.c.}}

\newcommand{\kEh}{{\hat k_E}}
\newcommand{\qv}{{q}}
\newcommand{\Yb}{{\bs Y}}
\newcommand{\Zb}{{\bs Z}}
\newcommand{\Sm}{{\mathcal S}}
\newcommand{\Cm}{{\mathcal C}}

\newcommand{\et}{{1_\eta}}
\newcommand{\Nb}{{\bs N}}
\newcommand{\Rb}{{\bs R}}
\newcommand{\Q}{{\mathcal{Q}}}
\newcommand{\U}{{\mathcal{U}}}

\newcommand{\D}{\mathcal{D}}
\newcommand{\Av}{{Q}}
\newcommand{\At}{{q}}
\newcommand{\av}{{q}}

\newcommand{\M}{{{\mathcal{M}}}}
\newcommand{\rh}{\varrho}

\newcommand{\Db}{{\bs D}}
\newcommand{\Lb}{{\bs L}}

\newcommand{\Pb}{{\bs P}}
\newcommand{\A}{\mathcal{A}}
\newcommand{\KK}{\tx{KK}}

\renewcommand{\[}{\begin{equation}}
\renewcommand{\]}{\end{equation}}
\newcommand{\mkk}{M_\mathrm{KK}}

\newcommand{\Y}{\boldsymbol{Y}}

\newcommand{\cA}{\boldsymbol{c}_A}

\newcommand{\pE}{\kEh}
\newcommand{\R}[4]{\boldsymbol{L}_{#4}^{#1}(#2, #3)}

\newcommand{\Rmq}[2]{\boldsymbol{R}_{#1}(#2)}

\newcommand{\Zm}{\boldsymbol{Z}}
\newcommand{\Xm}{\boldsymbol{X}}

\usetikzlibrary{decorations.pathmorphing}
\usetikzlibrary{decorations.markings}
\usetikzlibrary{arrows,shapes}

\tikzstyle{every picture}+=[remember picture]
\tikzstyle{na} = [baseline=-.5ex]
\tikzset{
vector/.style={decorate, decoration={snake,amplitude=1.2pt}, draw,segment length=4pt},
fermion/.style={draw=black, postaction={decorate}, decoration={markings,mark=at position .55 with {\arrow[draw=black]{>}}}},
fermionbar/.style={draw=black, postaction={decorate}, decoration={markings,mark=at position .55 with {\arrow[draw=black]{<}}}},
fermionnoarrow/.style={draw=black},
scalararrow/.style={dashed,draw=black, postaction={decorate}, decoration={markings,mark=at position .55 with {\arrow[draw=black]{>}}}},
scalar/.style={dashed}, 
gluon/.style={decorate, draw=black, decoration={coil,amplitude=2pt, mirror,segment length=4pt,aspect=0.8}},
vertex/.style={draw,circle,fill=black,inner sep=0pt,minimum size=1mm},  
comp/.style={line width=5pt, draw=blue!20},
elem/.style={line width=1pt, draw=black},
comp2/.style={line width=5pt, draw=blue!20},
elem2/.style={line width=1pt, draw=black},
ghost/.style={dotted,draw=black},
cross/.style={cross out, draw=black, minimum size = 2mm,inner sep=0,outer sep=0}
}

\begin{document}

\begin{titlepage}

\begin{flushright}
\normalsize
MITP/15-070\\
September 8, 2015
% arXiv:1509.nnnnn
\end{flushright}

\vspace{0.3cm}
\begin{center}
\Large\bf 
Impact of Warped Extra Dimensions on the\\ 
Dipole Coefficients in $\boldsymbol{b\to s\gamma}$ Transitions
\end{center}

\vspace{0.8cm}
\begin{center}
Raoul Malm$^a$, Matthias Neubert$^{a,b,c}$ and Christoph Schmell$^a$\\
\vspace{0.7cm} 
{\sl ${}^a$PRISMA Cluster of Excellence \& Mainz Institute for Theoretical Physics\\
Johannes Gutenberg University, 55099 Mainz, Germany\\[3mm]
${}^b$Institut f\"ur Theoretische Physik\\ 
Universit\"at Heidelberg, Philosophenweg 16, 69120 Heidelberg, Germany\\[3mm]
${}^c$Department of Physics, LEPP, Cornell University, Ithaca, NY 14853, U.S.A.}
\end{center}

\vspace{0.8cm}

\begin{abstract}   
We calculate the electro- and chromomagnetic dipole coefficients $C_{7\gamma,8g}$ and $\tilde C_{7\gamma,8g}$ in the context of the minimal Randall-Sundrum (RS) model with a Higgs sector localized on the IR brane using the five-dimensional (5D) approach, where the coefficients are expressed in terms of integrals over 5D propagators. Since we keep the full dependence on the Yukawa matrices, the integral expressions are formally valid to all orders in $v^2/M_{\rm KK}^2$. In addition we relate our results to the expressions obtained in the Kaluza-Klein (KK) decomposed theory and show the consistency in both pictures analytically and numerically, which presents a non-trivial cross-check. In Feynman-'t Hooft gauge, the dominant corrections from virtual KK modes arise from the scalar parts of the $W^\pm$-boson penguin diagrams, including the contributions from the scalar component of the 5D gauge-boson field and from the charged Goldstone bosons in the Higgs sector. The size of the KK corrections depends on the parameter $y_*$, which sets the upper bound for the anarchic 5D Yukawa matrices. We find that for $y_*\gtrsim1$ the KK corrections are proportional to $y_*^2$. We discuss the phenomenological implications of our results for the branching ratio ${\rm Br}(\bar B\to X_s\gamma)$, the time-dependent CP asymmetry $S_{K^*\gamma}$, the direct CP asymmetry $A_{\rm CP}^{b\to s\gamma}$ and the CP asymmetry difference $\Delta A_{\rm CP}^{b\to s\gamma}$. We can derive a lower bound on the first KK gluon resonance of $3.8\,$TeV for $y_*=3$, requiring that at least $10\%$ of the RS parameter space covers the experimental $2\sigma$ error margins. We further discuss the branching ratio ${\rm Br}(\bar B\to X_sl^+l^-)$ and compare our predictions for $C_{7\gamma,9,10}$ and $\tilde C_{7\gamma,9,10}$ with phenomenological results derived from model-independent analyses.
\end{abstract}
\vfil

\end{titlepage}

\section{Introduction}

In July 2012 the Higgs boson, the last missing piece of the Standard Model (SM), was discovered at the Large Hadron Collider (LHC) at CERN \cite{ATLAS:2012gk,CMS:2012gu}. Since then the hierarchy problem, i.e.\ the question about the mechanism that stabilizes the Higgs mass near the electroweak scale, is no longer a hypothetical issue. A promising possibility to solve the hierarchy problem is offered by Randall-Sundrum (RS) models \cite{Randall:1999ee}, in which the SM is embedded in a slice of anti-de Sitter space while the Higgs sector is localized on the ``infra-red (IR) brane'', one of two sub-manifolds bounding the extra dimension. The smallness of the electroweak scale can then be explained by the fundamental ultra-violet (UV) cutoff given by the warped Planck scale, whose value near the IR brane lies in the TeV range. Moreover, by allowing the fermion fields to propagate in the bulk, these models provide a natural explanation for the hierarchies observed in the flavor sector \cite{Grossman:1999ra,Gherghetta:2000qt,Huber:2000ie} and the smallness of flavor-changing neutral currents (FCNCs) \cite{Agashe:2004ay,Agashe:2004cp,Csaki:2008zd,Casagrande:2008hr,Blanke:2008zb,Blanke:2008yr,Bauer:2009cf}.

In this paper we investigate the FCNC process $b\to s\ga$ in the minimal RS model with a brane-localized Higgs sector. For two reasons this transition is very interesting in order to search for new physics. In the SM the dipole coefficients are one-loop suppressed and the transition is logarithmically suppressed by the GIM mechanism \cite{Haisch:2008ar}. In order to include the effects of the RS model on the transition $b\to s\ga$ we implement an effective Lagrangian, in which the heavy Kaluza-Klein (KK) quarks and bosons are integrated out. The most important operators are the electromagnetic dipole operators 
\begin{align}
   Q_{7\ga} = - \frac{e\,m_b}{4\pi^2}\,\bar s\,\sigma_{\mu\nu} F^{\mu\nu} P_R\,b \,, \qquad 
   \tilde Q_{7\ga} = - \frac{e\,m_b}{4\pi^2}\,\bar s\,\sigma_{\mu\nu} F^{\mu\nu} P_L\,b \,,
\end{align}
with $\sigma_{\mu\nu}=\frac{i}{2}\,[\ga_\mu,\ga_\nu]$ and the projection operators $P_{R,L}=\frac12 (1\pm\ga_5)$. Due to operator mixing we also consider the chromomagnetic dipole operators 
\begin{align}
   Q_{8g} = - \frac{g_s m_b}{4\pi^2}\,\bar s\,\sigma_{\mu\nu}\,G_a^{\mu\nu} t_a P_R\,b \,, \qquad
   \tilde Q_{8g} = - \frac{g_s m_b}{4\pi^2}\,\bar s\,\sigma_{\mu\nu}\,G_a^{\mu\nu} t_a P_L\,b \,,
\end{align}
where $t_a$ are the generators of $SU(3)_c$. The main focus of our paper lies on the derivation of integral expressions for the dipole coefficients at the one-loop level using five-dimensional (5D) propagators in the mixed position-momentum space and with the full dependence on the Yukawa interactions imposed by the mixed boundary condition at the IR brane. 

In the literature, the first discussions on $b\to s\ga$ in the RS model can be found in \cite{Agashe:2004ay,Agashe:2004cp,Agashe:2006iy}. There, the authors claimed that the penguin diagrams with the exchange of charged Higgs scalars (Goldstone bosons of the $W^\pm$ boson) along with KK fermions gives the dominant contribution to the dipole coefficients. The diagram with the exchange of KK gluons was found to be approximately aligned with the 4D down-type Yukawa matrix and therefore subleading. Furthermore, the authors claimed that the dipole coefficients in the brane-localized Higgs scenario were logarithmically divergent and sensitive to the UV cutoff. It was shown in \cite{Csaki:2010aj} that the diagrams contributing to the leptonic decay $\mu\to e\ga$ at one-loop are indeed finite. With the same technique the authors of \cite{Blanke:2012tv} investigated the process $b\to s\gamma$ working with 5D propagators and treating the Yukawa interactions as perturbations.~In \cite{Schmell:2014lka}, one of us discussed the $b\to s\ga$ process in the minimal RS model with a brane-localized Higgs sector working in the KK-decomposed theory, where the dipole coefficients are expressed via infinite sums over the contributions from different levels of KK modes. In \cite{Biancofiore:2014wpa} the authors calculated the dipole coefficients in the custodial RS model with a brane-localized Higgs sector, focusing only on the diagrams with an exchange of the first level of KK fermions along with gluons and charged Goldstone bosons. Recently, the authors of \cite{Beneke:2015lba} studied lepton flavor violation in RS models in the 5D framework, where they discussed the electromagnetic (leptonic) dipole operator in RS models with a brane-localized or nearly brane-localized Higgs and treated the Yukawa interactions as perturbations. In the present work, we perform a complete calculation of the electro- and chromomagnetic (quark) dipole coefficients including all contributions at one-loop order in the minimal RS model with a brane-localized Higgs sector.~We derive expressions for the dipole coefficients using 5D propagators computed by retaining the full dependence on the Yukawa interactions. In contrast to \cite{Blanke:2012tv,Beneke:2015lba}, we derive 5D expressions that are formally valid to all orders in $v^2/\MKK^2$. In contrast to \cite{Beneke:2015lba}, we focus on the quark dipole coefficients, including the contributions of the chromomagnetic dipole operator. In addition, we derive formulas in the KK-decomposed (4D) theory including the contributions from all KK levels and show the consistency with the results obtained in the 5D framework.

After introducing the model and setting up the notation in Section~\ref{sec:Prelbsga} we derive formulas for the dipole coefficients in the 5D framework in Section~\ref{sec:Wils5Dbsga}. In Section~\ref{sec:AnaWilsbsga} we compare our results with the expressions in the KK-decomposed theory and analyze the different KK contributions to the dipole coefficients.~After implementing the renormalization-group (RG) evolution from the KK scale down to the $B$-meson scale we discuss the phenomenological implications in Section~\ref{sec:Phenobsga}.~Our main results are summarized in the conclusions. 

\section{Theoretical setup}
\label{sec:Prelbsga}

We focus on RS models where the electroweak symmetry-breaking sector is localized on or near the IR brane. The extra dimension is chosen to be an $S^1/Z_2$ orbifold parametrized by a coordinate $\phi \in [-\pi, \pi]$, with two 3-branes localized on the orbifold fixed-points $\phi=0$ (UV brane) and $|\phi|=\pi$ (IR brane). The RS metric reads \cite{Randall:1999ee}
\begin{equation}\label{eqn:RSmetric}
   ds^2 = e^{-2\sigma(\phi)}\,\eta_{\mu\nu}\,dx^\mu dx^\nu - r^2 d\phi^2
   = \frac{\epsilon^2}{t^2} \left( \eta_{\mu\nu}\,dx^\mu dx^\nu
    - \frac{1}{M_{\rm KK}^2}\,dt^2 \right) ,
\end{equation}
where $e^{-\sigma(\phi)}$, with $\sigma(\phi)=kr|\phi|$, is referred to as the warp factor. The size $r$ and curvature $k$ of the extra dimension are assumed to be of Planck size, $k\sim 1/r\sim M_{\rm Pl}$.~The quantity $L=\sigma(\pi)=kr\pi$ measures the size of the extra dimension and is chosen to be $L\approx 33-34$ in order to explain the hierarchy between the Planck scale $M_{\rm Pl}$ and the TeV scale. We define the KK scale $M_{\rm KK}=k\epsilon$, with $\epsilon=e^{-\sigma(\pi)}$, which sets the mass scale for the low-lying KK excitations of the SM particles. On the right-hand side of \eqref{eqn:RSmetric} we have introduced a new coordinate $t=\epsilon\,e^{\sigma(\phi)}$, whose values on the UV and IR branes are $\epsilon$ and 1, respectively.\footnote{The dimensionless variable $t$ is related to the conformal coordinate $z$ frequently used in the literature by the simple rescaling $z=t/M_{\rm KK}\equiv R'\,t$.} In our analysis we consider the minimal RS model, adopting the conventions and notations of \cite{Casagrande:2008hr}. The gauge group is taken to be $SU(3)_c\times SU(2)_L\times U(1)_Y$ like in the SM, and it is broken to $SU(3)_c\times\mbox U(1)_{\rm em}$ by the Higgs vacuum expectation value (vev). 

A tree-level analysis of electroweak precision observables, mainly the $T$ parameter, implies that the mass of the first KK gluon resonance is pushed to values $M_{g^{(1)}}>11.3\,\TeV$ at 95\% confidence level (CL) \cite{Carena:2004zn}, where we have used the most up-to-date values from \cite{Baak:2014ora}.~Loop corrections could potentially change this bound in a significant way. Also, it is conceivable that new-physics contributions arising in a UV completion of the RS model bring $S$ and $T$ back into the phenomenologically favored region.~Nevertheless, one usually considers RS models with a built-in protection for the $T$ parameter by implementing a custodial symmetry via the gauge group $SU(3)_c\times SU(2)_L\times SU(2)_R\times U(1)_X\times P_{LR}$ \cite{Agashe:2003zs,Csaki:2003zu,Agashe:2006at}. Then the bound from electroweak precision observables reduces to $M_{g^{(1)}}>4.8\,\TeV$ at $95\%$ CL \cite{Malm:2013jia}. But on the other hand, the sensitivity of Higgs physics on virtual effects from heavy KK excitations is strongly increased in the model with custodial symmetry, due to the enlarged fermion multiplicity for each KK level. Comparing predictions for the signal rates of the Higgs decaying into pairs of electroweak gauge bosons with data from the LHC excludes KK gluon resonances lighter than $(15-20)\,\TeV\times(y_\ast/3)$ at 95\% CL \cite{Malm:2014gha}, where the precise value depends on the details of the localization of the Higgs sector near the IR brane. Here $y_\ast$ sets the upper bound for the entries of the anarchic 5D Yukawa matrices, $|(\Yb_q)_{ij}|\leq y_\ast$. In the minimal RS model the resulting bounds are much weaker. For values of $y_\ast\gtrsim\ord(1)$, the custodial RS model thus loses its main advantage of allowing for lighter KK resonances such that the minimal RS model is just as promising nowadays. Note also that the parameter $\e_K$ measuring CP violation in kaon mixing requires KK gluon masses in the range of 10\,TeV (with moderate fine tuning) irrespective of whether the minimal or the custodial RS models are considered \cite{Csaki:2008zd,Bauer:2011ah}.

\subsubsection*{Higgs localization}

In this work we focus on the RS model with a brane-localized Higgs field, where the inverse characteristic width $\Delta_h$ of the Higgs field along the extra dimension is assumed to be much larger than the inherent UV cutoff near the IR brane, i.e.\ $\Delta_h\gg\Lambda_\TeV\sim\mbox{several}\,\mkk$ \cite{Hahn:2013nza}. It is well known that quantum fields can be strictly localized on orbifold fixed points, and in such a scenario the quantity $\Delta_h$ can indeed be infinite or arbitrarily large. Similar to the case of Higgs production via gluon fusion \cite{Djouadi:2007fm,Falkowski:2007hz,Cacciapaglia:2009ky,Bhattacharyya:2009nb,Bouchart:2009vq,Casagrande:2010si,Azatov:2010pf,Azatov:2011qy,Goertz:2011hj,Carena:2012fk,Malm:2013jia,Archer:2014jca}, we will find that our results for the Higgs contributions to the $b\to s\gamma$ and $b\to sg$ dipole coefficients are sensitive to details of the localization mechanism. For these contributions we sometimes extend our results to the case of a so-called \textit{narrow bulk-Higgs} scenario, where the Higgs field lives in the bulk with an inverse width such that $\mkk\ll\Delta_h\ll\Lambda_\TeV$ \cite{Malm:2013jia}. This is a special case of a general bulk-Higgs with $\Delta_h\sim v$, where $v$ is the Higgs vev. 

The authors of \cite{Agashe:2014jca,Delaunay:2012cz} have calculated the RS contributions to the dipole operators in the KK-decomposed theory for a general bulk-Higgs field, where the localization parameter is taken to be $\beta\sim1$ (in our notation $\beta\sim\Delta_h/v$).~Numerically, they also discuss the quasi IR-localized limit by increasing $\beta$, i.e.\ by pushing the Higgs profile towards the IR brane. They find that heavy KK fermion modes with masses $m_{q_n}\sim\beta\MKK$ yield unsuppressed contributions %, hinting at a UV sensitivity
 in the case where the Higgs inverse width is of order the UV cutoff, $\beta\sim\Lambda_\TeV/v$ ($\Delta_h\sim\Lambda_\TeV$).~In this case, there are still some high-momentum KK modes that can probe the ``bulky nature" of the Higgs field.~In addition, the authors of \cite{Agashe:2014jca,Beneke:2015lba} have observed the non-decoupling of heavy KK excitations of the Higgs boson itself in the quasi IR-localized limit of large $\beta$.~%These findings hint at the fact that there is a certain amount of UV model dependence in the calculation of Higgs-induced contributions to the dipole operators (see \cite{Beneke:2015lba} for a nice recent discussion of this point).
These findings show that the results of Higgs-induced contributions to the dipole operators depend on the implementation of the Higgs sector (see \cite{Beneke:2015lba} for a nice recent discussion of this point).
In our brane-localized Higgs scenario, where $\Delta_h\gg\Lambda_\TeV$, we observe that the 5D description of the dipole coefficients gives results which are consistent with the description in the KK-decomposed theory when summing over the first few KK levels.~This shows that heavy KK modes near the UV cutoff decouple. Furthermore, KK excitations of the Higgs doublet do not arise in this version of the model.

\subsubsection*{Gauge sector}

In the minimal RS model the SM gauge group lives in the bulk and is broken to $U(1)_{\rm EM}$ on the IR brane, where the Higgs field develops a vev.~Details for the implementation of the Higgs, gauge-boson, and gauge-fixing sectors in the context of this model (and using our notations) have been given in \cite{Casagrande:2008hr}.~The KK-decomposition for the 5D gauge-boson field $B_M(x,t)$ is in general given by (with $B=A,G,W,Z$)
\begin{align}
   B_\mu(x,t) = \frac{1}{\sqrt r} \sum_n B_\mu^{(n)}(x)\,\chi_n^B(t) \,, \qquad
   B_5(x,t) = \frac{1}{\sqrt r} \sum_n \left( \frac{-kt}{m_{B_n}} \right) 
    \varphi_B^{(n)}(x)\,\pa_t\chi_n^B(t) \,,
\end{align}
where $B_\mu^{(n)}$ are the KK modes of the gauge bosons with masses $m_n^B$. The scalar particles $\varphi_{W}^{\pm(n)},\varphi_{Z}^{(n)}$ are ``unphysical'' in the sense that they provide the longitudinal degrees of freedom of the $W,Z$ bosons $(n=0)$ and their KK modes ($n\ge 1$), and thus they can be gauged away. Similarly the scalar particles $\varphi_{A}^{(n)},\varphi_{G}^{(n)}$ provide the longitudinal degrees of freedom for the photon and gluon KK modes. The scalar fields $W_5^\pm$ and $Z_5$ mix with the charged Goldstone bosons arising from the Higgs sector. Assuming for the time being that the scalar sector is localized on the IR brane, we parameterize the Higgs doublet after electroweak symmetry breaking in the usual form
\begin{equation}\label{eqn:PhiDoublet}
   \Phi(x) = \frac{1}{\sqrt{2}} 
   \begin{pmatrix}
    -i\sqrt{2} \varphi^+(x) \\
    v + h(x) + i\varphi^3(x) &	
   \end{pmatrix}\,,
\end{equation}
where $v$ denotes the Higgs vev in the RS model.~We determine the vev $v$ from the shift to the Fermi constant $G_F$, which can be derived in the RS model by considering (at tree level) the effect of the exchange of the infinite tower of KK $W$ bosons on the rate for muon decay.~We find that $v =(\sqrt2 G_F)^{-1/2}\left[1+\frac{L m_W^2}{4\MKK^2}+\ord\left(\frac{v^4}{\MKK^4}\right)\right]$ \cite{Malm:2013jia} with the experimentally measured value $(\sqrt2G_F)^{-1/2} \approx 246.2\,\GeV$.~The decomposition of the scalar fields $\varphi^\pm,\, \varphi^3$ in \eqref{eqn:PhiDoublet} into the mass eigenstates $\varphi_W^{\pm(n)}$, $\varphi_Z^{(n)}$ reads \cite{Casagrande:2008hr}
\begin{equation}
\begin{aligned}\label{varphidecomp}
   \varphi^\pm(x) &= \sum_n \frac{\tilde m_W}{m_n^W}\,\sqrt{2\pi}\,\chi_n^W(1)\,
    \varphi_W^{\pm(n)}(x) \,, & \quad
   \tilde m_W &= \frac{g_5}{\sqrt{2\pi r}}\,\frac{v}{2} \,, \\
   \varphi^3(x) &= \sum_n \frac{\tilde m_Z}{m_n^Z}\,\sqrt{2\pi}\,
    \chi_n^Z(1)\,\varphi_Z^{(n)}(x) \,, &
   \tilde m_Z &= \frac{g_5/c_w}{\sqrt{2\pi r}}\,\frac{v}{2} \,,
\end{aligned}
\end{equation}
where $c_w\equiv\cos\theta_w$ is the cosine of the weak mixing angle, and $\tilde m_W,\tilde m_Z$ are the leading contributions to the $W^\pm$- and $Z$-boson masses in an expansion in powers of $v^2/\mkk^2$, such that $m_{W,Z}=\tilde m_{W,Z}\left[1 - \frac{\tilde m_{W,Z}^2}{4\MKK^2}\left(L-1+\frac{1}{2L}\right) + \ord\left(\frac{v^4}{\MKK^4}\right)\right]$ \cite{Casagrande:2008hr}.~The $SU(2)_L$ and hypercharge 5D gauge couplings are denoted by $g_5$ and $g_5'$. In the context of RS models the weak mixing angle can be expressed as $s_w^2\equiv \sin^2\theta_w=g_5'^2/(g_5^2+g_5'^2)$, which can be studied experimentally via the $Z$-pole polarization asymmetries observed at LEP.~We take $s_w^2$ and $m_W$ as input values which implies that $m_Z$ is a derived quantity given in the RS model by $m_Z(m_W,s_w^2) = \frac{m_W}{c_w}\left[1- \frac{m_W^2}{4\MKK^2} \frac{s_w^2}{c_w^2} \left(L-1+\frac{1}{2L}\right) + \ord\left(\frac{v^4}{\MKK^4}\right)\right]$.~Since the profile of the zero mode is flat up to corrections of order $v^2/\mkk^2$, it follows that $\sqrt{2\pi}\,\chi_n^{W,Z}(1)$ in \eqref{varphidecomp} is close to~1, and hence the fields $\varphi^\pm$, $\varphi^3$ coincide with $\varphi_W^{\pm(0)}$, $\varphi_Z^{(0)}$ to leading order.~%Mixing effects arise at order $v^2/\mkk^2$ and higher. 
We mention that one can adjust the gauge-fixing Lagrangian so as to cancel any mixings between the vector and scalar fields \cite{Casagrande:2008hr}.

\subsubsection*{Quark sector}

In the quark sector the minimal RS model contains an $SU(2)_L$ doublet field $Q(x,t)$ and two $SU(2)_L$ singlet fields $u(x,t)$ and $d(x,t)$ in the 5D Lagrangian, each of which are three-component vectors in generation space.~The 5D fermion states can be described by four-component Dirac spinors \cite{Grossman:1999ra,Gherghetta:2000qt}.~We use a compact notation, where we collect the left- and right-handed components of the up- and down-type states into six-component vectors ${\cal U}_A=(U_A,u_A)^T$ and ${\cal D}_A=(D_A,d_A)^T$ with $A=L,R$, which are collectively referred to as ${\cal Q}_{L,R}$. Their decomposition into 4D KK modes reads 
\begin{equation}\label{KK-decomp}
   {\cal Q}_A(x,t) = \sum_n\,\,{\cal Q}_A^{(n)}(t)\,q_A^{(n)}(x) \,; \quad A=L,R \,.
\end{equation}
The superscript $n$ labels the different mass eigenstates in the 4D effective theory, such that $n=1,2,3$ refer to the SM quarks, while $n=4,\dots,9$ label the six fermion modes of the first KK level, and so on. The functions ${\cal Q}_{L,R}^{(n)}(t)$ denote the wave functions of the left- and right-handed components of the $n^{\rm th}$ KK mass eigenstate along the extra dimension.~The upper (lower) components of ${\cal Q}_{L,R}^{(n)}(t)$ include the profiles of the $SU(2)_L$ doublet (singlet) quark fields.

\section{Calculation of the dipole coefficients}
\label{sec:Wils5Dbsga}

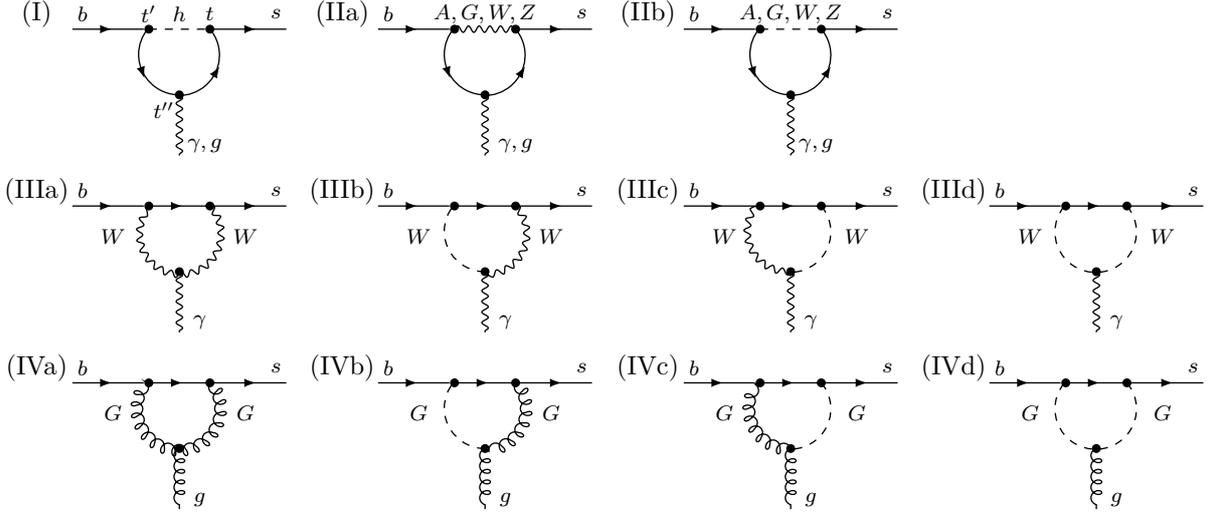
\begin{figure}[t!]
\begin{center}
\begin{tikzpicture}[line width=0.5pt,>=latex,scale=0.67]
\begin{scope}[xshift=0cm]
\draw (-2.4,0.3) node {\footnotesize (I)};
\draw (1,0.3) node {\scriptsize $t$};
\draw (-0.2,0.3) node {\scriptsize $t'$};
\draw (0.1,-1.6) node {\scriptsize $t''$};
\draw[fermion] (-1.7,0) -- (-0.3,0);
\draw[fermionbar] (2.5,0) -- (1,0);
\draw[scalar] (-0.2,0) -- (1,0);
\draw[vector] (0.4,-1.3) -- (0.4,-2.5);
\draw (-0.2,0)  node [vertex]  {};
\draw (1,0)  node [vertex]  {};
\draw (0.4,-1.3)  node [vertex]  {};
\draw (-1.5,0.3) node {\scriptsize $b$};
\draw (2.3,0.3) node {\scriptsize $s$};
\draw (0.4,0.3) node {\scriptsize $h$};
\draw (0.9,-2.3) node {\scriptsize $\ga,g$};
\draw[fermionbar] (1,0) arc (40:-90:0.8);
\draw[fermionbar] (0.4,-1.3) arc (-90:-220:0.8);
\end{scope}
\begin{scope}[xshift=6cm]
\draw (-2.4,0.3) node {\footnotesize (IIa)};
\draw[fermion] (-1.7,0) -- (-0.3,0);
\draw[fermionbar] (2.5,0) -- (1,0);
\draw[vector] (-0.2,0) -- (1,0);
\draw[vector] (0.4,-1.3) -- (0.4,-2.5);
\draw (-0.2,0)  node [vertex]  {};
\draw (1,0)  node [vertex]  {};
\draw (0.4,-1.3)  node [vertex]  {};
\draw (-1.5,0.3) node {\scriptsize $b$};
\draw (2.3,0.3) node {\scriptsize $s$};
\draw (0.4,0.3) node {\scriptsize $A,G,W,Z$};
\draw (1,-2.3) node {\scriptsize $\ga,g$};
\draw[fermionbar] (1,0) arc (40:-90:0.8);
\draw[fermionbar] (0.4,-1.3) arc (-90:-220:0.8);
\end{scope}
\begin{scope}[xshift=12cm]
\draw (-2.4,0.3) node {\footnotesize (IIb)};
\draw[fermion] (-1.7,0) -- (-0.3,0);
\draw[fermionbar] (2.5,0) -- (1,0);
\draw[scalar] (-0.2,0) -- (1,0);
\draw[vector] (0.4,-1.3) -- (0.4,-2.5);
\draw (-0.2,0)  node [vertex]  {};
\draw (1,0)  node [vertex]  {};
\draw (0.4,-1.3)  node [vertex]  {};
\draw (-1.5,0.3) node {\scriptsize $b$};
\draw (2.3,0.3) node {\scriptsize $s$};
\draw (0.4,0.3) node {\scriptsize $A,G,W,Z$};
\draw (0.9,-2.3) node {\scriptsize $\ga,g$};
\draw[fermionbar] (1,0) arc (40:-90:0.8);
\draw[fermionbar] (0.4,-1.3) arc (-90:-220:0.8);
\end{scope}
\begin{scope}[xshift=0cm,yshift = -3.5cm]
\draw (-2.4,0.3) node {\footnotesize (IIIa)};
\draw[fermion] (-1.7,0) -- (-0.3,0);
\draw[fermionbar] (2.5,0) -- (1,0);
\draw[fermion] (-0.2,0) -- (1,0);
\draw[vector] (0.4,-1.3) -- (0.4,-2.5);
\draw (-0.2,0)  node [vertex]  {};
\draw (1,0)  node [vertex]  {};
\draw (0.4,-1.3)  node [vertex]  {};
\draw (-1.5,0.3) node {\scriptsize $b$};
\draw (2.3,0.3) node {\scriptsize $s$};
\draw (0.4,0.3) node {\scriptsize $$};
\draw (1.7,-0.6) node {\scriptsize $W$};
\draw (-0.9,-0.6) node {\scriptsize $W$};
\draw (0.8,-2.3) node {\scriptsize $\ga$};
\draw[vector] (1,0) arc (40:-220:0.8);
\end{scope}
\begin{scope}[xshift=6cm,yshift = -3.5cm]
\draw (-2.4,0.3) node {\footnotesize (IIIb)};
\draw[fermion] (-1.7,0) -- (-0.3,0);
\draw[fermionbar] (2.5,0) -- (1,0);
\draw[fermion] (-0.2,0) -- (1,0);
\draw[vector] (0.4,-1.3) -- (0.4,-2.5);
\draw (-0.2,0)  node [vertex]  {};
\draw (1,0)  node [vertex]  {};
\draw (0.4,-1.3)  node [vertex]  {};
\draw (-1.5,0.3) node {\scriptsize $b$};
\draw (2.3,0.3) node {\scriptsize $s$};
\draw (0.4,0.3) node {\scriptsize $$};
\draw (1.7,-0.6) node {\scriptsize $W$};
\draw (-0.9,-0.6) node {\scriptsize $W$};
\draw (0.8,-2.3) node {\scriptsize $\ga$};
\draw[vector] (1,0) arc (40:-90:0.8);
\draw[scalar] (0.4,-1.3) arc (-90:-220:0.8);
\end{scope}
\begin{scope}[xshift=12cm,yshift = -3.5cm]
\draw (-2.4,0.3) node {\footnotesize (IIIc)};
\draw[fermion] (-1.7,0) -- (-0.3,0);
\draw[fermionbar] (2.5,0) -- (1,0);
\draw[fermion] (-0.2,0) -- (1,0);
\draw[vector] (0.4,-1.3) -- (0.4,-2.5);
\draw (-0.2,0)  node [vertex]  {};
\draw (1,0)  node [vertex]  {};
\draw (0.4,-1.3)  node [vertex]  {};
\draw (-1.5,0.3) node {\scriptsize $b$};
\draw (2.3,0.3) node {\scriptsize $s$};
\draw (0.4,0.3) node {\scriptsize $$};
\draw (1.7,-0.6) node {\scriptsize $W$};
\draw (-0.9,-0.6) node {\scriptsize $W$};
\draw (0.8,-2.3) node {\scriptsize $\ga$};
\draw[scalar] (1,0) arc (40:-90:0.8);
\draw[vector] (0.4,-1.3) arc (-90:-220:0.8);
\end{scope}
\begin{scope}[xshift=18cm,yshift = -3.5cm]
\draw (-2.4,0.3) node {\footnotesize (IIId)};
\draw[fermion] (-1.7,0) -- (-0.3,0);
\draw[fermionbar] (2.5,0) -- (1,0);
\draw[fermion] (-0.2,0) -- (1,0);
\draw[vector] (0.4,-1.3) -- (0.4,-2.5);
\draw (-0.2,0)  node [vertex]  {};
\draw (1,0)  node [vertex]  {};
\draw (0.4,-1.3)  node [vertex]  {};
\draw (-1.5,0.3) node {\scriptsize $b$};
\draw (2.3,0.3) node {\scriptsize $s$};
\draw (0.4,0.3) node {\scriptsize $$};
\draw (1.7,-0.6) node {\scriptsize $W$};
\draw (-0.9,-0.6) node {\scriptsize $W$};
\draw (0.8,-2.3) node {\scriptsize $\ga$};
\draw[scalar] (1,0) arc (40:-220:0.8);
\end{scope}
\begin{scope}[yshift=-7cm]
\begin{scope}[xshift=0cm,yshift = 0]
\draw (-2.4,0.3) node {\footnotesize (IVa)};
\draw[fermion] (-1.7,0) -- (-0.3,0);
\draw[fermionbar] (2.5,0) -- (1,0);
\draw[fermion] (-0.2,0) -- (1,0);
\draw[gluon] (0.4,-1.3) -- (0.4,-2.5);
\draw (-0.2,0)  node [vertex]  {};
\draw (1,0)  node [vertex]  {};
\draw (0.4,-1.3)  node [vertex]  {};
\draw (-1.5,0.3) node {\scriptsize $b$};
\draw (2.3,0.3) node {\scriptsize $s$};
\draw (0.4,0.3) node {\scriptsize $$};
\draw (1.7,-0.6) node {\scriptsize $G$};
\draw (-0.9,-0.6) node {\scriptsize $G$};
\draw (0.8,-2.3) node {\scriptsize $g$};
\draw[gluon] (1,0) arc (40:-220:0.8);
\end{scope}
\begin{scope}[xshift=6cm,yshift = 0]
\draw (-2.4,0.3) node {\footnotesize (IVb)};
\draw[fermion] (-1.7,0) -- (-0.3,0);
\draw[fermionbar] (2.5,0) -- (1,0);
\draw[fermion] (-0.2,0) -- (1,0);
\draw[gluon] (0.4,-1.3) -- (0.4,-2.5);
\draw (-0.2,0)  node [vertex]  {};
\draw (1,0)  node [vertex]  {};
\draw (0.4,-1.3)  node [vertex]  {};
\draw (-1.5,0.3) node {\scriptsize $b$};
\draw (2.3,0.3) node {\scriptsize $s$};
\draw (0.4,0.3) node {\scriptsize $$};
\draw (1.7,-0.6) node {\scriptsize $G$};
\draw (-0.9,-0.6) node {\scriptsize $G$};
\draw (0.8,-2.3) node {\scriptsize $g$};
\draw[gluon] (1,0) arc (40:-90:0.8);
\draw[scalar] (0.4,-1.3) arc (-90:-220:0.8);
\end{scope}
\begin{scope}[xshift=12cm,yshift = 0]
\draw (-2.4,0.3) node {\footnotesize (IVc)};
\draw[fermion] (-1.7,0) -- (-0.3,0);
\draw[fermionbar] (2.5,0) -- (1,0);
\draw[fermion] (-0.2,0) -- (1,0);
\draw[gluon] (0.4,-1.3) -- (0.4,-2.5);
\draw (-0.2,0)  node [vertex]  {};
\draw (1,0)  node [vertex]  {};
\draw (0.4,-1.3)  node [vertex]  {};
\draw (-1.5,0.3) node {\scriptsize $b$};
\draw (2.3,0.3) node {\scriptsize $s$};
\draw (0.4,0.3) node {\scriptsize $$};
\draw (1.7,-0.6) node {\scriptsize $G$};
\draw (-0.9,-0.6) node {\scriptsize $G$};
\draw (0.8,-2.3) node {\scriptsize $g$};
\draw[scalar] (1,0) arc (40:-90:0.8);
\draw[gluon] (0.4,-1.3) arc (-90:-220:0.8);
\end{scope}
\begin{scope}[xshift=18cm,yshift = 0]
\draw (-2.4,0.3) node {\footnotesize (IVd)};
\draw[fermion] (-1.7,0) -- (-0.3,0);
\draw[fermionbar] (2.5,0) -- (1,0);
\draw[fermion] (-0.2,0) -- (1,0);
\draw[gluon] (0.4,-1.3) -- (0.4,-2.5);
\draw (-0.2,0)  node [vertex]  {};
\draw (1,0)  node [vertex]  {};
\draw (0.4,-1.3)  node [vertex]  {};
\draw (-1.5,0.3) node {\scriptsize $b$};
\draw (2.3,0.3) node {\scriptsize $s$};
\draw (0.4,0.3) node {\scriptsize $$};
\draw (1.7,-0.6) node {\scriptsize $G$};
\draw (-0.9,-0.6) node {\scriptsize $G$};
\draw (0.8,-2.3) node {\scriptsize $g$};
\draw[scalar] (1,0) arc (40:-220:0.8);
\end{scope}
\end{scope}
\end{tikzpicture}
\parbox{15.5cm}
{\caption{\label{fig:C78}Diagrams contributing in the minimal RS model to the transitions $b\to s \ga$ and $b\to s g$ at the one-loop level. Solid lines denote the exchange of up- or down-type quarks while wavy or curled lines denote the exchange of (vector) gauge-bosons. Apart from diagram (I) a scalar (dashed) line includes the contribution from the fifth component of the gauge boson and the corresponding contribution from the Goldstone bosons in the Higgs sector.~The extra-dimensional coordinates of the vertices are labelled according to diagram (I).}}
\end{center}\vspace{-2mm}
\end{figure}

Like in the SM, the leading-order contributions to the $b\to s\ga$ and $b\to sg$ dipole coefficients in the RS model are loop suppressed, because there are no flavor-changing couplings that can induce a chirality flip.~But in contrast to the SM there are more one-loop diagrams to be considered.~Besides the additional exchange of KK $W^\pm$ bosons, new topologies appear due to the flavor-changing couplings of the Higgs boson, the $Z$ boson and its KK modes, and the photon and gluon KK modes. Figure~\ref{fig:C78} shows all relevant Feynman diagrams contributing in a general $R_\xi$ gauge. Internal scalar lines of the diagrams (II), (III) and (IV) include the contributions from the scalar component of the 5D gauge bosons and from the corresponding Goldstone bosons in the Higgs sector. In this section the Wilson coefficients $C_{7\ga,8g}$ and $\tilde C_{7\ga,8g}$ are defined via the general parametrization of the transition amplitude
\begin{align}\label{eqn:AC7ga}
   \A_{7\ga,8g} = i\frac{G_F}{\sqrt2}\,\lambda_t 
    \left[ C_{7\ga,8g}\,\langle s\ga|Q_{7\ga,8g}|b\rangle 
    + \tilde C_{7\ga,8g}\,\langle s\ga|\tilde Q_{7\ga,8g}|b\rangle \right] ,
\end{align}
where $G_F$ is the Fermi constant, and $\lambda_t\equiv V_{ts}^* V_{tb}$ is the relevant product of entries of the CKM matrix. The matrix elements in \eqref{eqn:AC7ga} are given by $\langle Q_{7\ga}\rangle=(e m_b/2\pi^2)\,\e^{*\mu}(q)\,\bar u(p_s)\,i\sigma_{\mu\nu}\,q^\nu P_R\,u(p_b)$ and $\langle Q_{8g}\rangle=(g_s m_b/2\pi^2)\,\e^{*\mu}(q)\,\bar u(p_s)\,i\sigma_{\mu\nu}\,q^\nu P_R\,u(p_b)$, where the outgoing photon (gluon) momentum is $q=p_b-p_s$. The chirality-flipped matrix elements $\langle\tilde Q_{7\ga,8g}\rangle$ are given by analogous expressions with $P_R\to P_L$. Working in Feynman-'t Hooft gauge ($\xi=1$), we compute each amplitude in Figure~\ref{fig:C78} using the Feynman rules of the 5D theory collected in Appendix~\ref{app:FR}. 

As an example we consider the penguin diagram (IIa) in Figure~\ref{fig:C78}, in which a 5D $W^\pm$-boson propagator and two 5D quark propagators arise. The corresponding amplitude with an external photon is given by
\begin{align}\label{eqn:AmpW}\notag
\A^{W,\tx{vector}}_{7\ga} &= \frac{4\pi Q_u e_5 g_5^2}{(2\pi r)^{3/2}}\int\!\frac{d^4k}{(2\pi)^4} \! \int_\e^1\!dt dt' dt''\,\e_{\mu}^*(q)\,D_{W,\al\beta}^{\xi=1}(t',t;k)\,\bar u(p_s) \Big[ \mathcal{D}_L^{(2)\da}(t) P_R\! +\! \mathcal{D}_R^{(2)\da}(t) P_L \Big] \\*
&\quad\mbox{}\hspace{0mm}\times\!\Pb_W \gamma^\alpha \Sb^u(t,t'';p_s\!-\!k)\ga^\mu \Sb^u(t'',t';p_b\!-\!k)\ga^\beta \Pb_W\! \Big[ \mathcal{D}_L^{(3)}(t') P_L \!+\! \mathcal{D}_R^{(3)}(t') P_R\Big]u(p_b) \,,
\end{align}
where $Q_u=2/3$ is the electric charge of the exchanged up-type quarks in the loop. The functions $\mathcal{D}_A^{(2)}(t)$ and $\mathcal{D}_A^{(3)}(t)$ with $A=L,R$ denote the profiles of the physical strange- and bottom-quark mass eigenstates, respectively, as defined in (\ref{KK-decomp}). In the above equation $e_5$ is the 5D electromagnetic coupling, while $g_5$ represents the 5D $SU(2)_L$ gauge coupling. The 4D electromagnetic coupling can be obtained by $e=e_5/\sqrt{2\pi r}$. The $2\times2$ matrix $\Pb_W\equiv\Pb_+=\tx{diag(1,0)}$ originates from the 5D Feynman rule for the $W_\mu^{+}\bar U_A^{} D_A^{}$ vertices (with $A=L,R$) in \eqref{eqn:ABmuFR} and projects out the profiles of the $SU(2)_L$ doublet quark fields. The 5D $W^\pm$-boson propagator in \eqref{eqn:AmpW} can be decomposed as
\begin{align}\label{eqn:5DPropKK}
\begin{split}
   D^\xi_{W,\mu\nu}(t,t';k) 
   &= B_W(t,t';-k^2-i0) \left( \eta_{\mu\nu}-\frac{k_\mu k_\nu}{k^2} \right) 
    + B_W(t,t';-k^2/\xi-i0)\,\frac{k_\mu k_\nu}{k^2} \,,
\end{split}
\end{align}
with the KK representation
\begin{align}\label{eqn:BBKKrep}
   B_W(t,t';-k^2) = \sum_{n}\,\frac{\chi^W_n(t)\,\chi^W_n(t')}{m_{W_n}^2-k^2-i0} \,.
\end{align}
The propagator function \eqref{eqn:BBKKrep} can be calculated in closed form, see \cite{Hahn:2013nza} for more details on the derivation and the solution. The 5D quark propagators in \eqref{eqn:AmpW} can be decomposed into four functions differing in chirality and Lorentz structure \cite{Randall:2001gb,Puchwein:2003jq,Contino:2004vy,Carena:2004zn,Csaki:2010aj},
\begin{equation}\label{eqn:Sprop}
\begin{split}
i\boldsymbol{S}^{q}(t,t^\prime;k) &=\left[\bm{\Delta}_{LL}^{q} (t,t^\prime;-k^2) \, \slashed k  +  \bm{\Delta}_{RL}^{q} (t,t^\prime;-k^2) \right]\, P_R + (L \leftrightarrow R)\,,
\end{split} 
\end{equation}
where $q=u,d$. The KK representations of the propagator functions read
\begin{align}\label{eqn:KK delta}
\begin{split}
   \bm{\Delta}_{LL}^{q} (t,t^\prime;-k^2) 
   &= \sum_n\,\frac{1}{k^2-m_{q_n}^2}\,\Q_L^{(n)}(t)\,\Q_L^{(n)\dagger}(t^\prime) \,, \\
   \bm{\Delta}_{RL}^{q} (t,t^\prime;-k^2) 
   &= \sum_n\,\frac{m_{q_n}}{k^2-m_{q_n}^2}\,\Q_R^{(n)}(t)\,\Q_L^{(n)\dagger}(t^\prime) \,,
\end{split}
\end{align}
and analogously for $\bs\Delta_{RR}^q$ and $\bs\Delta_{LR}^q$. Each propagator function is a $6\times 6$ matrix. The subscripts denote the handedness of the incoming and outgoing quark fields, such that the propagator function $\bs\Delta_{RL}^q$ implies a chirality flip.~Explicit expressions for these functions are given in Appendix~\ref{app:5DFermionProp} for the brane-localized Higgs scenario and the case of a narrow bulk-Higgs.

Next we outline some of the basic steps needed to extract the dipole coefficients from the diagrams in Figure \ref{fig:C78}: 

\begin{itemize}
\item We perform a Taylor expansion of each 5D propagator in the external momenta $p_s,p_b$ and keep the terms up to second order, since higher orders would contribute to higher-dimensional operators and yield suppressed contributions. For instance, for a 5D quark propagator function we apply the expansion ($q=u,d$)
\begin{align}\label{eqn:PropExp}
   \bs\Delta_{AB}^q(t,t';-(p_i-k)^2) 
   = \bigg[1 - 2 (p_i\cdot k)\,\frac{\pa}{\pa k^2} + 2\,(p_i\cdot k)^2
    \left( \frac{\pa}{\pa k^2}\right)^2 \mp \dots \bigg]\,\bs\Delta_{AB}^q(t,t';-k^2) \,,
\end{align}
where $p_i=p_{s,b}$, $k$ is the loop momentum, and $A,B\in\{L,R\}$.~We need to expand up to second order in the external momenta in order to obtain the leading effects of the dipole Wilson coefficients, since the matrix elements of the dipole operators contain the bottom mass $m_b$ and the momentum difference $q=p_b-p_s$.~In fact, the term linear in $p_i$  in \eqref{eqn:PropExp} contributes only in the RS model, and not in the SM, to the dipole Wilson coefficients.~Analogously we can expand the 5D vector-boson propagator function $B_B(t,t';-(p_i-k)^2)$ with subscript $B=A,G,W,Z$. 
\item The extra-dimensional integration of the vertex with the external photon or gluon can be performed analytically by using the flatness of their profiles $\chi_0^{A,G}=1/\sqrt{2\pi}$ and the orthonormality conditions for fermion and boson profiles
\begin{align}\label{eqn:OrthFermBos}
   \int_\e^1 dt\,\Q^{(n)}_A(t)\,\Q^{(n')\dagger}_A(t) = \delta_{nn'}\,\bm{1}_{6\times6} \,, \qquad
   \frac{2\pi}{L} \int_\e^1 \frac{dt}{t}\,\chi_n^B(t)\,\chi_{n'}^B(t) = \delta_{nn'} \,,
\end{align}
where $\Q=\U,\D$ and $A=L,R$ in the left equation and $B=A,G,W,Z$ on the right. 
\item The previous two bullets allow us to combine two 5D propagators of the same type when we expand them in the external momenta $p_s,p_b$ and perform the extra-dimensional integration of the vertex that couples to the external photon or gluon. For instance we can apply ($q=u,d$)
\begin{align}\label{eqn:Comb5DProps}
\begin{split}
   &\int_\e^1 dt''\,\bs\Delta_{RR}^q(t,t'';-(p_s-k)^2)\,\bs\Delta_{RL}^q(t'',t';-(p_b-k)^2) \\
   &\quad= \bigg\{ - \frac{\pa}{\pa{k^2}} + k\cdot(p_s + p_b) \left( \frac{\pa}{\pa{k^2}} \right)^2 \\
   &\quad\qquad\mbox{}- \frac{2}{3} \Big[ k\cdot p_s\,k \cdot p_b + (p_s\cdot k)^2 + (p_b\cdot k)^2 \Big]  
    \left( \frac{\pa}{\pa{k^2}} \right)^3 + \dots \bigg\}\,\bs\Delta_{RL}^q(t,t';-k^2) \,,
\end{split}
\end{align}
where we neglect terms of order $(k\cdot p_{s})^n (k\cdot p_{b})^{n'} $ with $n+n' \geq 3$. Analogous relations can be derived for products of different fermion and boson propagator functions. Equation~\eqref{eqn:Comb5DProps} can be used to reduce each amplitude by one extra-dimensional integration and one 5D propagator. 
\item We perform a Wick rotation to Euclidean momenta with $k^0=ik_E^0$ and $k_E=\sqrt{-k^2}$.
\item For the matching procedure on the dipole operators we first use that the photon or gluon is on-shell, $q_\mu\e^{\mu*}(q)=0$, which allows us to rewrite $p_{s,b}^\mu \e_\mu^*(q) = \frac{1}{2}(p_b+p_s)^\mu \e_\mu^*(q)$.~Then, we can use the Dirac equation $\s p_b u(p_b)=m_b \,u(p_b)$ and apply the Gordon identity
\begin{align}
   \bar u(p_s)\,i\sigma^{\mu\nu} q_\nu P_{L,R}\,u(p_b) 
   = \bar u(p_s)\,\Big[ (p_s + p_b)^\mu P_{L,R} 
    - \ga^\mu \left( m_s P_{L,R}  + m_b P_{R,L} \right) \Big]\,u(p_b)\,,
\end{align}
in order to extract the Wilson coefficients. 
\end{itemize}
In the following three subsections we discuss the gauge-invariant subsets of the diagrams shown in Figure \ref{fig:C78}.

\subsection{Higgs contribution}
\label{sec:HiggsContrbsga}

We begin with the first diagram (I) in Figure~\ref{fig:C78}, in which the Higgs boson and two down-type 5D quark propagators are exchanged. The Yukawa interactions of the Higgs boson with two down-type quarks are given by 
\begin{align}\label{eqn:Lahqq}
   \La_{hdd}(x) = - \int_{\e}^{1}\!dt\,\delta^\eta(t-1)\,\frac{h(x)}{\sqrt2}
    \left[ \bar D_L(x,t)\,\bs{Y}_d\,d_R(x,t) + \bar D_R(x,t)\,\Yb_d\,d_L(x,t) + \hc \right] ,
\end{align}
where the first term is often referred to as the ``correct-chirality Higgs coupling'' in the literature, since it is also present in the SM. On the other hand, the second term couples a right-handed $SU(2)_L$ doublet quark field to a left-handed $SU(2)_L$ singlet, which is not allowed in the SM and is thus called the ``wrong-chirality Higgs coupling''. The function $\delta^\eta(t-1)$ denotes the normalized Higgs profile along the extra dimension, which we take to be the regularized $\delta$-function. For the calculations we use a square box of width $\eta$ and height $1/\eta$, such that 
\begin{align}\label{eqn:deltaReg}
   \delta^\eta(t-1)\to \frac{1}{\eta}\,\theta(t-1+\eta) \,, \qquad \tx{with} \qquad 
   \eta\ll\frac{y_\ast\,v}{\MKK} \,,
\end{align}
where $\eta$ is related to the inverse Higgs width by $\Delta_h^{-1}\sim\eta/v$. The brane-localized Higgs scenario corresponds to values of $\eta\ll y_\ast\,v /\Lambda_\TeV$, while the narrow bulk-Higgs scenario implies values in the range $y_\ast\,v/\Lambda_\TeV\ll\eta \ll y_\ast\,v/\MKK$. Note that the shape of the regularized profile is irrelevant as long as $\eta\ll 1$.

With the Feynman rules in Appendix~\ref{app:FR} and the basic steps outlined in the beginning of this section we can derive an expression for the Wilson coefficient and find
\begin{align}\label{eqn:C785Dqhq}
\begin{split}
C^h_{7\ga,8g} &= \frac{\kappa_h^{7\ga,8g}}{4G_F\lambda_t}\, \frac{1}{v}
\int_0^\infty\!\frac{dk_E}{k_E^2+m_h^2}\,\bigg[ 
\left( \frac{k_E^2}{8}\,\pa_{k_E} - \frac{k_E^3}{8}\,\pa_{k_E}^2 \right)
\frac{T^d_{RL}(k_E^2)}{m_b} \\
&\hspace{47mm}\mbox{}+ \left( \frac{k_E^2}{32}\,\pa_{k_E} - \frac{k_E^3}{32}\,\pa_{k_E}^2 
- \frac{k_E^4}{96}\,\pa_{k_E}^3 \right)\frac{T^d_{RR}(k_E^2)}{\MKK}\bigg] \,,
\end{split}
\end{align}
where $m_h$ is the Higgs mass.~Concerning the derivatives we use the notation $\pa_{k_E} \equiv \pa/\pa{k_E}$.~Due to the parametrization of the amplitude in \eqref{eqn:AC7ga} we have to divide the dipole coefficient by $\lambda_t$ and $G_F$.~The couplings are given by $\kappa_h^{7\ga}  = Q_d $ and $\kappa_h^{8g} = 1$, where $Q_d=-1/3$ is the electric charge of the exchanged down-type quarks. The dimensionless propagator functions in \eqref{eqn:C785Dqhq} are defined via
\begin{align}\label{eqn:TdRRRL}
\begin{split}
   T^d_{RL}(k_E^2) &= \frac{-v}{\sqrt2} \int_\e^1\!dtdt'\,\delta^\eta(t-1)\,\delta^\eta(t'-1)\, 
    \D_L^{(2)\da}(t)\,\M_d^Y\,\bs\Delta_{RL}^d(t,t';k_E^2)\,\M_d^Y\,\D_R^{(3)}(t') \,, \\
   T^d_{RR}(k_E^2) &= \frac{-v\MKK}{\sqrt2} \int_\e^1\!dtdt'\, \delta^\eta(t-1)\,\delta^\eta(t'-1)\,\D_L^{(2)\da}(t)\,\M_d^Y\,\bs\Delta_{RR}^d(t,t';k_E^2)\,\M_d^{Y\da}\D_L^{(3)}(t') \,, 
\end{split}
\end{align}
including the regularized $\delta$-functions \eqref{eqn:deltaReg} and the matrix $\M_d^Y=\bs P_{12}\,\bs Y_d+\bs P_{21}\,\bs Y_d^{\da}$. The projector $\Pb_{ij}$ for $i,j=1,2$ is a $2\times2$ matrix with zero entries except for the $ij$-component, which equals 1. In order to perform the integrations over $t$ and $t'$ we need the solutions for the external quark profiles and the 5D quark propagators in the region $t,t'\in[1-\eta,1]$. 

The presence of the $\delta$-function regulator \eqref{eqn:deltaReg} implies that in the region near the IR brane, for $t\in[1-\eta,1]$, the quark profiles are determined by the coupled differential equations (with $\Q=\U,\D$ and $q=u,d$) \cite{Casagrande:2010si}
\begin{align}\label{eqn:FermProfDiff}
\begin{split}
   \left( \pa_t + \frac{1}{t}\,\M_q^c + \frac{\varrho}{\eta}\,\M_q^Y \right) \Q_{R}^{(n)}(t) 
   &= x_{q_n} \Q_L^{(n)}(t) \,, \\
   \left( \pa_t - \frac{1}{t}\,\M_q^c - \frac{\varrho}{\eta}\,\M_q^{Y} \right) \Q_{L}^{(n)}(t) 
   &= - x_{q_n} \Q_R^{(n)}(t) \,,
\end{split}
\end{align}
with $x_{q_n}\equiv m_{q_n}/\MKK$ and $\rh \equiv v/(\sqrt2 \MKK)$.~These equations can be simplified for the considered limit $\eta\ll y_\ast\,v/\MKK$. First, the term $\M_q^c=\Pb_{+}\,\bs c_Q-\Pb_{-}\,\bs c_q$, which contains the bulk mass parameters $\bs c_Q$ and $\bs c_q$, is parametrically suppressed for $\eta\,c_{Q_i,d_i}\ll y_\ast\rh$ and can therefore be neglected. The projector $\Pb_-$ projects on the lower components and is given by $\Pb_{-} = \tx{diag}(0,1)$. Secondly, the mass-dependent terms on the right side of \eqref{eqn:FermProfDiff} are suppressed for the SM quarks, since $\eta x_{q_n}\ll y_\ast\rh $. With these approximations, the basic solutions are given by the trigonometric functions (for $q=u,d$)
\begin{align}\label{eqn:SqtCqt}
   {\cal S}_{}(t) = \sinh\frac{\bs X_{q}\,(1-t)}{\eta} \,, \qquad 
   {\cal C}_{}(t) = \cosh\frac{\bs X_{q}\,(1-t)}{\eta} \,,
\end{align}
with the hermitian matrix $\Xb_q=\rh(\Yb_q\Yb_q^\da)^{1/2}$. The basic ansatz for the solution consists of four unknown coefficients each for $\Q_L^{(n)}(t)$ and $\Q_R^{(n)}(t)$. We can fix two coefficients by implementing the Dirichlet boundary conditions $(0\;\;1)\Q_L^{(n)}(1)=0$ and $(1\;\;0)\Q_R^{(n)}(1)=0$ for the orbifold-odd profiles on the IR brane. Two more coefficients can be eliminated by imposing the Neumann boundary conditions $(\pa_t\;\;0)\Q_L^{(n)}(t)|_{t=1}=0$ and $(0\;\;\pa_t)\Q_R^{(n)}(t)|_{t=1}=0$ for the orbifold-even profiles. Finally, the solutions for $t\in[1-\eta,1]$ are given by ($\Q=\U,\D$) 
\begin{align}\label{eqn:QLRsliver}
   \Q_L^{(n)}(t) = \begin{pmatrix} \frac{\mathcal{C}_{}(t)}{\mathcal{C}_{}(1_\eta)} & 0 \\ 
    0 & \frac{\bar{\cal S}_{}(t)}{\bar{\cal S}(1_\eta)}  \end{pmatrix} \Q_L^{(n)}(1_\eta) \,, 
   \qquad
  \Q_R^{(n)}(t) = \begin{pmatrix} \frac{\mathcal{S}(t)}{\mathcal{S}(1_\eta)} & 0 \\ 
   0 & \frac{\bar{\cal C}(t)}{\bar{\cal C}(1_\eta)} \end{pmatrix} \Q_R^{(n)}(1_\eta) \,,
\end{align}
where we use the short-hand notation $1_\eta\equiv 1-\eta$. The functions $\bar {\cal S}(t),\bar {\cal C}(t)$ are given by \eqref{eqn:SqtCqt} with $\Xb_q$ replaced by $\bar\Xb_q=\rh(\Yb_q^\da\Yb_q)^{1/2}$.

The derivation of the 5D quark propagator in the region near the IR brane has been discussed in detail in \cite{Malm:2013jia}. Appendix~\ref{app:5DFermionProp} contains all solutions that are relevant for the present work. Here we just comment that the basic solutions for the 5D propagator functions $\bs\Delta_{AB}^q(t,t';k_E^2)$ with $A,B\in\{L,R\}$ for $t,t'\in[1-\eta,1]$ are given in terms of the trigonometric functions ${\cal C}(t)$ and ${\cal S}(t)$ in \eqref{eqn:SqtCqt} with $\Xb_q$ replaced by \begin{align}\label{eqn:Sq}
   \Sb_q = \sqrt{\Xb_q^2 + \eta^2 \hat k_E^2} \,,
\end{align}
where $\hat k_E=k_E/\MKK$ is the Euclidean momentum normalized to the KK scale. The $\eta$ dependence of the propagator enters only via the product $\eta\kEh$. As we will see below, this leads to a different  behavior of the propagator depending on whether $\eta\kEh\ll y_\ast\rh$ or $\eta\kEh\gg y_\ast \rh$.

\subsubsection*{Calculation of the propagator functions $\bs{T_{RL}^d(k_E^2)}$ and $\bs{T_{RR}^d(k_E^2)}$}

It is instructive to discuss the calculation of the function $T_{RL}^d(k_E^2)$ in more detail, since it exhibits a sensitivity on the regulator $\eta$, which is similar to that observed in the calculation of the loop-induced Higgs coupling to two gluons \cite{Djouadi:2007fm,Falkowski:2007hz,Cacciapaglia:2009ky,Bhattacharyya:2009nb,Bouchart:2009vq,Casagrande:2010si,Azatov:2010pf,Azatov:2011qy,Goertz:2011hj,Carena:2012fk,Malm:2013jia,Archer:2014jca}. Applying the $\delta$-function regulator \eqref{eqn:deltaReg} and inserting the solutions for the external quark profiles \eqref{eqn:QLRsliver} into \eqref{eqn:TdRRRL}, we obtain
\begin{align}\label{eqn:TqRL1}\notag
%\begin{split}
   T_{RL}^{d}(k_E^2) &= \frac{-v}{\sqrt2} \D_L^{(2)\da}(1_\eta)\!
    \int_{1_\eta}^1\!\frac{dt dt'}{\eta^2} \!\Bigg[\!
    \left(\! \frac{\Cm(t)}{\Cm(1_\eta)} \Yb_d\,\bs\Delta_{RL}^{d,21}(t,t';k_E^2) 
    - \frac{\Xb_d}{\rh} \frac{\Sm(t)}{\Cm(1_\eta)} \bs\Delta_{RL}^{d,11}(t,t';k_E^2) \!\right) 
   \!\frac{\Cm(t')}{\Cm(1_\eta)}\Yb_d \\*
   &\quad\mbox{}+\!\left(\!\frac{\Cm(t)}{\Cm(1_\eta)}\Yb_d\,\bs\Delta_{RL}^{d,22}(t,t';k_E^2) -\frac{\Xb_d}{\rh}\frac{\Sm(t)}{\Cm(1_\eta)}\bs\Delta_{RL}^{d,12}(t,t';k_E^2)\!\right)
    \!\frac{\bar\Xb_d}{\rh}\frac{\bar\Sm(t')}{\bar\Cm(1_\eta)} \!\Bigg]\!\bs P_{12}\D_R^{(3)}(1_\eta)\,,
%\end{split}
\end{align} 
where ${\cal C}(t),{\cal S}(t)$ are defined in \eqref{eqn:SqtCqt} with $q=d$. The propagator functions in the region near the IR brane for $t,t'\in[1_\eta,1]$ can be found in Appendix~\ref{app:5DFermionProp}. 

We are not interested in the full dependence of $T_{RL}^d(k_E^2)$ on $\eta$, since in the end of the calculation we will always remove the regulator ($\eta\to 0$). However, since $T_{RL}^d(k_E^2)$ depends on the product $\eta \hat k_E$ via the 5D quark propagator functions and we integrate the function in \eqref{eqn:C785Dqhq} from zero to infinite Euclidean momentum, we have to investigate whether the momentum integration commutes with the limit $\eta\to 0$. If we implement a momentum cutoff $k_E\leq \Lambda_\tx{cut}$ for the integral, the question can be reformulated as whether \eqref{eqn:C785Dqhq} yields the same results when imposing the constraints $\eta\gg y_\ast v/\Lambda_\tx{cut}$ or $\eta\ll y_\ast v/\Lambda_\tx{cut}$.~Thus we need to investigate the ultra-violet (UV) behavior of $T^d_{RL}(k_E^2)$ for large Euclidean momenta near the cutoff $k_E\sim\Lambda_\tx{cut}$.

Let us begin with the first scenario $\eta \gg y_\ast v/\Lambda_\tx{cut}$, where $\eta$ is bounded from below. In fact, we also have to impose an upper bound $\eta\ll y_\ast v/\MKK$, which is required in order to find reliable solutions for the 5D propagator functions in the region $t,t'\in[1_\eta,1]$ \cite{Malm:2013jia}. When we consider large Euclidean momenta near the UV cutoff ($k_E\sim\Lambda_\tx{cut}$), the allowed range of $\eta$ implies the hierarchy $\kEh\gg y_\ast\rh/\eta$.~Consequently, the function $\Sb_d=(\Xb_d^2+\eta^2\hat k_E^2)^{1/2}$, which is contained in the 5D propagator solutions, becomes approximately independent of the Yukawa-dependent term, such that $\Sb_d\approx\eta\kEh$. In this limit, we find that ($\eta\hat k_E\gg y_\ast \rh$) 
\begin{align}\label{eqn:TRLnarrbulk}
   T_{RL}^d(k_E^2)\sim (\eta\kEh)^{-3}
\end{align}
falls off with the third inverse power of the product $\eta\hat k_E$.~An analogous analysis for $T_{RR}^d(k_E^2)$ shows that it exhibits the same behavior as in \eqref{eqn:TRLnarrbulk}. Since the imposed cutoff can be identified with the effective UV cutoff of the theory near the IR brane, $\Lambda_\tx{cut}\approx\Lambda_\TeV$, the behavior in \eqref{eqn:TRLnarrbulk} refers to the case of a narrow bulk-Higgs scenario.

We continue with the second scenario, where the $\delta$-function regulator is bounded from above by $\eta\ll y_\ast v/\Lambda_\tx{cut}$.~This case represents the brane-localized Higgs scenario for $\Lambda_\tx{cut}\approx \Lambda_\TeV$.~Consequently, the product $\eta\kEh$ is much smaller than $y_\ast \rh$ implying that $\Sb_d$ in \eqref{eqn:Sq} becomes approximately independent of the regulator, $\bs S_d\approx \Xb_d$.~In this limit, we find $(\eta\kEh\ll y_\ast\rh)$
\begin{align}\label{eqn:TdRLbrane}
\begin{split}
   T_{RL}^d(k_E^2) &= \D_L^{(2)\da}(1_\eta) \int_{1_\eta}^1\!\frac{dtdt'}{\eta^2}\,\bigg[ 
    \bigg( \frac{2\Xb_d}{\sinh2\Xb_d}\,\frac{\Zb_d(k_E^2)}{1+\Zb_d(k_E^2)}\,
    \frac{\Cm(t')}{\Cm(1_\eta)} \\
   &\hspace{43mm}\mbox{}+ \frac{\Xb_d}{\coth\Xb_d}\,\frac{\Cm(t')}{\Cm(1_\eta)} - \theta(t-t')\,\Xb_d\,\frac{\Sm(t')}{\Cm(1_\eta)} \bigg)\,
    \frac{\Cm(t')}{\Cm(1_\eta)}\,\Yb_d \\
   &\hspace{20mm}\mbox{}- \bigg( \frac{2\Xb_d}{\sinh2\Xb_d}\,\frac{\Zb_d(k_E^2)}{1+\Zb_d(k_E^2)}\,
    \frac{\Sm(t')}{\Cm(1_\eta)} \\
   &\hspace{27mm}\mbox{}+ \frac{\Xb_d}{\coth\Xb_d}\,\frac{\Sm(t')}{\Cm(1_\eta)} 
    - \theta(t-t')\,\Xb_d\,\frac{\Cm(t')}{\Cm(1_\eta)} \bigg)\, 
    \frac{\Sm(t')}{\Cm(1_\eta)}\Yb_d \bigg]\,\bs P_{12}\,\D_R^{(3)}(1_\eta) \,.
\end{split}
\end{align}
Here we have introduced the structure \cite{Malm:2013jia} (for $q=u,d$)
\begin{align}\label{eqn:Zq}
   \Zm_{q}^{}(k^2_E) = \varrho^2\,\tilde{\Y}_{{q}}\,\Rmq{{{q}}}{\hat k_E}
    \tilde{\Y}^{\dagger}_{{q}} \Rmq{{{Q}}}{\kEh} \,, 
\end{align}
with the modified Yukawa matrix $\tilde\Yb_q\equiv (\tanh\Xb_q/\Xb_q) \Yb_q$. We further need the ratio \begin{equation}\label{eqn:RAdef}
   \bm{R}_A(\hat k_E) 
   = \frac{I_{-\bm{c}_A-\frac12}(\epsilon\hat k_E)\,I_{\bm{c}_A-\frac12}(\hat k_E)
           - I_{\bm{c}_A+\frac12}(\epsilon\hat k_E)\,I_{-\bm{c}_A+\frac12}(\hat k_E)}%
          {I_{-\bm{c}_A-\frac12}(\epsilon\hat k_E)\,I_{\bm{c}_A+\frac12}(\hat k_E)
           - I_{\bm{c}_A+\frac12}(\epsilon\hat k_E)\,I_{-\bm{c}_A-\frac12}(\hat k_E)} \,;
    \quad A=Q,q
\end{equation}
of modified Bessel functions, where $\bs c_{Q},\bs c_{u,d}$ are the bulk mass parameters of the 5D quark fields \cite{Grossman:1999ra,Gherghetta:2000qt}. In \eqref{eqn:TdRLbrane} we have not yet combined the terms inside the two round brackets.~But when combining them we find that the $t,t'$-dependence completely cancels and the $t,t'$ integrations become trivial, an analogous observation was made for the Higgs production process via gluon fusion in \cite{Malm:2013jia}. In addition the remaining $\eta$-dependence completely cancels. Finally, in the brane-localized Higgs scenario we find the results ($\eta \kEh \ll y_\ast \rh$)
\begin{align}\label{eqn:TRLRRbrane}
\begin{split}
   T^d_{RL}(k_E^2) &= \D_L^{(2)\da}(1^-)\,\left[
    \frac{2\Xb_d}{\sinh2\Xb_d}\,\frac{\Zb_d(k_E^2)}{1+\Zb_d(k_E^2)}\,\frac{2\Xb_d}{\sinh2\Xb_d}
    + \frac{\Xb_d^2}{\cosh^2\Xb_d} \right]\tilde\Yb_d\,\bs P_{12}\,\D_R^{(3)}(1^-) \,, \\
   T^d_{RR}(k_E^2) &= \D_L^{(2)\da}(1^-)\,\frac{1}{\kEh}\,\frac{2\Xb_d}{\sinh2\Xb_d}\,
    \frac{\Zb_d(k_E^2)}{1+\Zb_d(k_E^2)}\,\frac{1}{\bs R_Q(\kEh)}\,\frac{2\Xb_d}{\sinh2\Xb_d}\,
    \Pb_{+}\,\D_L^{(3)}(1^-) \,,
\end{split}
\end{align}
which are independent of the $\delta$-function regulator.~We have also included the final result for $T_{RR}^d(k_E^2)$, which can be obtained by an analogous calculation.~For large Euclidean momenta $k_E\gg\MKK$ the structure $\Zb_q(k_E^2)$ in \eqref{eqn:Zq} can be expanded as $\Zb_d(k_E^2)\approx\rh^2\tilde \Yb_d\,\tilde\Yb_d^\da+\ord(\kEh^{-2})$. We observe that $T_{RL}^d(k_E^2)$ reaches a non-zero plateau in this limit, which is in contrast with relation \eqref{eqn:TRLnarrbulk} valid in the narrow bulk-Higgs scenario. Consequently, the contribution of $T_{RL}^d(k_E^2)$ to the dipole coefficient \eqref{eqn:C785Dqhq} exhibits a dependence on the model under consideration. On the other hand, the function $T_{RR}^d(k_E^2)$ vanishes also in the narrow bulk-Higgs scenario and does not lead to a model-dependent contribution.

Interestingly we could have obtained the same results for $T_{RL}^d(k_E^2)$ and $T_{RR}^d(k_E^2)$ in \eqref{eqn:TRLRRbrane} if we had naively evaluated the extra-dimensional coordinates at $t=t'=1^-$ instead of using the regularized $\delta$-function in \eqref{eqn:TdRRRL}. We have explicitly confirmed that ($\eta\kEh\ll y_\ast\rh$)
\begin{align}
\begin{split}
   T^d_{RL}(k_E^2) &= - \frac{v}{\sqrt2}\,\D_L^{(2)\da}(1^-)\,\M_d^Y\,
    \bs\Delta_{RL}^d(1^-,1^-;k_E^2)\,\M_d^Y\,\D_R^{(3)}(1^-) \,, \\
   T^d_{RR}(k_E^2) &= - \frac{v\MKK}{\sqrt2}\,\D_L^{(2)\da}(1^-)\,\M_d^Y\,
    \bs\Delta_{RR}^d(1^-,1^-;k_E^2)\,\M_d^{Y\da}\,\D_L^{(3)}(1^-) \,,
\end{split}
\end{align}
lead to the results \eqref{eqn:TRLRRbrane}. An analogous situation was encountered for the calculation of the propagator functions $T_\pm(k_E^2)$ in the case of the Higgs production process via gluon fusion~\cite{Malm:2013jia}. 

\subsubsection*{Final result for the Wilson coefficient $\bs{C_{7\ga,8g}^{h}}$}

The above analysis shows that the integrand of the dipole coefficient in \eqref{eqn:C785Dqhq} falls off with at least two inverse powers of Euclidean momenta $\sim k_E^{-2}$ in the UV, which implies the finiteness of the integral. Thus, we are allowed to perform partial momentum integrations in \eqref{eqn:C785Dqhq} and find
\begin{align}\label{eqn:C7hga2}
C^h_{7\ga,8g} &= \frac{\kappa_h^{7\ga,8g}}{4G_F\lambda_t}\frac{1}{v}\!\left[ 
    \lim_{k_E\rightarrow\infty}\!\frac{T^d_{RL}(k_E^2)}{4m_b} 
    - \int_0^\infty\!\frac{dk_E\,k_E m_h^4}{(k_E^2+m_h^2)^3}\!\left(\frac{T^{d}_{RL}(k_E^2)}{m_b}+ \frac{k_E^2}{k_E^2+m_h^2} \frac{T^d_{RR}(k_E^2)}{2\MKK}\right)\!\right]\,,
\end{align}
where all boundary terms at $k_E^2=0$ vanish.~Based on the previous analysis only for large Euclidean momenta we can have a non-zero boundary term in case of the brane-localized Higgs scenario, where $T_{RL}^d(k_E^2)$ approaches a non-zero plateau.~This is accounted for by the first term in the outer bracket of \eqref{eqn:C7hga2}.~We can insert our results for $T_{RL}^d(k_E^2)$ and $T_{RR}^d(k_E^2)$ in \eqref{eqn:TRLRRbrane} into \eqref{eqn:C7hga2} and find
\begin{align}\label{eqn:C785Dqhq2}\notag
C^h_{7\ga,8g} = \frac{\kappa_h^{7\ga,8g} }{4G_F\lambda_t}\frac{1}{v}\,
\D_L^{(2)\da}(1^-)\,\Bigg[ & \bs P_{12}\,\frac{g(\Xb_d,\tilde\Yb_d)}{4 m_b}\,\D_R^{(3)}(1^-) 
- \frac{2\Xb_d}{\sinh2\Xb_d} \int_0^\infty\!\frac{dk_E\,k_E m_h^4}{(k_E^2+m_h^2)^3}\\*
&\mbox{}\times \bigg(\frac{\bs P_{12}}{m_b} \frac{\Zb_d(k_E^2)}{1+\Zb_d(k_E^2)}\frac{2\Xb_d}{\sinh2\Xb_d}\tilde\Yb_d\,\D_R^{(3)}(1^-) \\*\notag
&\hspace{8mm}\mbox{} + \frac{\bs P_{+}}{2}\frac{k_E}{k_E^2+m_h^2}\frac{\Zb_d(k_E^2)}{1+\Zb_d(k_E^2)} \frac{1}{\Rb_Q(\hat k_E)} \frac{2\Xb_d}{\sinh2\Xb_d} \D_L^{(3)}(1^-)\!\bigg)\!\Bigg] \,,
\end{align}
with 
\begin{align}\label{eqn:gfuncbsga}
\begin{split}
   g(\Xb_q,\tilde\Yb_q)\Big|_\tx{brane Higgs} 
   &= \frac{2\Xb_q}{\sinh 2\Xb_q}\,
    \frac{\rh^2\tilde\Yb_q\tilde\Yb_q^\da}{1+\rh^2\tilde\Yb_q\tilde\Yb_q^\da}\,
    \frac{2\Xb_q}{\sinh 2\Xb_q}\,\tilde\Yb_q 
    = + \rh^2\Yb_q\Yb_q^{\da}\Yb_q + {\cal O}(\rh^4) \,, \\
   g(\Xb_q,\tilde\Yb_q)\Big|_\tx{narrow bulk-Higgs} 
   &= - \frac{\Xb_q^2}{\cosh^2\Xb_q}\,\tilde \Yb_q 
    = - \rh^2\Yb_q\Yb_q^{\da}\Yb_q + {\cal O}(\rh^4) \,.
\end{split}
\end{align}
The function $g(\Xb_q,\tilde \Yb_q)$ is model dependent.~To leading order in $v^2/\MKK^2$ it only differs in the relative sign for a brane-localized and narrow bulk-Higgs.~A similar observation was made for the KK tower contribution in case of Higgs production via gluon fusion \cite{Malm:2013jia}.~We will see numerically in Section~\eqref{sec:AppWilsCoeffbsga} that this term emerges from the penguin diagrams exchanging KK quarks. Note that in this present paper we limit our analysis of the narrow bulk-Higgs model to the contributions involving the zero modes of the scalar doublet (see in particular Section~\ref{sec:NarrowBulk}). The contributions of scalar KK excitations have been studied in \cite{Agashe:2014jca,Beneke:2015lba}.

We can generalize the results obtained in the brane-localized Higgs sector by allowing for two different Yukawa matrices $\Yb_q^C$ and $\Yb_q^S$ associated with orbifold-even and -odd quark profiles \cite{Azatov:2010pf,Azatov:2009na}.~In other words, we associate the correct-chirality Higgs coupling with $\Yb_q^C$ and the wrong-chirality coupling with $\Yb_q^S$, i.e.\ we replace in the Lagrangian for the Higgs coupling to down-type quarks $\Yb_d\to\Yb_d^C$ in the first term and $\Yb_d\to\Yb_d^S$ in the second term of \eqref{eqn:Lahqq}. We will refer to this model as the ``type-II brane-Higgs'' scenario \cite{Malm:2013jia}. We find that our previous analysis still holds, provided we use  $\tilde\Yb_q=(\tanh\Xb_q/\Xb_q)\Yb_q^C$ for the modified Yukawa matrix and $\Xb_q=\rh(\Yb_q^C\Yb_q^{S\da})^{1/2}$. We then obtain
\begin{align}\label{eqn:gfunctypeII}
   g(\Xb_q,\tilde \Yb_q)\Big|_\tx{brane Higgs}^\tx{type-II} 
   &= \frac{2\Xb_d}{\sinh 2\Xb_d}\,\frac{\rh^2\tilde\Yb_q\tilde\Yb_q^\da}{1+\rh^2\Yb_q\tilde\Yb_q^\da}\,
    \frac{2\Xb_d}{\sinh 2\Xb_d}\,\tilde\Yb_q = + \rh^2\Yb_q^C\Yb_q^{C\da}\Yb_q^C 
    + \ord(\rh^4) \,,
\end{align}
where to leading order in $v^2/\MKK^2$ the KK contribution emerges from the correct-chirality Higgs coupling. At this order there is no difference between the original result \eqref{eqn:gfuncbsga} and \eqref{eqn:gfunctypeII}. The wrong-chirality Higgs coupling only contributes at order $v^4/\MKK^4$.

In the following our paper concentrates on the RS model with a brane-localized Higgs sector and we set for simplicity $\Yb_q^C=\Yb_q^S\equiv \Yb_q$. The only exception is Section~\ref{sec:NarrowBulk}, where we discuss some results in the narrow bulk-Higgs scenario.

\subsection{Gauge-boson contribution} 
\label{sec:GaugeContrbsga}

We continue with the diagrams (IIa) and (IIb) in Figure~\ref{fig:C78}, where two internal quarks and one gauge boson are exchanged.~The Wilson coefficients for the vector and scalar contributions are given by (with $B=A,G,W,Z$)
\begin{align}\label{eqn:C78B}\notag
C_{7\ga,8g}^{{B,\tx{vector}}}  &= \frac{\kappa_B^{7\ga,8g}}{4\sqrt2G_F\lambda_t}\,2\pi\int_0^\infty dk_E \int_\e^1 dt dt'\,B_B(t',t;k_E^2)\,\D_L^{(2)\da}(t)\,\Pb_B \\*\notag
&\quad\mbox{}\times \Bigg[ \! \left( \frac{11k_E^2}{16}\pa_{k_E} + \frac{5k_E^3}{16}\,\pa_{k_E}^2 + \frac{k_E^4}{48}\,\pa_{k_E}^3 \right) \bs\Delta^q_{LL}(t,t';k_E^2)\,\Pb_B\,\D_L^{(3)}(t') \\*\notag
&\quad\mbox{}\hspace{7mm}+ \left( -\frac{3k_E^2}{2}\,\pa_{k_E} - \frac{k_E^3}{2}\,\pa_{k_E}^2 \right)\frac{\bs\Delta^q_{LR}(t,t';k_E^2)}{m_b}\,\Pb_B\,\D_R^{(3)}(t') \Bigg] \,,\\ \notag
C_{7\ga,8g}^{B,\tx{scalar}} &= \frac{\kappa_B^{7\ga,8g}}{4\sqrt2G_F\lambda_t}\,2\pi\int_0^\infty dk_E \int_\e^1 dt dt'\,B^\tx{scalar}_B(t',t;k_E^2)\,\D_L^{(2)\da}(t)\,\tilde\Vb_{B_5^-}(t)  \\*\notag
&\quad\mbox{}\times \Bigg[ \! \left( \frac{k_E^2}{32}\,\pa_{k_E} - \frac{k_E^3}{32}\,\pa_{k_E}^2
- \frac{k_E^4}{96}\,\pa_{k_E}^3 \right) \bs\Delta^q_{RR}(t,t';k_E^2)\,\Vb_{B^+_5}(t')\,\D_L^{(3)}(t')\\*
&\quad\mbox{}\hspace{7mm}+ \left( \frac{k_E^2}{8}\,\pa_{k_E} - \frac{k_E^3}{8}\,\pa_{k_E}^2 \right)\frac{\bs\Delta^q_{RL}(t,t';k_E^2)}{m_b}\,\tilde\Vb_{B^+_5}(t')\,\D_R^{(3)}(t') \Bigg]\,,
\end{align}
where we introduced the matrices $\Pb_A=\Pb_G\equiv\one_{2\times2}$ and $\Pb_Z\equiv(\Pb_+ + g_R^d/g_L^d\,\Pb_-)$, with $g_L^d\equiv T_d^3-Q_d\,s_w^2$ and $g_R^d\equiv -Q_d\,s_w^2$. In case of the $W^\pm$-boson loop up-type 5D quark propagator functions ($q=u$) arise, otherwise we need to set $q=d$ in \eqref{eqn:C78B}. The quark propagator functions $\bs\Delta_{AB}^q(t,t';k_E^2)$ are given explicitly in Appendix~\ref{app:5DFermionProp} for the brane-localized Higgs scenario. We remark that in case of the photon and gluon contributions to the Wilson coefficients ($B=A,G$) only KK resonances can contribute, therefore we have to subtract the zero mode 4D propagator $(2\pi k_E^2)^{-1}$ from  $B_B(t',t;k_E^2)$ in \eqref{eqn:C78B}.~The structures\footnote{The $\pm$ labels on the subscripts of ${\bs V}_{B_5^\pm}(t)$ and $\tilde{\bs V}_{B_5^\pm}(t)$ are only relevant for $B=W$ and can be ignored otherwise.} ${\bs V}_{B_5^\pm}(t)$ and $\tilde{\bs V}_{B_5^\pm}(t)$ can be found in Appendix~\ref{app:FR}.~The coefficients $\kappa_B^{7\ga,8g}$ are given by
\begin{equation}\label{eqn:kappasqBq}
\begin{aligned}
   \kappa_A^{7\ga} &= 2 Q_d^3\,e^2 \,, &\quad \kappa_G^{7\ga} &= 2 Q_d\,C_F\,g_s^2 \,, 
    &\quad \kappa_W^{7\ga} &= Q_u\,{\frac{g_5^2}{2\pi r}} \,, &\quad \kappa_Z^{7\ga} &= 2 Q_d\,(g_L^d)^2\,{\frac{g_5^2/c_w^2}{2\pi r}} \,, \\
   \kappa_A^{8g} &= 2 Q_d^2\,e^2 \,, &\quad \kappa_G^{8g} &= - \frac{1}{N_c}\,g_s^2\,, 
    &\quad \kappa_W^{8g} &= {\frac{g_5^2}{2\pi r}} \,, &\quad \kappa_Z^{8g} &= 2 (g_L^d)^2\,{\frac{g_5^2/c_w^2}{2\pi r}}\,,
\end{aligned}
\end{equation}
where $e$ is the 4D electromagnetic and $g_s$ the QCD 4D gauge coupling.~The  5D gauge coupling of $SU(2)_L$ can be obtained from $\frac{g_5^2}{{2\pi r}} = \frac{4\tilde m_W^2}{v^2} = 4\sqrt2G_F m_W^2 \left[1 - \frac{m_W^2}{2\MKK^2} \left(1-\frac{1}{2L} \right)  + \ord\left(\frac{v^4}{\MKK^4}\right)\right]$, which can be derived from the expansions of $\tilde m_W$ and $v$ to leading order in $v^2/\MKK^2$ given in the text below \eqref{varphidecomp}.~Furthermore $Q_u=2/3$, $Q_d=-1/3$ and $C_F=(N_c^2-1)/(2N_c)=4/3$ with $N_c=3$ being the color factor for quarks.~The largest factors occur in case of the penguin diagrams exchanging KK gluons and $W^\pm$-boson modes. 

The scalar Wilson coefficient in \eqref{eqn:C78B} contains the propagator function $B_B^\tx{scalar}(t,t';-k^2/\xi)$, which is related to the 5D scalar propagator in general $R_\xi$ gauge via $D^{\tx{scalar},\xi}_B(t,t';k) = -\frac{1}{\xi} B_B^\tx{scalar}(t,t';-k^2/\xi)$. Its KK decomposition ($B=A,G,W,Z$),
\begin{align}\label{eqn:BBssKKrep}
   B_B^\tx{scalar}(t,t';-k^2) = \sum_n \frac{\MKK^2}{m_{B_n}^2}\frac{tt'}{\e^2}\,
    \frac{\pa_t\chi_n^B(t)\,\pa_{t'}\chi_n^B(t')}{m_{B_n}^2-k^2-i0} \,,
\end{align}
can be used to express it in terms of the vector-boson propagator \eqref{eqn:BBKKrep} by means of the relation
\begin{align}\label{eqn:B55Bvec}
   B_{B}^\tx{scalar}(t,t';k_E^2) = \frac{\MKK^2}{\e^2}\,\frac{tt'}{k_E^2}\,\pa_t\,\pa_{t'} 
    \Big[ B_B(t,t';0) - B_B(t,t';k_E^2) \Big] \,.
\end{align}
We can use this equation to eliminate the brane-localized terms inside the structures $\bs V_{B_5^\pm}(t)$ and $\tilde {\bs V}_{B_5^\pm}(t)$ in case of the massive gauge bosons ($B=W,Z$). For example, the 5D Feynman rule for the $W_5^-\bar D_L U_R$ vertex given in \eqref{eqn:VBs} contains the term
\begin{align}\label{eqn:VtWsm}
\tilde\Vb_{W_5^-}(t) = -\frac{\e}{t}\,\bigg[ \Pb_W - \frac{\rh\MKK^2}{L\tilde m^2_W}\,\delta(t-1)\,\M^{Y}_{ud} \bigg] \,, 
\end{align}
where $\M_{ud}^Y=\Yb_u\,\Pb_{12} - \Yb_d^\da\,\Pb_{21}$. The first term originates from the fifth component of the gauge-boson coupling to quarks, while the second brane-localized term is due to the Yukawa coupling of the $W^\pm$-Goldstone boson. We now insert \eqref{eqn:B55Bvec} into \eqref{eqn:C78B} and perform partial integrations for the $t,t'$ coordinates, taking into account that all terms on the boundary are orbifold-odd and therefore vanish. The partial integrations lead to derivatives acting on fermion profiles and propagators. We can use the equation of motions for the fermion profiles and the differential equations satisfied by the 5D propagators to show that all brane-localized terms contained in $\bs V_{W_5^\pm}(t)$ and $\tilde {\bs V}_{W_5^\pm}(t)$ cancel. For example, the partial $t$-integration of the scalar Wilson coefficient in \eqref{eqn:C78B} leads to the term (for $B=W$)
\begin{align}\label{eqn:tderProfProp}
   \pa_t \left[ \D_L^{(2)\da}(t) \bs P_W  \bs\Delta_{RR}^u(t,t';k_E^2) \right] 
   & = \D_L^{(2)\da}(t) \bs P_W \frac{\bs\Delta_{LR}^u(t,t';k_E^2)}{\MKK}
    - \frac{m_s}{\MKK}\,\D_R^{(2)\da}(t) \bs P_W \bs\Delta_{RR}^u(t,t';k_E^2) \notag\\*
   &\quad\mbox{}-\rh\,\delta(t-1) \D_L^{(2)\da}(t)\M^{Y}_{ud}\,\bs\Delta_{RR}^u(t,t';k_E^2) \,,
\end{align}
where we have used that
\begin{align}
\begin{split}
   \pa_t\Q_L^{(n)}(t) &= - \frac{m_{q_n}}{\MKK}\,\Q_R^{(n)}(t) + \M_q(t)\Q_L^{(n)}(t) \,, \\
   \pa_t\bs\Delta_{RR}^q(t,t';k_E^2) &= \frac{1}{\MKK}\,\bs\Delta^q_{LR}(t,t';k_E^2)
    - \M_q(t)\,\bs\Delta_{RR}^q(t,t';k_E^2) \,.
\end{split}
\end{align}
The last term in \eqref{eqn:tderProfProp} cancels with the remaining contribution from the brane-localized term in \eqref{eqn:VtWsm}. Furthermore, we will discard contributions that are suppressed by the strange-quark mass.

In the last step we can perform partial integrations of the Euclidean momentum variable and finally obtain $(B=A,G,W,Z)$
\begin{align}\label{eqn:C785DqBq}
\begin{split}
C_{7\ga,8g}^{B} &= \frac{\kappa_B^{7\ga,8g}}{4\sqrt2 G_F\lambda_t}\,\Bigg\{\frac{5}{24}\,R^B_{LL} - \frac{1}{4}\,R^B_{LR} + 2\pi \int_0^\infty dk_E\int_\e^1 dtdt'\,\D_L^{(2)\da}(t)\,\Pb_B \\
&\quad\times\Bigg[ \frac{\bs\Delta^q_{LR}(t,t';k_E^2)}{m_b}\,\Pb_B\,\D_R^{(3)}(t')\left( -\frac{9 k_E^2}{8}\,\pa_{k_E} - \frac{3k_E^3}{8}\,\pa_{k_E}^2 \right) \\
&\qquad\mbox{}\hspace{1mm}+ \bs\Delta^q_{LL}(t,t';k_E^2)\,\Pb_B\,\D_L^{(3)}(t')\left( \frac{3k_E^2}{32}\,\pa_{k_E} - \frac{3k_E^3}{32}\,\pa_{k_E}^2-\frac{k_E^4}{32}\,\pa_{k_E}^3 \right) \! \Bigg]\it B_B^{}(t',t;k_E^2)\Bigg\} \,,
\end{split}
\end{align}
where we have combined both the vector and scalar Wilson coefficients. The partial momentum integrations are required for numerical reasons, since momentum derivatives acting on fermion propagators lead to complicated expressions that are very inefficient to evaluate. Note that the Wilson coefficients for $B=A,G$ differ only in the factors $\kappa_A^{7\ga,8g}$ and $\kappa_G^{7\ga,8g}$, see \eqref{eqn:kappasqBq}. Due to the partial momentum integrations we encounter non-zero boundary terms for large Euclidean momenta in \eqref{eqn:C785DqBq}, they are defined by
\begin{align}\label{eqn:RLLRLR}
\begin{split}
R^B_{LL} &= - 2\pi \lim_{k_E\to\infty} k_E^2 \int_\e^1 dt dt'\,B_B(t',t; 0)\,\D_L^{(2)\da}(t)\,\Pb_B\,\bs\Delta^q_{LL}(t,t';k_E^2)\,\Pb_B\,\D_L^{(3)}(t') \,, \\
R^B_{LR} &= - 2\pi  \lim_{k_E\to\infty} k_E^2 \int_\e^1 dt dt'\,B_B(t',t; 0)\,\D_L^{(2)\da}(t)\,\Pb_B\,\frac{\bs\Delta^q_{LR}(t,t';k_E^2)}{m_b}\,\Pb_B\,\D_R^{(3)}(t') \,,
\end{split}
\end{align}
where $q=u$ for $B=W$ and  $q=d$ for $B=A,G,Z$. In case of the penguin diagrams, in which photon (gluon) modes are exchanged, we have to subtract the zero-mode contribution from the full propagator function $B_A(t,t';0)$. The reason is that massless gauge bosons have constant profiles that lead to flavor-conserving interactions and therefore do not contribute to the Wilson coefficients.

\subsubsection*{Calculation of the boundary terms $\bs{R^B_{LL}}$ and $\bs{R^B_{LR}}$}

In order to determine the boundary terms in~\eqref{eqn:RLLRLR} we need to know the UV behavior of the boson and fermion propagator functions, which is worked out in Appendix~\ref{sec:UV5Dprop}. Using the results shown in equation~\eqref{eqn:BBf}, we can calculate the first boundary term in \eqref{eqn:RLLRLR} and obtain
\begin{align}\label{eqn:RLLres}
R^B_{LL} = 2\pi \int_\e^1\!dt\,B_B(t,t;0)\,\D_L^{(2)\da}(t) \, \Pb_{B}^2\,\D_L^{(3)}(t) \,,
\end{align}
where we have to remember to subtract the zero-mode propagator in case of $B=A,G$, since only KK photons and KK gluons can contribute. We can further simplify \eqref{eqn:RLLres} by using the explicit expressions for the propagator functions \cite{Casagrande:2008hr}
\begin{align}\label{eqn:BBprop}
\begin{split}
B_B(t,t;0) & =\frac{1}{2\pi\tilde m_B^2} + \frac{L(1-t^2)}{4\pi\MKK^2} \,; \quad B=W,Z \,,\\
B'_B(t,t;0) & = \frac{1}{4\pi\MKK^2} \left( L t^2 - t^2 (1-2\ln t) + \frac{1}{2L} \right)\,  \,;\quad B=A,G\,,
\end{split}
\end{align}
where $B'_{B}(t,t;0)$ in the second line includes only the KK modes. Inserting \eqref{eqn:BBprop} into \eqref{eqn:RLLres} and applying the orthonormality condition for the fermion profiles \eqref{eqn:OrthFermBos}, we obtain
\begin{align}\label{eqn:RLLres2}
\begin{split}
R_{LL}^A &= \frac{L}{2\MKK^2} \left[ (\bs\Delta_D)_{23} - \frac{2}{L} (\bs\Delta'_D)_{23} \right] , \\
R_{LL}^W &= -\frac{L}{2\MKK^2} \left[ (\bs\Delta_D)_{23}+ ({\bs\delta}_D)_{23}-({\bs\varepsilon}_D)_{23} \right]  - \frac{(\bs\delta_D)_{23}}{\tilde m_W^2} \,, \\
R_{LL}^Z &= - \frac{L}{2\MKK^2}\,\left[(\bs\Delta_D)_{23}+ \left(1-\frac{(g_R^d)^2}{(g_L^d)^2} \right) \bigg( ({\bs\delta}_D)_{23}-({\bs\varepsilon}_D)_{23} \bigg)\right] - \left(1-\frac{(g_R^d)^2}{(g_L^d)^2} \right)\frac{(\bs\delta_D)_{23}}{\tilde m_Z^2} \,,
\end{split}
\end{align}
where $R_{LL}^G=R_{LL}^A$. We recover the known overlap integrals
\begin{equation}\label{eqn:DeltaDefs}
\begin{aligned}
(\bs\Delta_D)_{nn'}& =\int_\e^1 dt\,t^2 \,\D_L^{(n)\da}(t)\,\D_L^{(n')}(t)\,, &
(\bs\Delta'_D)_{nn'} & =\int_\e^1 dt\,t^2\left(\frac{1}{2}-\ln t\right)\D_L^{(n)\da}(t)\,\D_L^{(n')}(t)\,,\\
(\bs\delta_D)_{nn'} & = \int_\e^1 dt\,\D_L^{(n)\da}(t)\,\Pb_-\,\D_L^{(n')}(t)\,, &
({\bs\varepsilon}_D)_{nn'} & =\int_\e^1 dt\,t^2\,\D_L^{(n)\da}(t) \,\Pb_-\,\D_L^{(n')}(t) \,,
\end{aligned}
\end{equation}
originally defined in \cite{Casagrande:2008hr}. For the other boundary term in \eqref{eqn:RLLRLR}, we can use relation \eqref{eqn:DeltaLRlim} and obtain
\begin{align}\label{eqn:RLRres}
\begin{split}
   R^B_{LR} &= - \frac{2\pi}{x_b}\,\Bigg[ 
    B_B(1^-,1^-;0)\,\D_L^{(2)\da}(1^-)\,\Pb_B\,\Bigg( \frac{\Pb_+}{1+\rh^2\tilde\Yb_q\tilde\Yb_q^\da} 
    + \frac{\Pb_-\,\rh^2\tilde\Yb_q^\da\tilde\Yb_q}{1+\rh^2\tilde\Yb_q^\da\tilde\Yb_q} \\
   &\hspace{25mm}\mbox{}- \frac{\Pb_{12}}{1+\rh^2\tilde\Yb_q\tilde\Yb_q^\da}\,\varrho\tilde{\bs Y}_q
    - \varrho\tilde{\bs Y}_q^\da\,\frac{\Pb_{21}}{1+\rh^2\tilde\Yb_q\tilde\Yb_q^\da} \Bigg)\,
    \Pb_B\,\D_R^{(3)}(1^-) \\
   &\hspace{25mm}\mbox{}- \int_\e^1\!dt\,\D_L^{(2)\da}(t)\,\Pb_B^2\,
    \left( \D_R^{(3)}(t)\,\frac{\pa_t}{2} + x_b\,\D_L^{(3)}(t) \right)B_B(t,t;0) \Bigg] \,, 
\end{split}
\end{align}
where $\Pb_{ij}$ is a $2\times2$ matrix with zero entries except for the $ij$-component, which equals~1. We have omitted the terms at $t=\e$, since the upper component of $\D_R^{(n)}(t)$ and the lower component of $\D_L^{(n)}(t)$ obey Dirichlet boundary conditions at the UV brane and therefore vanish. In order to obtain the last term we have used that the function $B_B(t,t';0)$ vanishes for $t' < t$ and we applied the equation of motion for the fermion profiles. We can further simplify \eqref{eqn:RLRres} by performing a partial $t$-integration of the term involving $\pa_t B_B(t,t;0)$ and by using the fermion equation of motions to show that 
\begin{align}\label{eqn:RLRres2}
\begin{split}
   R^B_{LR} &= \frac{1}{2}\,R_{LL}^B - \frac{2\pi}{x_b}\,B_B(1^-,1^-;0)\,
    \D_L^{(2)\da}(1^-)\,\Pb_B\,\Bigg[ 
    \frac{\Pb_+}{2}\,\frac{1-\rh^2\tilde\Yb_q\tilde\Yb_q^\da}{1+\rh^2\tilde\Yb_q\tilde\Yb_q^\da}
    - \frac{\Pb_-}{2}\,\frac{1-\rh^2\tilde\Yb_q^\da\tilde\Yb_q}{1+\rh^2\tilde\Yb_q^\da\tilde\Yb_q} \\
   &\quad\mbox{}- \frac{\Pb_{12}}{1+\rh^2\tilde\Yb_q\tilde\Yb_q^\da}\,\varrho\tilde{\bs Y}_q 
    - \varrho\tilde{\bs Y}_q^\da\,\frac{\Pb_{21}}{1+\rh^2\tilde\Yb_q\tilde\Yb_q^\da} 
    \Bigg]\,\Pb_B\,\D_R^{(3)}(1^-) \,,
\end{split}
\end{align}
where we neglected a term suppressed by $m_s/m_b$ and where we recovered the term $R_{LL}^B$. Applying the modified boundary conditions of the quark profiles\footnote{In the brane-localized Higgs scenario one can consistently calculate the fermion profiles by using modified boundary conditions at the IR brane without the notion of a regulator for the $\delta$-function \cite{Casagrande:2010si}. Here we can apply $(\rh\tilde\Yb_d^\da\;1)\D_L^{(n)}(1^-)=0$ and $(1\;-\rh\tilde\Yb_d)\D_R^{(n)}(1^-)=0$.}, we obtain
\begin{align}\label{eqn:RLRres3}
\begin{split}
R_{LR}^A &= \frac{1}{2}\,R_{LL}^A \,, \\
R_{LR}^W &= \frac{1}{2}\,R_{LL}^W - \frac{1}{2\tilde m_W^2}\frac{v}{\sqrt2m_b}\,\D_L^{(2)\da}(1^-)\,\Pb_{12}\,\frac{1-\rh^2\tilde\Yb_u\tilde\Yb_u^\da}{1+\rh^2\tilde\Yb_u\tilde\Yb_u^\da}\,\tilde\Yb_d\,\D_R^{(3)}(1^-) \,, \\
R_{LR}^Z &= \frac{1}{2}R_{LL}^Z - \frac{1}{2\tilde m_Z^2}\frac{v}{\sqrt2m_b} \left(1-\frac{g_R^d}{g_L^d}\right)^2\D_L^{(2)\da}(1^-)\,\Pb_{12}\,\frac{1-\rh^2\tilde\Yb_d\tilde\Yb_d^\da}{1+\rh^2\tilde\Yb_d\tilde\Yb_d^\da}\,\tilde\Yb_d\,\D_R^{(3)}(1^-) \,,
\end{split}
\end{align}
where $R_{LR}^G=R_{LR}^A$. Yukawa-dependent terms appear in case of the massive gauge bosons and originate from the Goldstone degrees of freedom, which are localized at the IR brane. 

Finally, we have succeeded in obtaining expressions for the boundary terms \eqref{eqn:RLLres2} and \eqref{eqn:RLRres3}, such that the complete Wilson coefficient $C_{7\ga,8g}^B$ in \eqref{eqn:C785DqBq} can be evaluated numerically. The chirality-flipped Wilson coefficients $\tilde C_{7\ga,8g}$ can be obtained from \eqref{eqn:C785DqBq} and \eqref{eqn:RLLRLR} by interchanging the label $L\leftrightarrow R$. The boundary terms can be calculated in analogy with the above steps, and we can use the results \eqref{eqn:RLLres} and \eqref{eqn:RLRres2}, for which we find $R_{RR}^B=R_{LL}^B|_{L\to R}$ and $R_{RL}^B = R_{LR}^B |_{L\leftrightarrow R}$. 

\subsection{Triple gauge-boson vertex contribution}\label{sec:TripleGaugeContrbsga}

Finally we discuss the diagrams exchanging two internal gauge bosons $(B = W,G)$ and one quark, see diagrams (IIIa)-(IIId) and (IVa)-(IVd) in Figure \ref{fig:C78}.~There are four diagrams each, involving vector and scalar components of the gauge-boson propagators.~We refrain from showing intermediate steps of the calculation but mention that we can proceed analogously as in the previous section and combine the vector- and scalar-boson contributions.~After some algebra we obtain the Wilson coefficients
\begin{align}\label{eqn:C785DBqB}\notag
C_{7\ga}^{WW} &= \frac{\kappa_{WW}}{4\sqrt2 G_F\lambda_t}\,\bigg\{\frac{1}{6}\,R_{LL}^{W} - \frac{1}{4}\,R^{W}_{LR} +  2\pi\int_0^\infty\!dk_E\int_\e^1\!dt dt'\,\D_L^{(2)\da}(t)\,\Pb_+ \\*\notag 
&\hspace{18mm}\mbox{}\times \bigg[ \bs\Delta^u_{LL}(t,t';k_E^2)\,\Pb_+\,\D_L^{(3)}(t')\left( \frac{3k_E^2}{32}\,\pa_{k_E} - \frac{3k_E^3}{32}\,\pa_{k_E}^2 + \frac{k_E^4}{32}\,\pa_{k_E}^3 \right) \\*\notag
&\hspace{25mm}\mbox{}+ \frac{\bs\Delta^u_{LR}(t,t';k_E^2)}{m_b}\,\Pb_+\,\D_R^{(3)}(t')\left( -\frac{3k_E^2}{8}\,\pa_{k_E} + \frac{3k_E^3}{8}\,\pa^2_{k_E} \right) \bigg]\,B_W(t',t;k_E^2) \bigg\} \,,\\\notag
C_{8g}^{GG} &= \frac{\kappa_{GG}}{4\sqrt2 G_F\lambda_t}\,\bigg\{\frac{1}{6}\,R_{LL}^{G} - \frac{1}{4}\,R^{G}_{LR} + 2\pi\int_0^\infty\!dk_E\int_\e^1\!dt dt'\,\D_L^{(2)\da}(t) \\*\notag
&\hspace{18mm}\mbox{}\times \bigg[ \bs\Delta^d_{LL}(t,t';k_E^2)\,\D_L^{(3)}(t')\left( -\frac{5k_E^2}{32}\,\pa_{k_E} + \frac{5k_E^3}{32}\,\pa_{k_E}^2 - \frac{k_E^4}{96}\,\pa_{k_E}^3 \right) \\*
&\hspace{25mm}\mbox{}+ \frac{\bs\Delta^d_{LR}(t,t';k_E^2)}{m_b}\,\D_R^{(3)}(t')\left( \frac{3k_E^2}{8}\,\pa_{k_E} - \frac{3k_E^3}{8}\,\pa^2_{k_E} \right) \bigg]\,B'_G(t',t;k_E^2) \bigg\} \,,
\end{align}
where $\kappa_{WW}=g_5^2/(2\pi r)$ and $\kappa_{GG}=N_c\,g_{s}^2 $. Note that the factor $\kappa_{GG}$ for the triple gluon vertex diagram is larger by $N_c^2=9$ compared to $\kappa_G^{8g}$ and comes with a relative sign.~We recover the same boundary terms that have already been calculated in Section~\ref{sec:GaugeContrbsga} apart from constant factors. 

\section{Analysis of the dipole coefficients}\label{sec:AnaWilsbsga}

\subsection{Finiteness of the integrals}

In order to show the finiteness of the dipole coefficients in \eqref{eqn:C785Dqhq}, \eqref{eqn:C785DqBq} and \eqref{eqn:C785DBqB} we need to know the UV behavior of the 5D boson and fermion propagators. We refer to Appendix~\ref{sec:UV5Dprop} for the corresponding derivations. For instance, the general behavior of the 5D (vector) gauge-boson propagator is given by (subscript $B=A,G,W,Z$)
\begin{align}\label{eqn:BBUVapp}
   B_B(t,t';k_E^2)\sim \frac{\sqrt{t t'}}{k_E}\,e^{-\kEh\,|t-t'|} \,. 
   \qquad (\kEh \gg 1/t,1/t') 
\end{align}
For large Euclidean momenta the propagator function is exponentially suppressed except for $|t-t'|\sim 1/\kEh$. Integrating \eqref{eqn:BBUVapp} along the coordinates $t$ and $t'$ we find
\begin{align}
   \int_\e^1\!dt dt'\,B_B(t,t';k_E^2)\sim \frac{1}{k_E^2} \,, 
   \qquad (\kEh \gg 1/\e) 
\end{align}
showing that the integral scales like $k_E^{-2}$ for Euclidean momenta $\kEh \gg 1/\e$. Based on this analysis, and extending it to the case of the fermion propagator functions, we can formulate a power counting for integrals, where each extra-dimensional coordinate is integrated over the full interval. Excluding brane-localized terms, the counting in terms of Euclidean momenta can be formulated as
\begin{align}\label{eqn:UVcounting}
   \bs\Delta_{AB}^q \to  (k_E)^{-1}, \qquad B_{W,Z,A,G} \to (k_E)^{-1} \,, \qquad  \int_\e^1\!dt \to (k_E)^{-1} \,,
\end{align}
where $A,B\in\{L,R\}$ and $q=u,d$ for the quark propagator functions and with the additional condition that the last $t$-integration is not counted. This condition can be traced back to the conservation of the total 5-momentum. We can apply the power-counting scheme \eqref{eqn:UVcounting} to the penguin loops \eqref{eqn:C785DqBq} and \eqref{eqn:C785DBqB}, showing that after the $t,t'$ integrations the integrands fall off like $k_E^{-2}$ for large Euclidean momenta. Thus the remaining momentum integration can be performed and yields a finite result. This is in agreement with the findings of \cite{Csaki:2010aj}, where the authors derived a power-counting scheme for the penguin diagrams treating the Yukawa interactions as small perturbations. The Higgs contribution contains two brane-localized vertices and our scheme \eqref{eqn:UVcounting} does not apply. In fact, the analysis of Section~\ref{sec:HiggsContrbsga} shows that the propagator functions in the brane-localized Higgs scenario scale like $T_{RR}^d(k_E^2)\sim \kEh^{-1}$ and $T_{RL}^d(k_E^2)\sim \tx{const} + \ord(\kEh^{-1})$ for large Euclidean momenta. Since the Higgs boson propagator scales like $k_E^{-2}$ the Wilson coefficient is finite which is in agreement with the results of \cite{Csaki:2010aj}. 

In summary, using our expressions for the 5D propagators with non-trivial boundary conditions at the IR brane we have confirmed the findings of \cite{Csaki:2010aj} that the dipole coefficients are finite and calculable. This conclusion is also consistent with our results \eqref{eqn:C78KK} derived in the KK-decomposed theory and discussed in the following section, where we can show that all dipole coefficients converge after summing up the KK towers. 

\subsection{Connection with the KK-decomposed theory} \label{sec:KKcalc}

We can express the dipole coefficients, as defined via the amplitude \eqref{eqn:AC7ga}, in terms of sums over zero-mode and KK-mode contributions. Starting from the expressions \eqref{eqn:C785Dqhq}, \eqref{eqn:C785DqBq} and \eqref{eqn:C785DBqB} in the 5D framework we replace the 5D propagator functions by their corresponding KK representations. The appearing momentum integrals can be performed analytically and we obtain the loop functions $I_{3,4}(x)$ and $I_\tx{6-11}(x)$, which are defined via their integral representations \eqref{eqn:LoopFuncsInt} in Appendix~\ref{sec:LoopFunctions}. We find the compact expressions
\begin{align}\label{eqn:C78KK}
\begin{split}
C_{7\ga,8g}^{B} &= \frac{\kappa_B^{7\ga,8g}}{4\sqrt2G_F\lambda_t} \sum_{n,m}\frac{1}{m_{B_m}^2}\left( \frac{m_{q_n}}{m_b}\,\frac{I_6(x^{q_n}_{B_m})}{2}\,V_{2mn}^{B^-}\,\tilde V_{nm3}^{B^+} + \frac{I_7(x^{q_n}_{B_m})}{2}\,V_{2mn}^{B^-}\,V_{nm3}^{B^+} \right) \,, \\ 
C_{7\ga}^{WW} &= \frac{\kappa_{WW}}{4\sqrt2G_F\lambda_t} \sum_{n,m}\frac{1}{m_{B_m}^2}\left( \frac{m_{u_n}}{m_b}\,I_{8}(x_{u_n}^{B_m})\,V^{W^-}_{2mn}\,\tilde V^{W^+}_{nm3} + I_{9}(x_{u_n}^{B_m})\,V^{W^-}_{2mn}\,V^{W^+}_{nm3}\right) \,, \\  
C_{8g}^{GG} &= \frac{\kappa_{GG}}{4\sqrt2G_F\lambda_t} \sum_{n,m}\frac{1}{m_{B_m}^2}\,\left(\frac{m_{d_n}}{m_b}\,I_{10}(x_{d_n}^{B_m})\,V^G_{2mn}\,\tilde V^G_{nm3} + I_{11}(x_{d_n}^{B_m})\,V^G_{2mn}\,V^G_{nm3} \right) \,, \\
C_{7\ga,8g}^{h} &= - \frac{\kappa_h^{7\ga,8g}}{4\sqrt2G_F\lambda_t} \sum_n\frac{1}{m_h^2}\,\left( \frac{m_{d_n}}{m_b}\,I_{3}(x_h^{d_n})\,(g_h^d)_{2n}\,(g_h^d)_{n3} + I_{4}(x_h^{d_n})\,(g_h^d)_{2n}\,(\tilde g_h^{d})_{n3} \right) \,,
\end{split}
\end{align}
where $x_{b}^{a}=m_{a}^2/m_{b}^2$, and $q=u$ for $B=W$ and $q=d$ for $B=A,G,Z$ in the first line.~The summation index $m$ counts the contributions from the gauge-boson zero  ($m=0$ for the SM gauge-bosons) and KK modes $(m\geq1)$, while $n$ counts the quark zero $(n=1,2,3$ for the SM quarks) and KK modes ($n=4,...,9$ for the first KK level and so on).~We mention that there are no contributions from the massless zero modes (the SM photon and gluon), which implies that the summation starts with $m=1$ in the first line for $B=A,G$ and in the third line of \eqref{eqn:C78KK}.~The $\pm$ superscripts on the overlap integrals $V^{B^\pm}_{nmk}$ and $\tilde V^{B^\pm}_{nmk}$ are only relevant in the case of $B=W$ and can be ignored otherwise.~The definitions of the overlap integrals can be found in Appendix~\ref{app:FR}, while explicit expressions for the loop functions $I_{3,4}(x)$ and $I_\tx{6-11}(x)$ are given in \eqref{eqn:LoopFuncsExpl}.~We note that when we insert the integral representations of the loop functions \eqref{eqn:LoopFuncsInt} into \eqref{eqn:C78KK} we can identify the boundary terms $R_{LL}^B$ and $R_{LR}^B$, defined in the 5D approach by \eqref{eqn:RLLRLR}, with the expressions $(B=A,G,W,Z)$
\begin{align}\label{eqn:RLLRLRKK}
R_{LL}^B & = \sum_{m,n} \frac{1}{m_{B_m}^2}\,V_{2mn}^{B^-}\,V_{nm3}^{B^+}\,, & 
R_{LR}^B & = \sum_{m,n} \frac{1}{m_{B_m}^2}\,\frac{m_{q_n}}{m_b}\,V_{2mn}^{B^-}\,\tilde V_{nm3}^{B^+}
\end{align}
in the KK-decomposed theory.~Those terms originate from penguin diagrams where scalar components of the 5D gauge bosons are exchanged.~In fact, we have also checked equation \eqref{eqn:C78KK} by using the 4D Feynman rules listed in Appendix~\ref{app:FR} and following the basic steps to obtain the dipole coefficients.~The chirality-flipped coefficients $\tilde C_{7\ga,8g}^B$ can be obtained by replacing $V_{nmk}^B \leftrightarrow\tilde V_{nmk}^B$ and $(g_h^d)_{nk}\leftrightarrow(\tilde g_h^d)_{nk}$. 

We emphasize that there are two terms in each round bracket for the Wilson coefficients in \eqref{eqn:C78KK}. In the SM only diagrams with a chirality flip on the external $b$-quark line contribute to $C_{7\ga}$, since the $W^\pm$ boson couples only to left-chiral quarks. Since in the RS model we can have also couplings to right-chiral quarks, there are additional contributions originating from diagrams where the chirality flip is performed on the internal quark line, which generates the factor $m_{q_n}/m_b$ in front of the first term in each of the brackets in \eqref{eqn:C78KK}. When exchanging KK quarks in the loop this factor is large and enhances the contributions.

We remark that we have numerically checked that all corrections in the RS model decouple with $\MKK^{-2}$, which is not directly obvious from the expressions \eqref{eqn:C78KK}.~For instance, let us discuss the KK contributions to $C_{7\ga,8g}^W$.~At first we stress that the loop functions can only take values from the compact intervals $I_6(x)\in[-2,-\frac{1}{2}]$ and $I_7(x)\in[\frac{5}{12},\frac{2}{3}]$, and are irrelevant for the discussion.~We begin with the contribution of penguin diagrams that exchange SM quarks $(n=1,2,3)$ with KK $W$ bosons $(m\geq1)$ in the first line of \eqref{eqn:C78KK}.~Obviously the suppression by the squared KK $W$-boson mass implies that the contribution decouples with $\MKK^{-2}$.~Next, we discuss the contribution from exchanging the $W$ boson $(m=0)$ with KK quarks $(n\geq 4)$.~The overlap integrals scale like $V_{20n}^{W^-}\sim V_{n03}^{W^+}\sim \tilde V_{n03}^{W^+} \sim \MKK^{-1}$ for $n\geq4$, which implies that the second term in the round bracket of the first line in \eqref{eqn:C78KK} decouples with $\MKK^{-2}$.~The first term in the round bracket is more subtle, since it is enhanced by the KK-quark mass $m_{q_n}\sim \MKK$.~However, numerically we observe that the summation over complete KK levels ($n=4,...,9$ for the first KK level and so on) leads to cancellations such that there appears an additional $\MKK^{-1}$ suppression.~Hence, also the first term in the round bracket, when summed over complete KK levels, decouples with $\MKK^{-2}$.~In a similar fashion we can proceed with the contributions from the penguin diagrams with KK $W$ bosons and KK quarks.~The discussion can also be extended for the remaining Wilson coefficients.~In fact the decoupling behavior with $\MKK^{-2}$ is apparent in the approximate expressions that will be given in Section~\ref{sec:AppWilsCoeffbsga}.

Finally, notice that the Wilson coefficients in the SM can be recovered from the second terms in $C_{7\ga,8g}^{W}$ and $C_{7\ga}^{WW}$ by summing only over the gauge-boson ($m=0$) and quark zero modes ($n=1,2,3$), and by replacing the overlap integrals with the CKM matrix elements $V_{20n}^{W^-}\to V^*_{u_n s}$ and $V^{W^+}_{n03}\to V_{u_nb}$ with $u_{1,2,3}=u,c,t$.

\subsection{Numerical evaluation} 
\label{sec:NumEval}

The first step is to generate anarchic 5D Yukawa matrices, where each entry is bounded from above by $y_\ast$, i.e.\ $|(\Yb_q)_{ij}|\leq y_\ast$.~The real and imaginary parts of each entry are randomized with a flat distribution.~If we expand the exact profiles of the $W$ boson and the SM quarks in $v^2/\MKK^2$ and only keep the leading terms, referred to as the zero-mode approximation (ZMA) in \cite{Casagrande:2008hr}, we can directly calculate the Wolfenstein parameters $\bar\rho$ and $\bar\eta$ solely from the 5D Yukawa matrices.~Next, we choose a random value for the bulk-mass parameter $c_{u_3}\in [-1/2,1]$, which corresponds to a localization of the right-chiral top quark near the IR brane.~Working at leading order in $v^2/\MKK^2$ we can determine the remaining eight bulk-mass parameters $c_{Q_{1,2,3}}$, $c_{u_{1,2}}$ and $c_{d_{1,2,3}}$ from the experimental values for the six quark masses evaluated at the scale $\mu = 1\,\TeV$ and from the Wolfenstein parameters $A$ and $\lambda$.~Then, we choose a random value for $\MKK\in[1,10]\,\TeV$ ($M_{g^{(1)}}\in[2.45,24.5]\,\TeV$) and calculate the whole set $x_\tx{theo}=\{m_u,m_d,m_s,m_c,m_b,m_t,A,\lambda,\bar\rho,\bar\eta\}$ using the exact expressions that are valid to all orders in $v^2/\MKK^2$.~Finally, we calculate the function $\chi^2(x)=\sum_{n=1}^{10} (x_\tx{exp}(n)-x_\tx{theo}(n))^2/\sigma^2_\tx{exp}(n)$, where $x_\tx{exp}$ contains the experimental values of the quark masses and Wolfenstein parameters with standard deviations given by $\sigma_\tx{exp}$.~Points with $\chi^2(x)/\tx{dof}>11.5/10$, corresponding to less than 68\% CL, are rejected.~Based on the procedure described above we generate six sets of 5000 RS points with different values for $y_\ast =0.5,\,1,\,1.5,\,2,\,2.5$ and $3$. The upper value on $y_\ast$ originates from requiring that the Yukawa sector remains in the perturbative regime \cite{Csaki:2008zd}.

\begin{table}[t!]
\begin{center}
\begin{tabular}{|c|c|cccccc|} \hline
\multirow{2}{*}{Picture}& \multirow{2}{*}{\parbox[c]{5cm}{\centering Average time to calculate $C_{7\ga,8g}$ for one RS point}} &\multicolumn{6}{|c|}{Average time fractions for each contribution} \\
& & $C_{7\ga,8g}^W$  & $C_{7\ga}^{WW}$ & $C_{7\ga,8g}^Z$ & $C_{7\ga,8g}^{A/G}$ & $C_{8g}^{GG}$ & $C_{7\ga,8g}^h$  \\ \hline
5D & 571\,\tx{min} &2\% & 4\% & 21\% & 32\% & 41\% & $\leq$ 0.01\%\\
4D & 9\,\tx{min} &  \multicolumn{2}{c}{16\%} & 46\% & \multicolumn{2}{c}{38\%} & $\leq$ 0.1\% \\\hline
\end{tabular}
\parbox{15.5cm}{\caption{\label{tab:Perf4D5D} 
Time performance for calculating the Wilson coefficients $C_{7\ga}$ and $C_{8g}$ in the 5D and 4D (including 5 KK levels) pictures. The first column contains the average time needed in order to calculate the Wilson coefficients for one RS point on a 2.4~GHz Intel Core i5 processor, such that the results are compatible at the few per mille level in both approaches (see Figure~\ref{fig:Comp} for more details). The additional columns show the relative fractions of time needed for the calculation of the six different contributions. Similar values are obtained in case of the chirality-flipped Wilson coefficients.}}
\end{center}\vspace{-2mm}
\end{table} 

We have implemented the integrals arising in the expressions for the dipole coefficients in \eqref{eqn:C785Dqhq}, \eqref{eqn:C785DqBq} and \eqref{eqn:C785DBqB} in {\sc Mathematica}. Since we expect the RS corrections to the SM Wilson coefficients to lie in the few percent range, we need to calculate the integrals to an accuracy of a few per mille. We therefore set {\sc PrecisonGoal} to~3 for the numerical integrations. Furthermore we use a UV momentum cutoff such that $k_E\le\Lambda_\tx{cut}=100\,\MKK$, which improves the time performance without losing the required precision. It turns out that the numerical integrations over $t$ and $t'$ can be made faster by making the substitution $t\to\phi/\pi=\ln(t/\e)/\ln(1/\e)$ and analogously for $t'$, which maps the integration region on the unit square. The first row of Table~\ref{tab:Perf4D5D} compares the time performance of calculating the Wilson coefficients $C_{7\ga}$ and $C_{8g}$ in the 5D and 4D pictures, averaged over many sets of RS parameter points. We need on average 571~minutes per RS point on a 2.4~GHz Intel Core i5 processor to calculate $C_{7\ga}$ and $C_{8g}$ in the 5D approach. In more detail, the calculation splits into six parts belonging to different amplitude topologies. The least amount of time is required for the calculation of the Higgs contribution, since the $t$ and $t'$ integrations can be performed analytically, leaving over one momentum integral to be evaluated, see \eqref{eqn:C785Dqhq}. Most of the computational time is needed for the KK contributions of the neutral gauge bosons, since the corresponding integrands involve all components of the 5D fermion propagator functions, in contrast to the $W^\pm$-boson Wilson coefficients in \eqref{eqn:C785DqBq} and \eqref{eqn:C785DBqB}. 

\begin{figure}[t!]
\begin{center}
\psfrag{a}[t]{\small $\displaystyle \left|\frac{\re (C_{7\ga}^{})_\tx{5D}}{\re (C_{7\ga}^{})_\tx{4D}} - 1\right|$ [\%]}
\psfrag{b}[]{\small Points}
\psfrag{c}[t]{\small $\displaystyle \left|\frac{\re [(C_{7\ga}^{})_\tx{5D}-C_{7\ga}^\tx{SM}(\mu_W)]}{\re [(C_{7\ga}^{})_\tx{4D}-C_{7\ga}^\tx{SM}(\mu_W)]} - 1\right|$ [\%]}
\psfrag{d}[]{\small Points}
\psfrag{e}[b]{\scriptsize \parbox{3.1cm}{$\MKK\in[1,10]\,\TeV$\\}}
\psfrag{f}[b]{\scriptsize \parbox{3.1cm}{$\MKK\in[1,10]\,\TeV$\\}}
\includegraphics[width=0.9\textwidth]{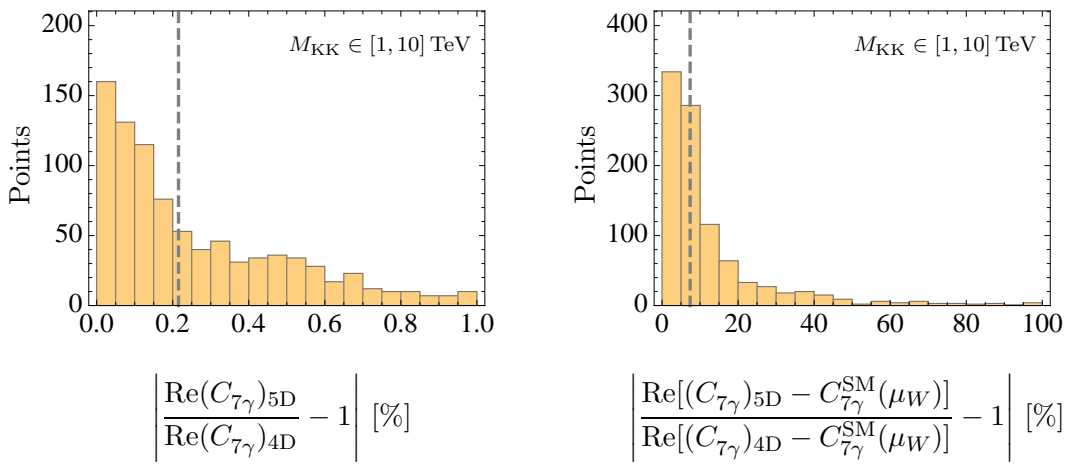} 
\parbox{15.5cm}{\vspace{9mm} \caption{\label{fig:Comp} 
Compatibility of the results for the Wilson coefficient $C_{7\ga}$ (left) and for the RS corrections relative to the SM Wilson coefficient $C_{7\ga}^\SM(\mu_W)\approx-0.20$ (right) calculated in the 5D and 4D (including 5 KK levels) pictures. Both histograms contain RS parameter points with different values for $y_\ast$ and $\MKK\in[1,10]\,\TeV$. The vertical dashed lines denote the median values of the corresponding distributions.}}
\end{center}\vspace{-2mm}
\end{figure}

We can compare our results from the 5D approach with the summation over zero and KK modes in the KK-decomposed theory, based on the results shown in \eqref{eqn:C78KK}. In order to achieve a consistency between both approaches at the few per mille level, we need to sum over five complete KK levels. The left plot in Figure \ref{fig:Comp} confirms that the results for $C_{7\ga}$ calculated in the 5D and 4D approaches are consistent at the few per mille level. The results for the RS corrections relative to the SM Wilson coefficient $C_{7\ga}^\SM$ are compatible at the $10\%$ level in both pictures, as shown in the right plot in Figure \ref{fig:Comp}. This presents a non-trivial cross-check of the formulas derived in Section~\ref{sec:Wils5Dbsga}. In both plots of Figure~\ref{fig:Comp} we have focused on the real parts of $C_{7\ga}$, but we have checked that the histograms look similar in case of the imaginary parts and also in case of $C_{8g}$ and the corresponding chirality-flipped Wilson coefficients. We need on average 9~minutes to calculate the KK quark and gauge-boson masses as well as the overlap integrals in the 4D formulation, see Table~\ref{tab:Perf4D5D}. Effectively there are only four different amplitude topologies, since $C_{7\ga,8g}^W$ and $C_{7\ga}^{WW}$ both depend on the same masses and overlap integrals, and analogously for $C_{7\ga,8g}^G$ and $C_{8g}^{GG}$. Therefore we present combined time fractions for those Wilson coefficients in Table~\ref{tab:Perf4D5D}. When we require a consistency of a few per mille for the calculation of the Wilson coefficients between the 4D and 5D approaches, we find that the summation over KK levels is faster by a factor of order $60$. As a consequence, after we have verified that the results in the 5D and 4D approaches agree at the required level of precision, we will implement the equations in \eqref{eqn:C78KK} for the numerical calculation of the Wilson coefficients for most of the RS points used in the phenomenological analysis in Section~\ref{sec:Phenobsga}. 

\subsection{Approximate expressions}\label{sec:AppWilsCoeffbsga}

We emphasize that the integral expressions \eqref{eqn:C785Dqhq}, \eqref{eqn:C785DqBq} and \eqref{eqn:C785DBqB} and the corresponding results in the KK-decomposed theory \eqref{eqn:C78KK} are formally valid to all orders in $v^2/\MKK^2$.~All numerical results that will be presented are calculated from those equations according to the procedure described in \eqref{sec:NumEval}.~However, in order to better understand the size of the different contributions from the diagrams in Figure~\ref{fig:C78} we will also derive some approximate formulas in this section.~The first step is to parametrize the RS corrections relative to the SM Wilson coefficients by
\begin{align}\label{eqn:RScorr}
\begin{split}
C_{7\ga}^{\tx{RS},0}(\mu_W) & = C_{7\ga}^{W,0}(\mu_W)+C_{7\ga}^{WW,0}(\mu_W) + C_{7\ga}^{Z,0}(\mu_W) + C_{7\ga}^{h,0}(\mu_W) - C_{7\ga}^{\SM} (\mu_W)\,,\\
C_{7\ga}^{\tx{RS},\KK}(\mu_\KK) & = C_{7\ga}^{W,\KK}(\mu_\KK) +C_{7\ga}^{WW,\KK}(\mu_\KK)  + C_{7\ga}^{A,\KK}(\mu_\KK) + C_{7\ga}^{G,\KK}(\mu_\KK) \\ 
& \quad\mbox{} + C_{7\ga}^{Z,\KK}(\mu_\KK) + C_{7\ga}^{h,\KK}(\mu_\KK) \,,
\end{split}
\end{align}
where we distinguish corrections that arise from the exchange of only zero modes $C_{7\ga}^{\RS,0}(\mu_W)$ and of loops including at least one virtual KK particle $C_{7\ga}^{\RS,\KK}(\mu_\KK)$.~The individual zero-mode contributions $C_{7\ga}^{B,0}(\mu_W)$ for $B=W,WW,Z,h$ are defined at the electroweak scale $\mu_W \sim m_W$ and are given simply by setting $m=0$ and summing over $n=1,2,3$ in \eqref{eqn:C78KK}.~On the other hand the KK contributions $C_{7\ga}^{B,\KK}(\mu_\KK)$ are defined at the KK scale $\muKK\sim\MKK$.~Analogous parametrizations hold for the chirality flipped Wilson coefficients and for $C_{8g}^{\tx{RS},0}(\mu_W)$ and $C_{8g}^{\tx{RS},\KK}(\mu_\KK)$, where the triple gluon-vertex contribution must be included. In the SM the contribution to the Wilson coefficients at leading order is given by the penguin diagrams (II) and (III) in Figure~\ref{fig:C78}, in which virtual $W^\pm$ bosons and up-type quarks are exchanged. The charm- and top-penguin diagrams yield comparable contributions, since the product of the CKM matrix elements are of similar size, $|\lambda_t|\approx|\lambda_c|$, where $\lambda_q=V^*_{qs} V_{qb}$. Making use of the unitarity of the CKM matrix, which implies $\lambda_u+\lambda_c+\lambda_t=0$, the SM Wilson coefficients at the electroweak scale are given by 
\begin{align}\label{eqn:SMWils}
   C_{7\ga}^{\SM}(\mu_W) = Q_u \left( \frac{I_7(x_t)}{2} - \frac{1}{3} \right) + I_9(x_t^{-1})
    - \frac{5}{12} \,, \qquad 
   C_{8g}^{\SM}(\mu_W) = \frac{I_7(x_t)}{2} - \frac{1}{3} \,,
\end{align}
where $x_t\equiv m_t^2/m_W^2$. The loop functions $I_7(x)$ and $I_9(x)$ can be found in Appendix~\ref{sec:LoopFunctions}. Using $m_{t,\tx{pole}}=176.7^{+4.0}_{-3.4}\,\GeV$ \cite{Agashe:2014kda} we find $C_{7\ga}^\SM(\mu_W) \approx -0.20$ and $C_{8g}^\SM(\mu_W)\approx -0.098$.

\begin{table}[t!]
\begin{center}
\begin{tabular}{|c|ccccccc|} \hline
\multirow{2}{*}{RS corrections} & \multicolumn{7}{|c|}{Median values of the distributions in $[\%]$} \\
& $W$  & $WW$ & $Z$ & $A$ & $G$ & $GG$ & $h$  \\ \hline
$|C_{7\ga}^{B,0}(\mu_W)|/\sum_B|C_{7\ga}^{B,0}(\mu_W)|$ & $34.8$ &$65.0$ & $0.07$ & - & - & - & 0.001\\
$|C_{8g}^{B,0}(\mu_W)|/\sum_B|C_{8g}^{B,0}(\mu_W)|$ & 99.6 & - & 0.4 & - & -& - & 0.007\\\hline
$|C_{7\ga}^{B,\KK}(\mu_\KK)|/\sum_B|C_{7\ga}^{B,\KK}(\mu_\KK)|$ & 33.9 & 51.2 & 7.7 & 0.0001& 0.1& - &  7.0 \\
$|C_{8g}^{B,\KK}(\mu_\KK)|/\sum_B |C_{8g}^{B,\KK}(\mu_\KK)|$ & 52.8 & - & 24.1 & 0.0004 & 0.005 & 0.6 & 22.0 \\\hline
\end{tabular}
\parbox{15.5cm}
{\caption{\label{tab:CiBdev} 
Median values of the distributions in the left column based on RS points with $y_\ast=3$ and $\MKK \in [1,10]\,\TeV$. The median values can be used to estimate the relative size of the RS corrections arising from the exchange of only zero modes $C_{7\ga,8g}^{B,0}(\mu_W)$ and of loops including at least one virtual KK particle $C_{7\ga,8g}^{B,\KK}(\mu_\KK)$. Similar values are obtained in case of the chirality-flipped Wilson coefficients.}}
\end{center}\vspace{-2mm}
\end{table} 

In a first step, we look at the relative size of the RS corrections $C_{7\ga,8g}^{B,0}(\mu_W)$ and $C_{7\ga,8g}^{B,\KK}(\mu_\KK)$ based on a set of RS parameter points.~To this end, we compare the median values of the distributions obtained from calculating $|C_{7\ga,8g}^{B,0}(\mu_W)|$ and $|C_{7\ga,8g}^{B,\KK}(\mu_\KK)|$ and normalizing them to the total sum of each (absolute) correction $\sum_B |C_{7\ga,8g}^{B,0}(\mu_W)|$ and $\sum_B |C_{7\ga,8g}^{B,\KK}(\mu_\KK)|$.~Table~\ref{tab:CiBdev} shows the results.~The general pattern is that the penguin loop diagrams with $W^\pm$-boson exchange give the largest corrections, which is also true for different values of $y_\ast$.~In case of the zero-mode contribution $C_{7\ga,8g}^{\RS,0}$ we find that the largest corrections are given by the deviations of the overlap integrals $V^{W^-}_{203}$ and $V^{W^+}_{303}$ in the RS model with respect to the CKM matrix elements $V^*_{ts}$ and $V_{tb}$ in the SM, and by the coupling of the $W$ boson to right-chiral quarks.~Those corrections stem from the non-flatness of the $W$-boson profile and from deviations of the exact ($Z_2$-even) quark profiles from the ZMA expressions \cite{Casagrande:2008hr}.~The zero-mode contributions from the $Z$ and Higgs bosons arise due to their flavor-changing couplings to quarks in the RS model, but they are suppressed by small down-type quark masses and can be neglected.~To leading order in $v^2/\MKK^2$ we find
\begin{align}\label{eqn:C7gaRS0app}\notag
C_{7\ga}^{\RS,0}(\mu_W) & = \frac{Q_u I_6(x_t) + 2I_8(x_t^{-1})}{4V_{tb}} \hat a_3^{(U)\da} \frac{m_t^2}{\MKK^2} \frac{2F^2(c_{Q_i})}{(1+2\bs c_Q)^2}\!\left[\!\frac{1}{3+2\bs c_{Q}} - \e^{1+2\bs c_Q} + \frac{\e^{2+4\bs c_Q}}{1-2\bs c_Q}\right]\!\hat a_3^{(D)} \\\notag
& \hspace{-16mm} + \frac{Q_u I_7(x_t)+2 I_9(x_t^{-1})}{4V_{ts}^*} \hat a_2^{(D)\da}\!\left[\!\frac{m_t^2}{\MKK^2}\frac{3+14\bs c_Q+8\bs c_Q^2-F^2(\bs c_Q)(4+4\bs c_Q)}{(1-4\bs c_Q^2)(3+2\bs c_Q)} - \frac{L m_W^2}{\MKK^2} \frac{F^2(\bs c_Q)}{3+2\bs c_Q}\!\right]\!\hat a_3^{(U)} \\
& \hspace{-16mm} + \frac{Q_u I_7(x_t)+2I_9(x_t^{-1})}{4V_{tb}} \hat a_3^{(U)\da}\!\left[\!\frac{m_t^2}{\MKK^2}\frac{3+14\bs c_Q+8\bs c_Q^2-F^2(\bs c_Q)(4+4\bs c_Q)}{(1-4\bs c_Q^2)(3+2\bs c_Q)} - \frac{L m_W^2}{\MKK^2} \frac{F^2(\bs c_Q)}{3+2\bs c_Q}\!\right]\!\hat a_3^{(D)},
\end{align}
with the function $F(c)\equiv \sgn(\cos\pi c) \sqrt{(1+2c)/(1-\e^{1+2c})}$ \cite{Casagrande:2008hr}.~The three-component vectors $\hat a_n^{(U)}$ and $\hat a_n^{(D)}$ for $n=1,2,3$ form the columns of the unitary matrices $\bs U_u$ and $\bs U_d$, which define the CKM matrix $\bs V_\tx{CKM}=\bs U_u^\da\,\bs U_d$ in the ZMA~\cite{Casagrande:2008hr}.~The corrections in the first line of \eqref{eqn:C7gaRS0app} stem from the $W^+ \bar u_Rd_R$ coupling, while the remaining terms include corrections to the CKM matrix elements $V_{tb}$ and $V_{ts}^*$ in the RS model.~Concerning the KK contributions we find, contrary to the observation made in \cite{Blanke:2012tv}, (independently of $y_\ast$) that the triple gluon vertex contribution is subdominant and does not enhance the chromomagnetic dipole coefficients. In general we find that the penguin diagrams with the exchange of photon and gluon KK modes yield very small corrections.~At last we can compare the relative magnitude between $C_{7\ga,8g}^{\RS,0}(\mu_W)$ and $C_{7\ga,8g}^{\RS,\KK}(\mu_\KK)$.~Numerically, we find that both contributions are similar in size for $y_\ast\approx 2$.~For larger values of $y_\ast$ the KK contributions dominate in size.

In the next step, we will take a closer look at the main KK contributions to the Wilson coefficients and derive approximate expressions.

\subsubsection*{$\bs{W}$- and $\bs{Z}$-boson contributions} 

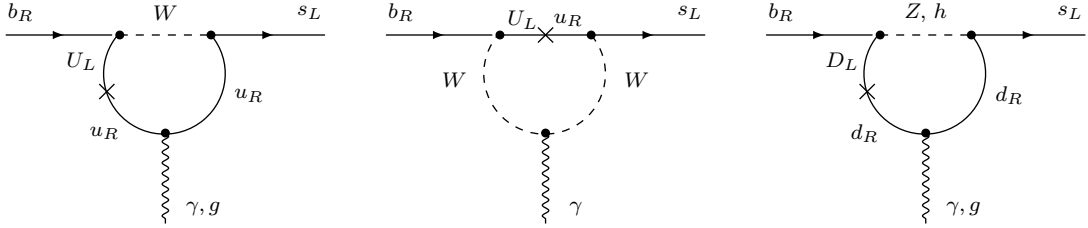
\begin{figure}[t!]
\begin{center}
\begin{tikzpicture}[line width=0.5pt,>=latex,scale=1]
\begin{scope}[xshift=6cm]
\draw[fermion] (-1.7,0) -- (-0.3,0);
\draw[fermionbar] (2.5,0) -- (1,0);
\draw[scalar] (-0.2,0) -- (1,0);
\draw[vector] (0.4,-1.3) -- (0.4,-2.5);
\draw (-0.2,0)  node [vertex]  {};
\draw (1,0)  node [vertex]  {};
\draw (0.4,-1.3)  node [vertex]  {};
\draw (-1.5,0.3) node {\scriptsize $b_R$};
\draw (2.3,0.3) node {\scriptsize $s_L$};
\draw (0.4,0.3) node {\scriptsize $W$};
\draw (1.5,-0.6) node {\scriptsize $$};
\draw (-0.7,-0.6) node {\scriptsize $$};
\draw (0.9,-2.3) node {\scriptsize $\ga,g$};
\draw[fermionnoarrow] (1,0) arc (40:-220:0.8);
\draw (-0.37,-0.75) node[cross] {};
\draw (1.5,-0.8) node {\scriptsize $u_R$};
\draw (-0.4,-1.3) node {\scriptsize $u_R$};
\draw (-0.7,-0.35) node {\scriptsize $U_L$};
\end{scope}
\begin{scope}[xshift=11cm,yshift = 0]
\draw[fermion] (-1.7,0) -- (-0.3,0);
\draw[fermionbar] (2.5,0) -- (1,0);
\draw[fermionnoarrow] (-0.2,0) -- (1,0);
\draw[vector] (0.4,-1.3) -- (0.4,-2.5);
\draw (-0.2,0)  node [vertex]  {};
\draw (1,0)  node [vertex]  {};
\draw (0.4,-1.3)  node [vertex]  {};
\draw (-1.5,0.3) node {\scriptsize $b_R$};
\draw (2.3,0.3) node {\scriptsize $s_L$};
\draw (0.4,0.3) node {\scriptsize $$};
\draw (1.6,-0.6) node {\scriptsize $W$};
\draw (-0.8,-0.6) node {\scriptsize $W$};
\draw (0.8,-2.3) node {\scriptsize $\gamma$};
\draw[scalar] (1,0) arc (40:-220:0.8);
\draw (0.4,0) node[cross] {};
\draw (0.1,0.2) node {\scriptsize $U_L$};
\draw (0.7,0.2) node {\scriptsize $u_R$};
\end{scope}
\begin{scope}[xshift=16cm]
\draw[fermion] (-1.7,0) -- (-0.3,0);
\draw[fermionbar] (2.5,0) -- (1,0);
\draw[scalar] (-0.2,0) -- (1,0);
\draw[vector] (0.4,-1.3) -- (0.4,-2.5);
\draw (-0.2,0)  node [vertex]  {};
\draw (1,0)  node [vertex]  {};
\draw (0.4,-1.3)  node [vertex]  {};
\draw (-1.5,0.3) node {\scriptsize $b_R$};
\draw (2.3,0.3) node {\scriptsize $s_L$};
\draw (0.4,0.3) node {\scriptsize $Z,\,h$};
\draw (1.5,-0.6) node {\scriptsize $$};
\draw (-0.7,-0.6) node {\scriptsize $$};
\draw (0.9,-2.3) node {\scriptsize $\ga,g$};
\draw[fermionnoarrow] (1,0) arc (40:-220:0.8);
\draw (-0.37,-0.75) node[cross] {};
\draw (1.5,-0.8) node {\scriptsize $d_R$};
\draw (-0.4,-1.3) node {\scriptsize $d_R$};
\draw (-0.7,-0.35) node {\scriptsize $D_L$};
\end{scope}
\end{tikzpicture}
\parbox{15.5cm}
{\caption{\label{fig:ScalarDiag}For $y_\ast\gtrsim 1$ those diagrams give the main KK corrections $C_{7\ga,8g}^{\RS,\KK}(\mu_\KK)$ for the transitions $b\to s \ga$ and $b\to s g$ at the one-loop level. Internal solid lines labelled by $u_R$ denotes the exchange of singlet up-type KK quarks, while $U_L,D_L$ imply the exchange of $SU(2)_L$ doublet KK quarks. Crosses denote a chirality flip on the internal KK quark lines. Here, dashed lines labelled with $W$ or $Z$ denote the contributions from the corresponding Goldstone bosons in the Higgs sector.}}
\end{center}\vspace{-2mm}
\end{figure}

We begin with the KK contribution of the $W^\pm$- and $Z$-boson penguin diagrams to the dipole coefficients.~{\color{black}The dominant contributions come from diagrams in which charged/neutral scalar zero modes (stemming from the fifth component of the 5D gauge-boson field and the Goldstone bosons in the Higgs sector) and KK quarks are exchanged, which are implicitly included in the first two expressions in \eqref{eqn:C78KK} for the Wilson coefficients.}\footnote{\color{black}If we would not include the contributions from the Goldstone bosons, the diagrams with gauge-boson zero-modes and KK quarks would be suppressed to leading order by $v^4/\MKK^4$.~In this case the contributions from KK gauge-bosons and KK quarks would be dominant, since they contribute already at order $v^2/\MKK^2$.}~In this case we are allowed to take the limits $x_{B_m}^{q_n}\gg 1$ and $x^{W_m}_{q_n}\ll 1$ for the loop functions in \eqref{eqn:C78KK}, leading to $I_6(x)\approx -1/2$, $I_7(x)\approx 5/12$, $I_8(x)\approx-1$ and $I_9(x)\approx 5/12$.~We find
\begin{align}\label{eqn:C7WZapp}\notag
C_{7\ga,8g}^{W,\tx{KK}}(\mu_\KK) &\approx \frac{\kappa_W^{7\ga,8g}}{4\sqrt2G_F\lambda_t}\,\bigg[\frac{5}{24}\,\bigg( R_{LL}^W - \sum_{n=1}^3 \frac{V_{20n}^{W^-}\,V_{n03}^{W^+}}{m_W^2}\bigg) - \frac{1}{4} \bigg( R_{LR}^W - \frac{m_t}{m_b}\,\frac{V_{203}^{W^-}\,\tilde V_{303}^{W^+}}{m_W^2} \bigg) \bigg] \,, \\\notag
C_{7\ga}^{WW,\tx{KK}} (\mu_\KK)&\approx \frac{\kappa_{WW}^{7\ga}}{4\sqrt2G_F\lambda_t}\,\bigg[ \frac{1}{6}\,\bigg( R_{LL}^W - \sum_{n=1}^3 \frac{V_{20n}^{W^-}\,V_{n03}^{W^+}}{m_W^2} \bigg) - \frac{1}{4}\,\bigg( R_{LR}^W - \frac{m_t}{m_b}\,\frac{V_{203}^{W^-}\,\tilde V_{303}^{W^+}}{m_W^2} \bigg) \bigg] \,, \\
C_{7\ga,8g}^{Z,\tx{KK}} (\mu_\KK) &\approx \frac{\kappa_Z^{7\ga,8g}}{4\sqrt2G_F\lambda_t}\,\bigg[ \frac{5}{24}\,\bigg( R_{LL}^Z-\sum_{n=1}^3 \frac{V_{20n}^{Z}\,V_{n03}^{Z}}{m_Z^2} \bigg) - \frac{1}{4}\,R_{LR}^Z\bigg] \,,
\end{align}
where the boundary terms $R_{LL}^B$ and $R_{LR}^B$ are given in \eqref{eqn:RLLres2} and \eqref{eqn:RLRres3}.~Since the limits of the loop functions we have taken are not valid in case of SM quarks, we have to subtract the contributions from the quark zero modes in \eqref{eqn:C7WZapp}.~We observe that the corrections of $C_{7\ga}^{W,\KK}(\mu_\KK)$ and $C_{7\ga}^{WW,\KK}(\mu_\KK)$ add up constructively, since $\kappa_{W}^{7\ga}=Q_u\,g_5^2/(2\pi r)$ and $\kappa_{WW}^{7\ga}=g_5^2/(2\pi r)$.~We can further simplify the boundary terms $R_{LL}^{W,Z}$ and $R_{LR}^{W,Z}$ in \eqref{eqn:C7WZapp} by expanding them in $v^2/\MKK^2$ and neglecting terms that are suppressed by $m_s/m_b$. We obtain
\begin{align}\label{eqn:RBapp}\notag
R_{LL}^W &\approx - \frac{L}{2\MKK^2}\,(\bs\Delta_D)_{23} - \frac{(\bs\delta_D)_{23}}{\tilde m_{W}^2} \,,\\\notag
R_{LR}^W &\approx - \frac{L}{4\MKK^2}\,(\bs\Delta_D)_{23}+ \frac{1}{2\tilde m_{W}^2}\,\frac{v}{\sqrt2 m_b} \frac{v^2}{\MKK^2} \D_L^{(2)\da}(1^-)\,\Pb_{12}\,\Yb_u\,\Yb_u^\da\,\Yb_d\,\D_R^{(3)}(1^-)  \,, \\\notag
R_{LL}^Z &\approx - \frac{L}{2\MKK^2}\,(\bs\Delta_D)_{23}-\left(1 - \frac{(g_R^d)^2}{(g_L^d)^2} \right)\frac{(\bs\delta_D)_{23}}{\tilde m_{Z}^2} \,, \\\notag
R_{LR}^Z &\approx - \frac{L}{4\MKK^2}\,(\bs\Delta_D)_{23}- \frac{g_R^d}{g_L^d}\,\left(1-\frac{g_R^d}{g_L^d}\right)\frac{(\bs\delta_D)_{23}}{\tilde m_{Z}^2} \\*
&\quad\mbox{}\,+\left(1-\frac{g_R^d}{g_L^d}\right)^2\,\frac{1}{2\tilde m_{Z}^2}\,\frac{v}{\sqrt2 m_b}\,\frac{v^2}{\MKK^2} \D_L^{(2)\da}(1^-)\,\Pb_{12}\,\Yb_d\,\Yb_d^\da\,\Yb_d\,\D_R^{(3)}(1^-) \,,
\end{align}
where in the case of $R_{LR}^{W}$ and $R_{LR}^Z$ we have implemented the relation \cite{Casagrande:2010si} 
\begin{align}
   \frac{1}{\sqrt2}\,\Q_L^{(m)\da}(1^-)\,\Pb_{12}\,\tilde\Yb_q\,\Q_R^{(n)}(1^-)
   = \delta_{mn}\,\frac{m_{q_n}}{v}-\frac{m_{q_m}}{v}\,(\bs\delta_q)_{mn}-\frac{m_{q_n}}{v}\,(\bs\delta_Q)_{mn} \,.\end{align} 
We checked numerically for different values of $y_\ast$ that the approximate formulas \eqref{eqn:C7WZapp} together with \eqref{eqn:RBapp} are accurate at the $10\%$ level compared with the exact expressions. We emphasize that the approximate expressions are independent of the masses and profiles of the KK quark and gauge-boson modes.~In (\ref{eqn:RBapp}) we encounter terms including products of three Yukawa matrices $\Yb_u\Yb_u^\da\Yb_d$ and $\Yb_d\Yb_d^\da\Yb_d$ originating from the IR brane-localized terms in $R_{LR}^{W}$ and $R_{LR}^{Z}$ in \eqref{eqn:RLRres3}. They originate from diagrams exchanging $W^\pm$ and $Z$ Goldstone bosons with a chirality flip on the internal KK quark line as shown in Figure~\ref{fig:ScalarDiag}.~Those terms yield the dominant KK corrections for not too small values of the Yukawa matrix entries, which is approximately fulfilled for RS points with $y_\ast\gtrsim1$.~In this case we can derive simpler expressions and find ($y_\ast\gtrsim 1$)
\begin{align}\label{eqn:C78KKappB}\notag
C_{7\ga}^{W,\tx{KK}}(\mu_\KK) &\approx \frac{Q_u}{\lambda_t}\bigg[-\frac{1}{8}\frac{v}{\sqrt2 m_b} \frac{v^2}{\MKK^2} \D_L^{(2)\da}(1^-)\,\Pb_{12}\,\Yb_u \Yb_u^\da \Yb_d\,\D_R^{(3)}(1^-) + \frac{1}{4}\,\frac{m_t}{m_b}\,\frac{V_{203}^{W^-}\,\tilde V_{303}^{W^+}}{m_W^2}\bigg]\,, \\ 
C_{7\ga}^{Z,\tx{KK}}(\mu_\KK) &\approx \frac{Q_d}{\lambda_t}\bigg[- \frac{1}{16} \frac{v}{\sqrt2 m_b} \frac{v^2}{\MKK^2} \D_L^{(2)\da}(1^-)\,\Pb_{12}\,\Yb_d \Yb_d^\da \Yb_d\,\D_R^{(3)}(1^-) \bigg]\,,
\end{align}
where the chromomagnetic dipole Wilson coefficients $C_{8g}^{W,\KK}$ ($C_{8g}^{Z,\KK}$) can be obtained from the first (second) line in \eqref{eqn:C78KKappB} by sending $Q_u\to1$ ($Q_d\to1$).~Moreover, $C_{7\ga}^{WW,\KK}$ is given analogously by the expression in the first line of \eqref{eqn:C78KKappB} with $Q_u$ set to 1.~We checked numerically that the approximate expressions are valid at the 10\% level with respect to the exact expressions for RS points with $y_\ast\gtrsim 1$.~Approximate formulas for the chirality flipped Wilson coefficients $\tilde C^{B,\KK}_{7\ga,8g}(\mu_\KK)$ can be obtained from \eqref{eqn:C78KKappB} by making the replacements $L\leftrightarrow R$, $\Yb_q \leftrightarrow \Yb_q^\da$, $V_{203}^{W^-}\to \tilde V_{203}^{W^-}$ and $\tilde V_{303}^{W^+}\to V_{303}^{W^+}$.

\subsubsection*{Higgs contribution}

The diagrams contributing to $C_{7\ga,8g}^{h,\KK}(\mu_\KK)$ involve the exchange of the Higgs boson with KK quark modes. For the exchange of KK quarks we can use that $x^{d_n}_h\gg1$, allowing us to take the limits  $I_3(x)=1/(2x) + \ord(x^{-2})$ and $I_4(x)=1/(12x) + \ord(x^{-2})$. The contribution associated with $I_3(x)$ dominates, since this loop function is less suppressed than $I_4(x)$ in the considered limit and the contribution is enhanced by $m_{d_n}/m_b$, which is a large factor for KK modes. If we only keep the corresponding contribution associated with $I_3(x)$, we obtain approximately 
\begin{align}\label{eqn:C78KKapph}
C_{7\ga}^{h,\tx{KK}} (\mu_\KK)&\approx \frac{Q_d}{\lambda_t} \bigg[\frac{1}{16}\,\frac{v}{\sqrt2 m_b}\,\frac{v^2}{\MKK^2}\,\D_L^{(2)\da}(1^-)\,\bs P_{12}\,\Yb_d \Yb_d^\da \Yb_d\,\D_R^{(3)}(1^-) - \frac{1}{8} (\bs\delta_D)_{23} \bigg] \,,
\end{align}
where we have expanded the expression to leading order in $v^2/\MKK^2$ and neglected $m_s/m_b$-suppressed terms.~The corresponding expression for $C_{8g}^{h,\tx{KK}}$ is given by \eqref{eqn:C78KKapph} with $Q_d\to1$.~Numerically we have checked that \eqref{eqn:C78KKapph} is accurate at the 10\% level with respect to the exact expressions.~Note that the $\Yb_d \Yb_d^\da \Yb_d$ structure in \eqref{eqn:C78KKapph} originates from the leading order expansion in $v^2/\MKK^2$ of the function $g(\Xb_d,\Yb_d)$ defined in \eqref{eqn:gfuncbsga} and gives the dominant contribution for $y_\ast\gtrsim 1$. In fact, this term exactly cancels the expression $C_{7\ga}^{Z,\KK}(\mu_\KK)$ in \eqref{eqn:C78KKappB}.~Consequently, for $y_\ast\gtrsim1$ the KK corrections from the $Z$-Goldstone boson and Higgs diagrams cancel to very good approximation.

\subsubsection*{Dependence on $\bs{M_{g^{(1)}}}$ and $\bs{y_\ast}$}

\begin{figure}[t!]
\begin{center}
\psfrag{a}[]{\small $|C^{\RS,\KK}_{7\ga}(\mu_\KK)|$}
\psfrag{b}[]{\small $|\tilde C^{\RS,\KK}_{7\ga}(\mu_\KK)|$}
\psfrag{c}[]{\small $|C^{\RS,\KK}_{8g}(\mu_\KK)|$}
\psfrag{d}[]{\small $|\tilde C^{\RS,\KK}_{8g}(\mu_\KK)|$}
\psfrag{y}[]{\small Points}
\psfrag{e}[b]{\scriptsize \quad \parbox{3cm}{$M_{g^{(1)}}=10\,\TeV$\\$y_\ast=3$}}
\includegraphics[width=0.9\textwidth]{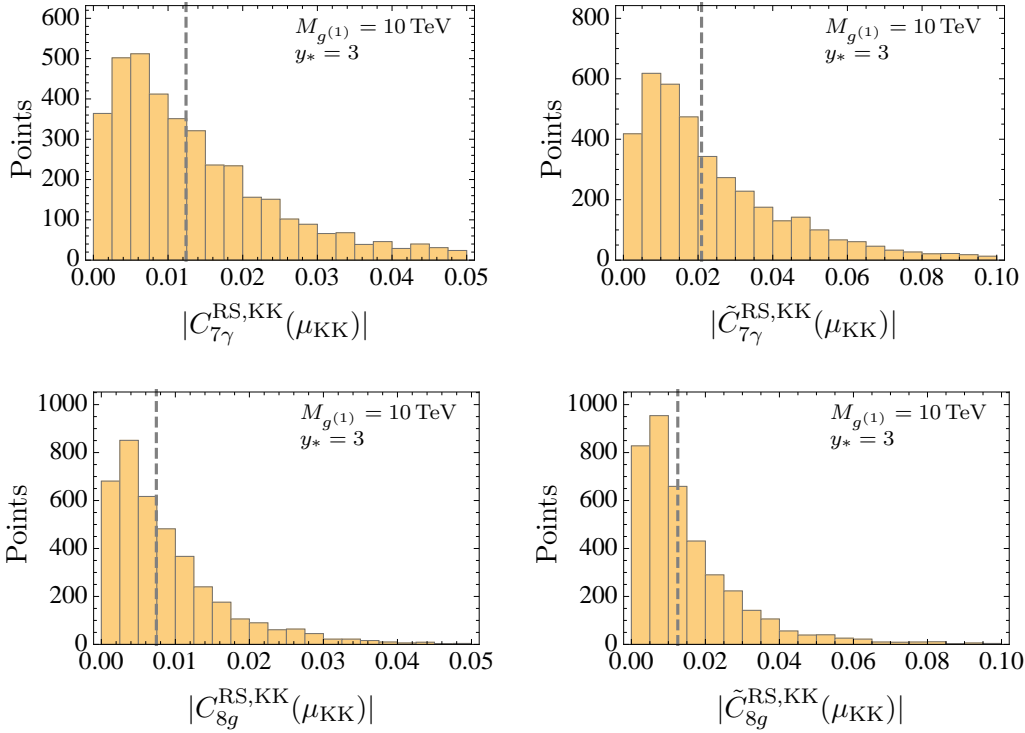} 
\parbox{15.5cm}
{\caption{\label{fig:C7C8muKKsize} 
Absolute corrections from KK modes to the Wilson coefficients at the KK scale for a set of RS points with $y_\ast = 3$ and $M_{g^{(1)}}= 10\,\TeV$. The vertical dashed lines denote the median values of the corresponding distributions. The size of the corrections are equally distributed among the real and imaginary parts of the Wilson coefficients.}}
\end{center}\vspace{-2mm}
\end{figure}

Figure~\ref{fig:C7C8muKKsize} shows histograms of the (absolute) KK corrections $|C_{7\ga,8g}^{\RS,\KK}(\mu_\KK)|$ and $|\tilde C_{7\ga,8g}^{\RS,\KK}(\mu_\KK)|$ for a set of RS parameter points with $y_\ast=3$ and $M_{g^{(1)}}=10\,\TeV$. We choose $y_\ast=3$ to obtain maximal effects, while still staying in the perturbative regime for the Yukawa sector. The value $M_{g^{(1)}}=10\,\TeV$ is close to the lowest KK gluon mass that is consistent with the tree-level analysis of electroweak precision data. The distributions can be well described by the approximate formulas given in \eqref{eqn:C78KKappB} and \eqref{eqn:C78KKapph}.~For different values of $M_{g^{(1)}}$ and $y_\ast$, the corresponding distributions can be obtained by the formula
\begin{align}
   C_{7\ga,8g}^{\RS,\KK}(\mu_\KK) 
   \approx C_{7\ga,8g}^{\RS,\KK}(\mu_\KK)\Big|_{M_{g^{(1)}}=10\,\TeV,\,y_\ast=3} 
   \times \left(\frac{10\,\TeV}{M_{g^{(1)}}}\right)^2 
   \times \left(\frac{y_\ast}{3}\right)^2 ,
\end{align}
which is a good approximation for $y_\ast \gtrsim 1$.  For smaller values of $y_\ast$ the KK contributions do not follow a simple scaling law with $y_\ast$. An analogous equation holds for the distributions of the chirality-flipped Wilson coefficients. In order to get a rough estimate for the typical size of the KK corrections, we can calculate the median values of the distributions of $|C_{7\ga,8g}^{\RS,\KK}(\mu_\KK)|$ and find ($y_\ast \gtrsim 1$)
\begin{align}
   \tx{Median}\left(|C_{7\ga,8g}^{\RS,\KK}(\mu_\KK)|\right) 
   \approx a_{7\ga,8g} \times \left(\frac{10\,\TeV}{M_{g^{(1)}}}\right)^2 
   \times \left(\frac{y_\ast}{3}\right)^2 , 
\end{align}
with $a_{7\ga}=0.012$ and $a_{8g}=0.0073$.~In case of the median values of $|\tilde C_{7\ga,8g}^{\RS,\KK}(\mu_\KK)|$ the coefficients read $\tilde a_{7\ga}=0.020$ and $\tilde a_{8g}= 0.012$. These coefficients represent the median values of the distributions shown in Figure \ref{fig:C7C8muKKsize}. We observe that the KK corrections to the chromomagnetic dipole coefficients are (approximately) smaller by the factor $\kappa_W^{8g}/(\kappa_W^{7\ga}+\kappa_{WW}^{7\ga})=3/5$ with respect to the electromagnetic dipole coefficients. Furthermore, we find the general pattern that the chirality-flipped Wilson coefficients are enhanced, which can be explained by the different localization of the left- and right-handed bottom-quark profiles. The left-handed bottom quark profile, which enters $\tilde C^{\RS,\KK}_{7\ga}(\muKK)$ and $\tilde C^{\RS,\KK}_{8g}(\muKK)$, is more localized towards the IR brane ($c_{b_L}=c_{Q_3}>c_{b_R}=c_{d_3}$) and is thus more sensitive to flavor-violating effects. This hierarchy of the bulk mass parameters is due to the large mass difference of the top and the bottom quark, which requires that $F(c_{b_L}) > F(c_{b_R})$. 

We can (approximately) relate our results with the numerical analysis of the Wilson coefficients performed in \cite{Blanke:2012tv}, where the case of $y_\ast=3$ and $M_{g^{(1)}}=2.5\,\TeV$ was discussed. When we consider $M_{g^{(1)}}=2.5\,\TeV$ we find that the corrections $C_{7\ga}^{\RS,\KK}(\MKK)$ are larger by a factor of roughly $5$ compared with \cite{Blanke:2012tv}. In case of $C_{8g}^{\RS,\KK}(\MKK)$ we find that the corrections are similar in size. Concerning the corrections to the chirality-flipped Wilson coefficients, we find that they are larger by a factor of $\sim2$ with respect to $C_{7\ga,8g}$, while \cite{Blanke:2012tv} reported a stronger enhancement by one order of magnitude.~While we have been unable to trace the origin of these discrepancies, the fact that we have performed our analysis using both the 5D and 4D formulations of the RS model and found consistent results in both approaches provides a highly non-trivial cross-check of our calculations.

\subsection{Comment on the narrow bulk-Higgs scenario} 
\label{sec:NarrowBulk}

We have observed that the sum of the KK contributions $C_{7\ga,8g}^{h,\KK}(\mu_\KK)$ and $C_{7\ga,8g}^{Z,\KK}(\mu_\KK)$ cancels to a very good approximation for $y_\ast\gtrsim 1$. In this section we investigate whether this cancellation still holds in the narrow bulk-Higgs model. 

The Higgs contribution was already calculated in Section~\ref{sec:HiggsContrbsga}, and the final result has been given in \eqref{eqn:C785Dqhq2} and \eqref{eqn:gfuncbsga}, including the case of a narrow bulk-Higgs. It follows that the first term (containing the $\Yb_d\Yb_d^\da\Yb_d$ structure) in the approximative formula for $C_{7\ga,8g}^{h,\KK}(\mu_\KK)$ in \eqref{eqn:C78KKapph} must be multiplied with a minus sign in the case of a narrow bulk-Higgs. Concerning the $Z$-boson contribution we focus on the scalar diagram (IIb).~For $y_\ast\gtrsim 1$ the dominant corrections are due to the exchange of the Goldstone $Z$ boson.~The corresponding terms can be extracted from $C_{7\ga,8g}^{Z,\tx{scalar}}$ in \eqref{eqn:C78B}. We can proceed analogously to the calculation of the Higgs contribution discussed in Section~\ref{sec:HiggsContrbsga}. Therefore, we refrain from giving more details here and quote the final result ($y_\ast\gtrsim1$)
\begin{align}\label{eqn:C78Z5v2}\notag
C_{7\ga,8g}^{Z,\tx{scalar}} &\approx \frac{\kappa_Z^{7\ga,8g}}{4G_F \lambda_t}\,\frac{1}{(g_L^d)^2}\,\frac{1}{2v}\,\D_L^{(2)\da}(1^-) \Bigg\{ \Pb_{12}\,\frac{h(\Xb_d,\tilde\Yb_d)}{4m_b}\,\D_R^{(3)}(1^-) \\
&\quad\mbox{}+ 2\pi \int_0^\infty dk_E\,\frac{\Zb_d(k_E^2)}{1+\Zb_d(k_E^2)}\Bigg[ \frac{1}{\rh} \frac{\Pb_+}{\Rb_Q} \D_L^{(3)}(1^-)\bigg( \frac{5k_E}{32}\pa_{k_E} + \frac{3k_E^2}{32}\pa_{k_E}^2 + \frac{k_E^3}{96}\pa_{k_E}^3 \bigg) \\\notag
&\quad\mbox{}+ \frac{\Pb_{12}}{m_b}\,\tilde\Yb_d\,\D_R^{(3)}(1^-)\bigg( k_E + \frac{7k_E^2}{8} \pa_{k_E} + \frac{k_E^3}{8}\pa_{k_E}^2 \bigg) \Bigg]\frac{\e^2\MKK^2}{L^2 \tilde m_Z^2}\,B_{Z}^\tx{scalar}(1^-,1^-;k_E^2) \Bigg\} \,.
\end{align}
The scalar $Z$-boson propagator behaves like $B_Z^\tx{scalar}(1^-,1^-;k_E^2) = L^2 \tilde m_Z^2 / (2\pi \e^2 k_E^2 \MKK^2) + \ord(k_E^{-4})$ for large Euclidean momenta, rendering the integral finite. The function $h(\Xb_q,\tilde\Yb_q)$ in the first line of \eqref{eqn:C78Z5v2} is given by
\begin{align}\label{eqn:hfuncbsga}
\begin{split}
   h(\Xb_q,\tilde\Yb_q)\Big|_\tx{brane Higgs} 
   &= - \frac{\rh^2\tilde\Yb_q\tilde\Yb_q^\da}{1+\rh^2\Yb_q\tilde\Yb_q^\da}\,\tilde\Yb_q 
    = - \rh^2\Yb_q\Yb_q^{\da}\Yb_q + {\cal O}(\rh^4) \,, \\
   h(\Xb_q,\tilde\Yb_q)\Big|_\tx{narrow bulk-Higgs} 
   &= - \frac{1}{2}\left(\frac{\Xb_d\coth\Xb_d}{\cosh^2\Xb_d} - 1 \right) \tilde\Yb_q 
   = \frac{\rh^2}{3} \Yb_q\Yb_q^{\da}\Yb_q + {\cal O}(\rh^4) \,,
\end{split}
\end{align}
where the difference between the brane-localized and narrow bulk-Higgs scenario is to leading order the relative factor $-1/3$. Thus, in the narrow bulk-Higgs scenario the approximate expression for $C_{7\ga,8g}^{Z,\tx{KK}}(\mu_\KK)$ in \eqref{eqn:C78KKappB} must be multiplied with $-1/3$. Finally, adding $C_{7\ga,8g}^{h,\KK}(\mu_\KK)$ and $C_{7\ga,8g}^{Z,\KK}(\mu_\KK)$ we obtain approximately (for $y_\ast\gtrsim1$)
\begin{align}\label{eqn:C78KKapph2}
\begin{split}
   C_{7\ga}^{h,\tx{KK}}(\mu_\KK) +  C_{7\ga}^{Z,\tx{KK}}(\mu_\KK) 
   &\approx \frac{Q_d}{2\lambda_t} \frac{v}{\sqrt2 m_b}\frac{v^2}{\MKK^2}\,\D_L^{(2)\da}(1^-)\,
    \bs P_{12}\,\Yb_d \Yb_d^\da \Yb_d\,\D_R^{(3)}(1^-) \\
    &\;\quad  \!\times\! \left\{\begin{array}{ll} 0 \,; &\quad \tx{brane Higgs\,,} \\ 
    -\frac{1}{12} \,; & \quad \tx{narrow bulk-Higgs\,.} \end{array}\right. 
    \end{split}
\end{align}
The corresponding expression for the coefficient of the chromomagnetic dipole operator is obtained by replacing $Q_d\to 1$.~The structure $\Yb_d\Yb_d^\da\Yb_d$ cancels in the brane-localized Higgs case, while there remains a non-zero contribution in case of the narrow bulk-Higgs scenario.~This observation and the factors $0$ and $-\frac{1}{12}$ in \eqref{eqn:C78KKapph2} were first encountered in \cite{Beneke:2012ie,Beneke:2014sta} for the case of lepton penguin loops.~In fact, we can exactly reproduce the result (26) in \cite{Beneke:2014sta} for the Higgs contribution in the lepton sector from equation \eqref{eqn:C78KKapph2} by replacing $Q_d\to Q_e=-1$ and by accounting for factors in the definition of the Wilson coefficient.~We note that while the contributions from the neutral scalars cancel for $y_\ast\gtrsim1$ in the brane-localized Higgs scenario, we still have left over the (dominant) contributions from the charged Goldstone bosons.~The latter contribution is absent in case of the leptonic dipole coefficient for the transition $l_i\to l_j\gamma$ in the minimal RS model, which does not include right-chiral $SU(2)_L$ singlet neutrinos.~However, a non-zero contribution from neutral scalars would be present in case of the RS model with custodial protection, which can be found in \cite{Moch:2014ofa,Beneke:2015lba}.

Finally, we remark that in order to calculate the contribution of the charged $W^\pm$ Goldstone-bosons in the narrow bulk-Higgs scenario, we need to perform $t,t'$ integrations over matrix-valued functions mixing $\Yb_u$ with $\Yb_d$. Since we could not handle those integrations in a semi-analytic way we will therefore confine our analysis to the brane-localized Higgs scenario in the remainder of this paper. 

\subsection{Renormalization-group running to the meson scale} 
\label{sec:RGbsga}

In the previous section we have analyzed the corrections to the SM Wilson coefficients from the zero modes $C_{7\ga,8g}^{\RS,0}(\mu_W)$ defined at the electroweak scale $\mu_W\sim m_W$ and from the KK particles $C_{7\ga,8g}^{\RS,\KK}(\mu_\KK)$ at the KK scale $\muKK\sim \MKK$. For the phenomenology we are interested in the Wilson coefficients $C_{7\ga,8g}(\mu_b)$ at the meson scale $\mu_b\sim m_b$. When running down from higher scales down to $\mu_b$, QCD effects generically lead to a mixing between dimension-6 operators. The general effective Lagrangian for a new-physics model at a high scale ($\mu_{\rm KK}$ in our case) can be written in the form
\begin{align}\label{eqn:LeffRSbsga}
\begin{split}
   \La_\text{eff} = \frac{G_F}{\sqrt2}\,\lambda_t\,\Bigg[ 
   &\sum_{q=u,c,t} \sum_{i=1,2}\,C_i^{(q)}\,Q_i^{(q)} + C_{7\ga}\,Q_{7\ga} + C_{8g}\,Q_{8g} \\
   &\mbox{}+ \sum_{i=1,2} \sum_{A=L,R} \bigg( \sum_{q=u,c,t,d,s,b} C_{i}^{(q),LA}\,Q_{i}^{(q)}(L,A) 
    + \hat C_{i}^{(d),LA}\,\hat Q_{i}^{(d)}(L,A) \bigg) \\
   &\mbox{}+ \left\{ Q\to\tilde Q,\,C\to\tilde C,\,L\leftrightarrow R \right\} \Bigg] \,,
\end{split}
\end{align}
where we adopt the notation of \cite{Buras:2011zb}. Here $Q_{1,2}^{(q)}$ are the charged current-current operators, $Q_{7\ga,8g}$ are the dipole operators, and $Q_{1,2}(A,B)$, $\hat Q_{1,2}(A,B)$ are neutral current-current operators (including the four-quark QCD and electroweak penguin operators of the SM). In the RS model such operators are induced by the exchange of the heavy KK modes of the $Z$ boson, photon and gluon.~For simplicity, however, they will be neglected in our analysis.~When running down from $\muKK\sim\MKK$ to $\mu_W\sim m_W$ we consider only the mixing between $Q_{7\ga}$ and $Q_{8g}$, which accounts for the dominant evolution effects.~Between the electroweak scale and the meson scale we further include the RS contribution to the charged current-current operator $Q_2^{(c)}=4(\bar s_j\ga_\mu P_L c_j)\,(\bar c_i\ga^\mu P_L b_i)$, which is also important in the SM calculation.~The corresponding Wilson coefficient in the SM reads $C_2^{(c)\SM}(\mu_W)=-\lambda_c/\lambda_t\approx 1$.~In the RS model the $W$-boson coupling to quarks receives corrections, which we include later in the analysis by defining the Wilson coefficient
\begin{align}
C_2^{(c)\RS,0}(\mu_W) = - \frac{1}{4\sqrt2 G_F\lambda_t}\,\frac{g_5^2}{2\pi r}\,\frac{1}{m_W^2}\,V_{202}^{W^-}\,V_{203}^{W^+} - C_2^{(c)\SM}(\mu_W) \,.
\end{align}
The overlap integrals $V_{nmk}^{W^\pm}$ are defined in relation \eqref{eqn:VW4D} of Appendix~\ref{app:FR}.

\begin{figure}[t!]
\begin{center}
\psfrag{a}[]{\small $\re C^\RS_{7\ga}(\mu_b)$}
\psfrag{b}[]{\small $\re \tilde C^\RS_{7\ga}(\mu_b)$}
\psfrag{c}[]{\small $\re C^\RS_{8g}(\mu_b)$}
\psfrag{d}[]{\small $\re\tilde C^\RS_{8g}(\mu_b)$}
\psfrag{y}[]{\small Points}
\psfrag{e}[b]{\scriptsize \qquad\qquad\qquad\qquad \parbox{3cm}{$M_{g^{(1)}}=10\,\TeV$\\$y_\ast=3$}}
\includegraphics[width=0.9\textwidth]{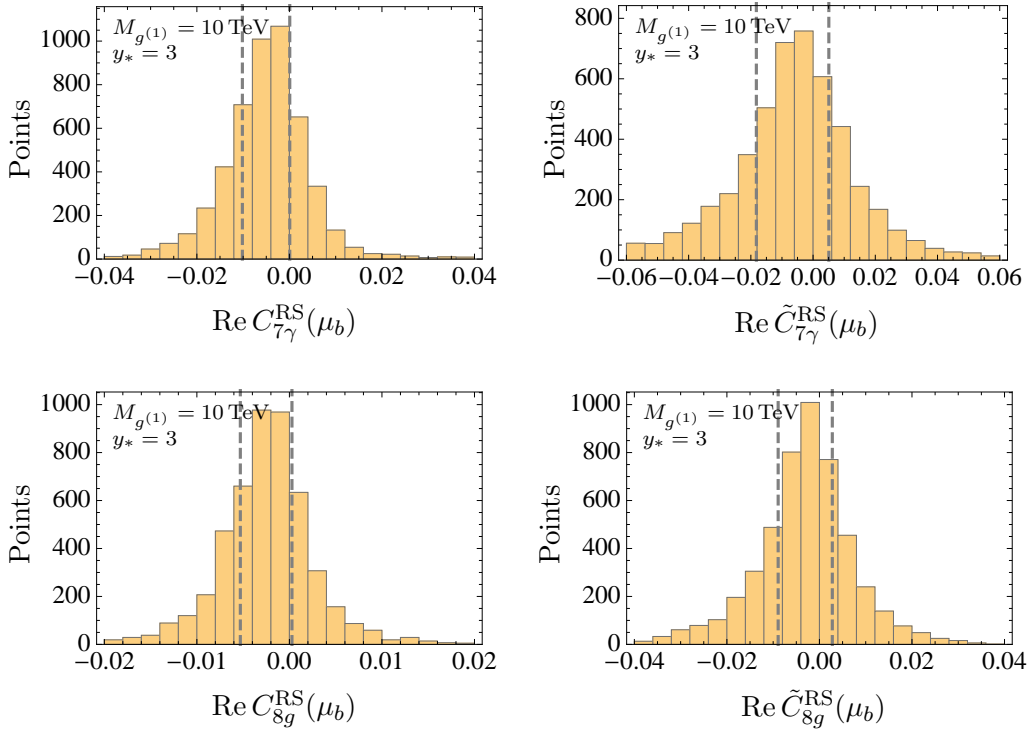} 
\parbox{15.5cm}
{\caption{\label{fig:C7C8mubsize} 
Corrections to the effective Wilson coefficients in the SM $C^\SM_{7\ga}(\mu_b)\approx -0.30$ and $C^\SM_{8g}\approx 0.15$ at the $B$-meson scale $\mu_b=4.8\,\GeV$ for a set of RS points with $y_\ast = 3$ and $M_{g^{(1)}}= 10\,\TeV$. The vertical dashed lines denote the lower ($25\%$) and upper ($75\%$) quartiles of the distributions. The distributions look similar for the imaginary parts of the Wilson coefficients.}}
\end{center}\vspace{-2mm}
\end{figure}

Let us outline the basic steps how we evolve the Wilson coefficients down to the meson scale. We need the evolution matrix $U(\mu_1,\mu_2)$ which can be calculated from the anomalous-dimension matrix $\hat\gamma$ of our operator basis, which is contained in the more general basis considered in \cite{Buras:2011zb}. The running between scales is accomplished at leading order by
\begin{align}
   U(\mu_1,\mu_2) = \hat V\left( \left[\frac{\alpha_s(\mu_2)}{\alpha_s(\mu_1)} 
    \right]^{\frac{\vec\ga^{(0)}}{2 \beta_0}}\right)_{\!D} \!\hat V^{{-1}} \,,
\end{align} 
where $\vec \ga^{(0)}$ includes the eigenvectors of the transposed anomalous-dimension matrix $\hat\ga^{(0)^T}$. The matrices $\hat V^{}$ diagonalize $\hat\ga^{(0)^T}$, such that $\hat V^{{-1}} \hat\ga^{(0)^T} \hat V^{}$ is diagonal. Note that $\hat\gamma^{(0)}$ and $\beta_0=(33-2f)/3$ depend on the number of active flavors $f$. Between the scales $\mu_\KK$ and $\mu_b$ we integrate out the top quark, such that the evolution matrix splits into two parts,
\begin{align}
   U(\mu_b,\muKK) = U^{(f=5)}(\mu_b,\mu_W)\,U^{(f=6)}(\mu_W,\muKK) \,.
\end{align}
The RS corrections at the KK scale, coming from integrating out heavy KK resonances, are contained in the coefficient $\vec C^{\RS,\KK}(\muKK)$. The evolution down to the electroweak scale is given by $\vec C^{\RS,\KK}(\mu_W)=U(\mu_W,\muKK)\,\vec C^{\RS,\KK}(\muKK)$. At the electroweak scale the $W$ boson and the top quark are integrated out. Matching on this new effective Lagrangian we include the contributions from the boson and fermion zero modes, which are given by $\vec C^{(0)}(\mu_W)=\vec C^\SM(\mu_W)+\vec C^{\RS,0}(\mu_W)$, where $\vec C^{\RS,0}(\mu_W)$ contains the zero-mode corrections to the SM coefficient. Next we evolve this contribution down to the meson scale. The effective Wilson coefficient reads
\begin{align}
   \vec C(\mu_b) = \vec C^\SM(\mu_b) + U(\mu_b,\muKK)\,\vec C^{\RS,\KK}(\muKK)
    + U(\mu_b,\mu_W)\,\vec C^{\RS,0}(\mu_W) \,,
\end{align}
where the SM Wilson coefficients are given by $C_{7\ga}^\SM(\mu_b)\approx -0.30$ and $C_{8g}^\SM (\mu_b) \approx -0.15$. Performing all steps including the dipole and the charged current-current operators, the RS corrections to the electro- and chromomagnetic dipole operators at the $B$-meson scale are given by
\begin{align}\label{eqn:C78mub}
\begin{split}
   C_{7\ga}^\RS(\mu_b) &= 0.475\,C_{7\ga}^{\RS,\KK}(\muKK) + 0.123\,C_{8g}^{\RS,\KK}(\muKK)
    + 0.667\,C_{7\ga}^{\RS,0}(\mu_W) \\
   &\quad\mbox{}+ 0.092\,C_{8g}^{\RS,0}(\mu_W) - 0.174\,C_2^{(c)\RS,0}(\mu_W) \,, \\
   C_{8g}^\RS(\mu_b) &= 0.522\,C_{8g}^{\RS,\KK}(\muKK) + 0.702\,C_{8g}^{\RS,0}(\mu_W)
    - 0.080\,C_2^{(c)\RS,0}(\mu_W) \,.
\end{split}
\end{align}
The numbers in front of the KK corrections $C_{7\ga,8g}^{\RS,\KK}(\muKK)$ have been calculated for $\muKK=1\,\TeV$, $\mu_b=4.8\,\GeV$ and $\mu_W=80.4\,\GeV$.~In our numerical analysis we set $\mu_\KK=\MKK$ for each RS point.~Relation~\eqref{eqn:C78mub} also holds for the chirality-flipped Wilson coefficients, since (massless) QCD is blind to the fermion chirality. 

\begin{figure}[t!]
\begin{center}
\psfrag{a}[]{\small $\re C^\RS_{7\ga}(\mu_b)$}
\psfrag{b}[]{\small $\re C^\RS_{8g}(\mu_b)$}
\psfrag{c}[]{\small $\re \tilde C^\RS_{7\ga}(\mu_b)$}
\psfrag{d}[]{\small $\re \tilde C^\RS_{8g}(\mu_b)$}
\psfrag{e}[]{\scriptsize \qquad\qquad\qquad\qquad \parbox{3cm}{$M_{g^{(1)}}=10\,\TeV$\\}}
\includegraphics[width=0.9\textwidth]{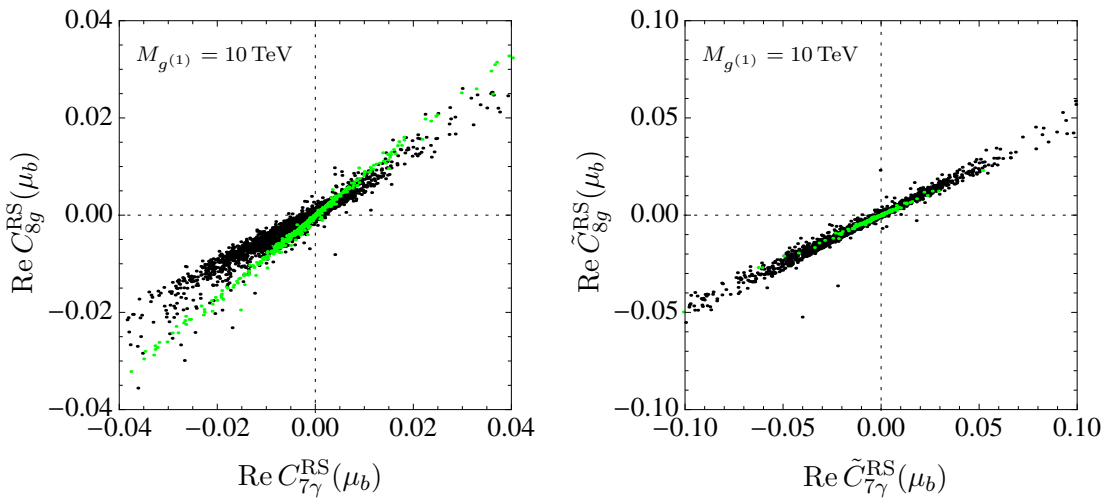}
\parbox{15.5cm}
{\caption{\label{fig:C7C8corr}
Left (right) is shown the approximate linear correlation between the RS corrections to the (flipped) electro- and chromomagnetic dipole coefficients for RS points with $y_\ast=3\,(0.5)$ and $M_{g^{(1)}}=10\,\TeV$.~A similar correlation is found for the imaginary parts of the Wilson coefficients. There is no correlation between $C_{7\ga,8g}^\RS(\mu_b)$ and their chirality-flipped counterparts.}}
\end{center}\vspace{-2mm}
\end{figure}

Figure~\ref{fig:C7C8mubsize} shows the RS corrections to the SM values of the dipole coefficients at the $B$-meson scale for RS points with $y_\ast=3$ and $M_{g^{(1)}}=10\,\TeV$. In general, the RS corrections to the SM Wilson coefficients $C^\SM_{7\ga}(\mu_b)$ and $C^\SM_{8g}(\mu_b)$ lie in the few percent region. On the other hand, the relative corrections to the chirality-flipped Wilson coefficients are large, since in the SM $\tilde C_{7\ga}^\SM(\mu_b)$ and $\tilde C_{8g}^\SM(\mu_b)$ are suppressed by $m_s/m_b$.~The dominant contributions to $C_{7\ga}^\RS(\mu_b)$ are given by the RG-evolved KK and zero-mode corrections $C_{7\ga}^{\RS,\KK}(\muKK)$ and $C_{7\ga}^{\RS,0}(\mu_W)$.~An analogous statement holds for $C_{8g}^\RS(\mu_b)$.~The mixing of the chromomagnetic dipole Wilson coefficients into $C_{7\ga}^\RS(\mu_b)$ yields a correction of roughly $10\%$ for $y_\ast=3$ and $7\%$ for $y_\ast=0.5$.

Finally, we remark that $C_{7\ga}^\RS(\mu_b)$ and $C_{8g}^\RS(\mu_b)$ are linearly correlated, which can be seen in Figure~\ref{fig:C7C8corr}.~This is expected, since the main contributions arise from penguin diagrams containing $W^\pm$-boson modes, and they only differ by the factor $(\kappa_{W}^{7\ga}+\kappa_{WW}^{7\ga})/\kappa_{W}^{8g}=1+Q_u=5/3$.~The coefficients $C_{7\ga}^\RS(\mu_b)$ and $\tilde C_{7\ga}^\RS(\mu_b)$, as well as $C_{8g}^\RS(\mu_b)$ and $\tilde C_{8g}^\RS(\mu_b)$, are however largely uncorrelated.

\section{Phenomenology}
\label{sec:Phenobsga}

\subsection{Branching ratio $\tx{Br}\bs(\bar{\bs B}\bs{\to X_s\ga)}$}

We begin with the CP- and isospin-averaged $\bar B\to X_s\gamma$ branching ratio, which is one of the cleanest observables in $B$ physics from a theoretical point of view. Measurements lead to the combined result $\text{Br}(\bar B\to X_s \gamma)_\tx{exp}=(3.43\pm 0.21\pm 0.07)\times 10^{-4}$ \cite{Amhis:2014hma} for the branching ratio define with a lower cut $E_\gamma>E_0=1.6\,\GeV$ on the photon energy in the meson rest frame. The SM prediction at NNLO reads $\text{Br}(\bar B\to X_s\gamma)_\tx{SM}=(3.36\pm 0.23)\times 10^{-4}$ \cite{Misiak:2015xwa} for $E_0=1.6\,\GeV$, showing that both values are compatible at the $1\sigma$ level. In order to estimate the effects of the RS model we use the approximate formula (for $E_\ga>1.6\,\GeV$) \cite{Kagan:1998bh}
\begin{align}\label{eqn:BrXsga}
   \text{Br}(\bar B\to X_s\ga) = \text{Br}(\bar B\to X_s\ga)_\SM 
    + 0.00247 \left( |C_{7\ga}^\RS|^2 + |\tilde C_{7\ga}^\RS|^2 - 0.706 \re C_{7\ga}^\RS \right) ,
\end{align}
where the Wilson coefficients have to be evaluated at the $B$-meson scale $\mu_b=4.8\,\GeV$. While all known non-perturbative contributions (see in particular \cite{Benzke:2010js} for an estimate of non-local hadronic effects) are taken into account, the RS corrections are included at leading order in $\alpha_s$. This is accurate enough to estimate their impact, because these effects are generally small. 

\begin{figure}[t]
\begin{center}
\begin{minipage}{0.49\linewidth}
\psfrag{x}[]{\small $\tx{Br}(\bar B\to X_s\ga)\;[10^{-4}]$}
\psfrag{y}[]{\small $\tx{Br}(\bar B \to X_s l^+ l^-)\;[10^{-7}] $}
\psfrag{a}[l]{\scriptsize \hspace{-2mm}\parbox{3cm}{$M_{g^{(1)}}=10\,\TeV$\\}}
\includegraphics[width=0.96\linewidth]{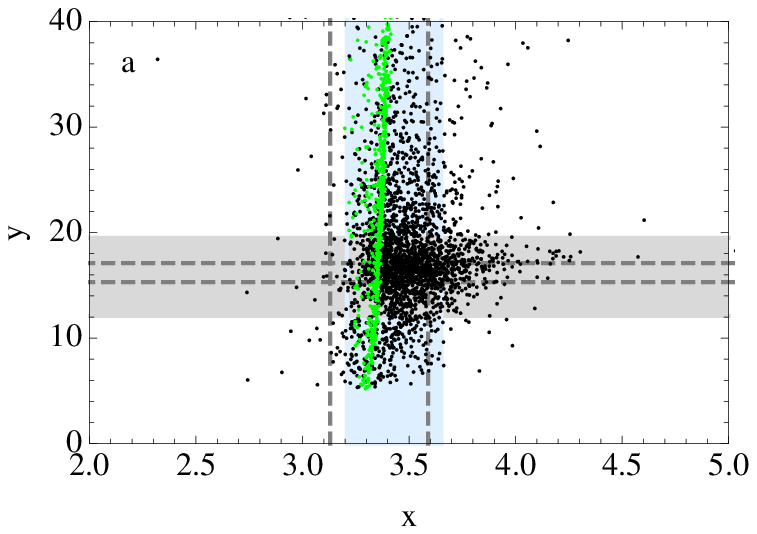} 
\end{minipage} \hfill
\begin{minipage}{0.49\linewidth}
\psfrag{x}[]{\small $\tx{Br}(\bar B\to X_s \ga)\; [10^{-4}]$}
\psfrag{y}[]{\small $S_{K^*\ga}\;[\%]$}
\psfrag{a}[l]{\scriptsize \hspace{-2mm} \parbox{3cm}{$M_{g^{(1)}}=10\,\TeV$\\}}
\includegraphics[width=1\linewidth]{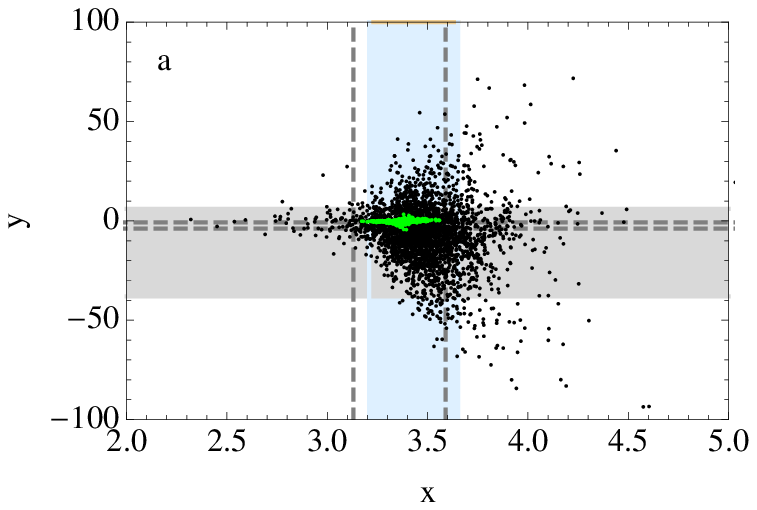} 
\end{minipage}
\parbox{15.5cm}
{\caption{\label{fig:Brbsga} 
Left is shown the branching ratio $\tx{Br}(\bar B\to X_s\,l^+l^-)$ with respect to the inclusive radiative decay $\tx{Br}(\bar B\to X_s\ga)$. The right plot shows the time-dependent CP asymmetry $S_{K^*\ga}$ as a function of the branching ratio for the decay $\bar B\to X_s\ga$. In both plots the light gray and blue bands show the $1\sigma$ experimental error margins while the area between the dashed lines contains the SM prediction with $1\sigma$ uncertainty. All black (green) points represent possible RS scenarios with $y_\ast=3\,(0.5)$  and $M_{g^{(1)}}=10\,\TeV$.}}
\end{center}\vspace{-2mm}
\end{figure}

The dominant corrections in \eqref{eqn:BrXsga} stem from the last term in the round bracket, which is proportional to $\re C_{7\ga}^\RS(\mu_b)$. The squared contributions (and in particular the chirality-flipped Wilson coefficient) have only a minor impact. Since the KK contributions are approximately proportional to $y_\ast^2$, the biggest effects can be expected for large values of $y_\ast$.~There exists an upper limit $y_\ast \leq y_\tx{max}$ when requiring that the Yukawa sector remains in the perturbative regime, and it is conventional to choose the value $y_\tx{max}\approx 3$ \cite{Csaki:2008zd}. We have generated RS parameter sets for different values of $y_\ast$ and $M_{g^{(1)}}$ (Yukawa matrices and quark bulk masses), which correctly reproduce the SM quark masses and the Wolfenstein parameters, see Section~\ref{sec:NumEval} for more details. In the left plot in Figure~\ref{fig:Brbsga}, we show predictions for the branching ratios $\tx{Br}(\bar{B}{\to X_s\ga)}$ and $\tx{Br}(\bar B\to  X_s\,l^+ l^-)$, which will be discussed in the next section, for a large set of RS model points with $M_{g^{(1)}}=10\,\TeV$. The black (green) points are obtained with $y_\ast=3\,(0.5)$.~We find that more than 90\% (99\%) of these points lie within the experimental $2\sigma$ bands.~The RS corrections to $\tx{Br}(\bar B\to X_s\ga)$ approximately scale with $y_\ast^2$ in the region where $y_\ast\gtrsim 2$.~For smaller values of $y_\ast$ there is no simple scaling dependence.~In general, we find that the size of the RS corrections to $\tx{Br}(\bar B\to X_s\ga)$ is strongly dependent on $y_\ast$, in contrast to the observation of \cite{Blanke:2012tv}, where no significant correlation in their numerical scan was reported. If we require that at least 10\% of the RS points lie within the $2\sigma$ error margin, we can derive the lower bound $M_{g^{(1)}} \geq 3.4\,\TeV$ for $y_\ast=3$. This bound cannot compete with the constraints from a tree-level analysis of electroweak precision data at 95\% CL. On the other hand, if we set $M_{g^{(1)}}=2.5\,\TeV$, which is the lowest value allowed from the direct search of resonances in the invariant mass spectrum of $t\bar t$ production by the ATLAS \cite{TheATLAScollaboration:2013kha} and CMS \cite{CMS:lhr} collaborations, the maximal Yukawa value can be constrained from above to $y_\ast\lesssim2$.

Let us comment on two further constraints on the RS parameter space.~First, we consider the CP-violating observable $\e_K$ in kaon mixing, which can receive large corrections in the RS model due to a strong chiral enhancement of the four-quark operator $Q_4=(\bar d_Rs_L)(\bar d_L s_R)$, after performing the RG running from $\MKK$ down to the kaon mass.~When we impose the constraint that the RS prediction for $\e_K$ lies in the $2\sigma$ region of the SM prediction we find that roughly $15\%$ ($0.7\%$) of the black (green) points in Figure~\ref{fig:Brbsga} survive.~The fraction of allowed points decreases with smaller values of $y_\ast$, since the RS corrections to $\e_K$ are approximately proportional to $1/y_\ast^2$.~Still, the shape of the distribution of points is not strongly affected, since $\e_K$ is uncorrelated with the observables discussed in this paper.~Secondly, we can discuss the impact of Higgs physics, where the strongest bounds arise from the signal rates of the Higgs decaying into pairs of electroweak gauge bosons.~Comparing with LHC data one finds the condition $M_{g^{(1)}} \geq  (15-20)\,\TeV \times (y_\ast/3)$ at $95\%$ CL \cite{Malm:2014gha}.~Applying this bound to the RS points with $M_{g^{(1)}}=10\,\TeV$ would exclude the black points ($y_\ast=3$) but still allow for the green points ($y_\ast=0.5$).

\subsection{Branching ratio $\tx{Br}\bs(\bar{\bs B}\bs{\to X_s\,l^+l^-)}$}

\begin{figure}[t!]
\begin{center}
\psfrag{a}[]{\small $\re C^\RS_{9}(\mu_b)$}
\psfrag{b}[]{\small $\re \tilde C^\RS_{9}(\mu_b)\;[10^{-3}]$}
\psfrag{c}[]{\small $\re C^\RS_{10}(\mu_b)$}
\psfrag{d}[]{\small $\re \tilde C^\RS_{10}(\mu_b)\;[10^{-3}]$ }
\psfrag{y}[]{\small Points}
\psfrag{e}[l]{\scriptsize \parbox{3cm}{$M_{g^{(1)}}=10\,\TeV$\\$y_\ast=3$}}
\includegraphics[width=0.9\textwidth]{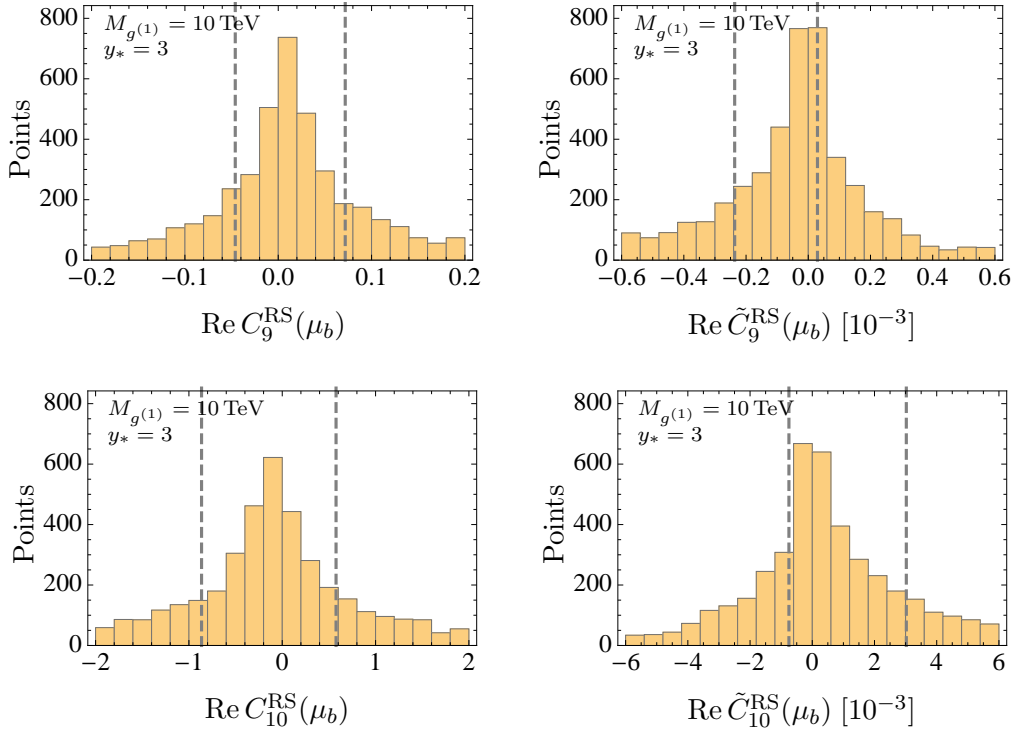} 
\parbox{15.5cm}
{\caption{\label{fig:C9C10size} 
Corrections to $C_{9,10}(\mu_b)$ and $\tilde C_{9,10}(\mu_b)$ in the RS model for parameter sets with $y_\ast = 3$ and $M_{g^{(1)}}=10\,\TeV$. The vertical dashed lines denote the lower and upper quartiles of the distributions, i.e.\ 50\% of the RS points are included in the region in between the dashed lines. In the SM we take $C_9^\SM(\mu_b)\approx 4.07$ and $C_{10}^\SM(\mu_b)\approx -4.31$. Similar plots are obtained for the imaginary parts of the Wilson coefficients.}}
\end{center}\vspace{-2mm}
\end{figure}

Next we consider the inclusive decay $\bar B\to X_s\,l^+l^-$ in the low $q^2$ region of the dilepton invariant mass, $1\,\GeV^2 < q^2 < 6\, \GeV^2$. From the latest measurements of the branching ratio in the low dilepton mass region from Belle $\tx{Br}(\bar B\to  X_s\,l^+l^-)_\tx{exp}^\tx{Belle}=(14.93 \pm 5.04^{+4.11}_{-3.21})\times 10^{-7}$ \cite{Iwasaki:2005sy} and Barbar $\tx{Br}(\bar B\to  X_s\,l^+ l^-)_\tx{exp}^\tx{Barbar} = ({16.0^{+4.1}_{-3.9}}\,^{+1.7}_{-1.3}\pm1.8)\times10^{-7}$ \cite{Lees:2013nxa} we take the combined value $\tx{Br}(\bar B\to  X_s\,l^+ l^-)_\tx{exp}=(15.8\pm 3.7)\times 10^{-7}$ \cite{Huber:2015sra}, which is in good agreement with the SM prediction $\tx{Br}(\bar B\to X_s\,l^+l^-)_\tx{SM}^{}=(16.2 \pm 0.9)\times 10^{-7}$ \cite{Huber:2015sra}. For the calculation of the branching ratio in the RS model we need to take into account the electroweak (leptonic) penguin operators $Q_{9,10}$ and $\tilde Q_{9,10}$ by adding the effective Lagrangian $\La_\tx{eff} = \frac{G_F}{\sqrt2} \lambda_t \sum_{i=9,10}\,(C_i Q_i+\tilde C_{i}\tilde Q_{i} )$ to \eqref{eqn:LeffRSbsga}. The operators are defined by
\begin{align}
Q_{9} &= \frac{e^2}{4\pi^2}\,(\bar s \ga_\mu P_L b)(\bar l \ga^\mu l) \,, &  
Q_{10} &= \frac{e^2}{4\pi^2}\,(\bar s \ga_\mu P_L b)(\bar l \ga^\mu \ga_5 l) \,,
\end{align}
where as always the chirality-flipped operators can be obtained by replacing $P_L\to P_R$. In the SM $C_9(\mu_W)$ and $C_{10}(\mu_W)$ are loop suppressed. In the RS model corrections are induced at tree-level due to the flavor-changing couplings of the $Z$ boson, the Higgs and the $Z$-boson and photon KK modes. The Higgs contributions are suppressed by lepton masses and can therefore be neglected. To leading order in $v^2/\MKK^2$, the RS corrections to the Wilson coefficients are given by \cite{Bauer:2009cf}  
\begin{align}\label{eqn:C9C10} \notag
C_9^\RS(\mu_\KK) &= \frac{1}{\lambda_t}\left[Q_d Q_l \frac{8\pi^2 v^2}{\MKK^2} (\bs\Delta_D')_{23}- \frac{8\pi}{\al_e} (g_R^l+g_L^l) \left((g_R^d-g_L^d) (\bs\delta_D)_{23}- g_L^d \frac{Lm_Z^2}{2\MKK^2} (\bs\Delta_D)_{23} \right) \right] , \\ \notag
C_{10}^\RS(\mu_\KK) &= \frac{1}{\lambda_t} \left[ - \frac{8\pi}{\al_e} \left(g_R^l - g_L^l \right)\left(  (g_R^d-g_L^d) (\bs\delta_D)_{23} - g_L^d \frac{Lm_Z^2}{2\MKK^2} (\bs\Delta_D)_{23} \right)\right] , \\ \notag
\tilde C_9^\RS(\mu_\KK) &= \frac{1}{\lambda_t} \left[Q_d Q_l \frac{8\pi^2 v^2}{\MKK^2} (\bs\Delta_d')_{23} - \frac{8\pi}{\al_e} (g_R^l+g_L^l)\left(  (g_R^d-g_L^d) (\bs\delta_d)_{23} - g_R^d \frac{Lm_Z^2}{2\MKK^2} (\bs\Delta_d)_{23} \right)\right] , \\
\tilde C_{10}^\RS(\mu_\KK) &= \frac{1}{\lambda_t }\left[ -\frac{8\pi}{\al_e}\left(g_R^l - g_L^l \right) \left((g_R^d-g_L^d) (\bs\delta_d)_{23}- g_R^d \frac{Lm_Z^2}{2\MKK^2}(\bs\Delta_d)_{23} \right) \right] ,
\end{align}
where $\al_e=e^2/4\pi$, $Q_d=-1/3$, $Q_l=-1$, $g_L^l=-1/2 + s_w^2$ and $g_R^l= s_w^2$. The functions $(\bs\Delta_{D})_{23}$, $(\bs\Delta_{D}')_{23}$ and $(\bs\delta_{D}^{})_{23}$ are defined in \eqref{eqn:DeltaDefs}, while $(\bs\Delta_{d})_{23}$ and $(\bs\Delta_{d}')_{23}$ can be found in \cite{Casagrande:2008hr}.~Figure~\ref{fig:C9C10size} shows the distributions of these Wilson coefficients at the $B$-meson scale.~The coefficients $C_{9,10}^\RS(\mu_b)$ are much larger than their chirality-flipped counterparts since the left-handed $b$-quark is more localized towards the IR brane than the right-handed one, $F(c_{b_L})\gg F(c_{b_R})$. 

We emphasize that the new-physics contribution $C_{10}^\RS(\mu_b)$ is largest for RS parameter points with small values of $y_\ast$. The reason is that the overlap integral $(\bs\Delta_D)_{23}$ in \eqref{eqn:C9C10} can be expressed approximately by $(\bs\Delta_D)_{23}\sim F(c_{s_L})\,F(c_{b_L})$ \cite{Casagrande:2008hr}, showing its sensitivity on the localization of the left-handed strange- and bottom-quark profiles. For smaller values of $y_\ast$, the profiles are shifted towards the IR brane, in order to reproduce the correct bottom and strange quark masses, and the overlap integral increases in magnitude. For the branching ratio in the low $q^2\in[1,6]\,\GeV^2$ region, we have implemented formula (3.9) in \cite{Guetta:1997fw} and find
\begin{align}\label{eqn:BrBXsll}
\begin{split}
   &\tx{Br}(\bar B\to X_s\,l^+l^-) = \tx{Br}(\bar B\to X_s\,l^+l^-)_\SM \\ 
   &\quad\mbox{}+ 10^{-7}\times\Big[ 1.41 \re C^\RS_{7\ga} - 0.74 \re \tilde C_{7\ga}^\RS 
    + 2.81 \re C_9^\RS - 0.059 \re \tilde C_9^\RS - 4.65\re C_{10}^\RS \\ 
   &\qquad\mbox{}+ 0.074 \re \tilde C_{10}^\RS + 30.18\,(|C^\RS_{7\ga}|^2+|\tilde C^\RS_{7\ga}|^2) 
    + 0.52\,(|C^\RS_{9}|^2+|\tilde C^\RS_{9}|^2) \\ 
   &\qquad\mbox{}+ 0.52\,(|C^\RS_{10}|^2+|\tilde C^\RS_{10}|^2) + 1.94 \re C^\RS_{7\ga} \tilde C^{\RS*}_{7\ga} 
    - 0.008 \re (C^\RS_{9} \tilde C_{9}^{\RS*} + C^\RS_{10} \tilde C_{10}^{\RS*}) \\ 
   &\qquad\mbox{}+ 2.42 \re(C^\RS_{7\ga}C_{9}^{\RS*} 
    + \tilde C^\RS_{7\ga}\tilde C_{9}^{\RS*}) - 0.017 \re(C^\RS_{7\ga}\tilde C_{9}^{\RS*} 
    + \tilde C^\RS_{7\ga} C_{9}^{\RS*}) \Big] \,,
\end{split}
\end{align}
where all Wilson coefficients have to be evaluated at the $B$-meson scale $\mu_b$. In our analysis we have used $C_{7\ga}^\SM(\mu_b)\approx - 0.30$, $C_9^\SM(\mu_b)\approx 4.07$ and $C_{10}^\SM(\mu_b)\approx -4.31$.~We note that the coefficient of $\re C_{7\ga}^\RS(\mu_b)$ is smaller than naively expected. The reason is that the corresponding coefficient results from two terms that are interfering destructively, and the difference is rather sensitive to the SM values of the Wilson coefficients $C_{7\ga}^\SM(\mu_b)$ and $C_{9}^\SM(\mu_b)$. We stress, however, that this 
sensitivity does not have a large impact on our analysis, since the corrections to the branching ratio in \eqref{eqn:BrBXsll} are dominated by the RS corrections to $C_{10}(\mu_b)$. 

The left plot in Figure~\ref{fig:Brbsga} shows the branching ratio $\tx{Br}(\bar B\to X_s\,l^+l^-)$ versus $\tx{Br}(\bar B\to X_s\ga)$ for RS parameter points with $y_\ast=3$ (0.5) and $M_{g^{(1)}}=10\,\TeV$ in black (green). More than 70\% (30\%) of the model points lie within the experimental $2\sigma$ region. In fact, those numbers do not significantly change when switching off the electroweak dipole corrections $C_{7\ga}^\RS(\mu_b)$ and $\tilde C_{7\ga}^\RS(\mu_b)$.~The decay $\tx{Br}(\bar B\to X_s\,l^+l^-)$ is mostly sensitive to the tree-level corrections $C_{10}^\RS(\mu_b)$ in the RS model, justifying the discussion of the branching ratio in \cite{Bauer:2009cf}, where the corrections from the dipole coefficients have been neglected. Requiring that at least 10\% of the RS parameter points lie inside the $2\sigma$ error margin yields the lower bound $M_{g^{(1)}}\geq 3.3\,\TeV$ (6.2\,\TeV) for $y_\ast=3$ (0.5).  

\subsection{Time-dependent CP asymmetry in $\bar {\bs B}^0\bs{\to {\bar K}^{*0} \ga}$}

In the SM, the left-handed structure of the weak interactions makes the emitted photon mainly left-handed in $b$ decays ($b\to s\ga_L$) and right-handed in $\bar b$ decays ($\bar b\to \bar s \ga_R$), since the chirality-flipped Wilson coefficients are suppressed by $m_s/m_b$ relative to the original ones. The helicity suppression of right-handed photons makes the time-dependent CP asymmetry dominated by $B$-meson mixing in the SM, irrespective of hadronic uncertainties. But in new-physics scenarios like the RS model there can be chirality flips on internal lines of a penguin diagram, such that the amplitude for a right-handed photon in $b$ decays is no longer suppressed by $m_s/m_b$. Experimentally, the photon helicity can be accessed indirectly by using the time-dependent CP asymmetry in $\bar B^0\to\bar K^{*0}\ga$ decays, defined as
\begin{align}
\frac{\Gamma(\bar B^0(t)\to \bar K^{*0}\ga) - \Gamma(B^0(t) \to K^{*0}\ga)}
        {\Gamma(\bar B^0(t) \to \bar K^{*0}\ga) + \Gamma(B^0(t) \to K^{*0}\ga)} 
    = S_{K^*\ga} \sin(\Delta m_B t) - C_{K^*\ga} \cos(\Delta m_B t ) \,,
\end{align}
where $\Delta m_B$ is the mass difference between the heavier and the lighter neutral $B$-meson mass eigenstate. The mesons $K^{*0}$ and $\bar K^{*0}$ are observed via their decay into the CP eigenstate $K_S \pi^0$. The helicity suppression can be measured by $S_{K^*\ga}$, which to leading order is given by \cite{Ball:2006cva,Altmannshofer:2011gn}
\begin{align}
   S_{K^*\ga}\approx \frac{2}{|C_{7\ga}(\mu_b)|^2 + |\tilde C_{7\ga}(\mu_b)|^2}\,
    \im\Big[ e^{-i \phi_d}\,C_{7\ga}(\mu_b)\,\tilde C_{7\ga}(\mu_b) \Big] \,.
\end{align}
This observable is sensitive to the chirality-flipped Wilson coefficient $\tilde C_{7\ga}(\mu_b)$.~The angle $\phi_d$ is the phase of $B^0-\bar B^0$ mixing and has been measured in $B\to J/\psi \, K_S$ decays to be $\sin\phi_d=0.682\pm 0.019$ \cite{Amhis:2014hma}. Due to the occurrence of $\tilde C_{7\ga}(\mu_b)$ in the numerator, the SM prediction for $S_{K^*\ga}$ is suppressed by the ratio $m_s/m_b$ and reads $S_{K^*\ga}^\SM=(-2.3\pm 1.6)\%$ \cite{Amhis:2012bh}. The current experimental value $S_{K^*\ga}^\tx{exp} = (-16\pm22)\%$ \cite{Amhis:2014hma} still suffers from large uncertainties. 

The right plot of Figure~\ref{fig:Brbsga} shows the RS contributions to $S_{K^*\ga}$ and the branching ratio $\tx{Br}(\bar B\to X_s\ga)$ by the black (green) points for $y_\ast=3\,(0.5)$ and $M_{g^{(1)}}=10\,\TeV$. Gray and blue bands denote the experimental values with the $1\sigma$ error margins for $S_{K^*\ga}$ and $\tx{Br}(\bar B \to X_s \ga)$, respectively. Compared with the SM prediction the RS corrections can be significant due to the sensitivity of $S_{K^*\ga}$ on the imaginary part of $\tilde C_{7\ga}(\mu_b)$, which can receive large corrections in the RS model. On the other hand, the corrections are not significant when compared with the experimental result due to the large uncertainty. We observe that more than $95\%\,(99\%)$ of the points lie within the experimental $2\sigma$ region for $y_\ast=3\,(0.5)$ and $M_{g^{(1)}}=10\,\TeV$. Requiring that at least 10\% of the RS points lie within the experimental $2\sigma$ regions of $S_{K^*\ga}$ and $\tx{Br}(\bar B\to X_s \ga)$, we can derive the lower bound $M_{g^{(1)}}\geq 3.8\,\TeV$.

\subsection{Direct CP asymmetry in $\bar {\bs B}\bs{\to X_s \ga}$}

The direct CP asymmetry measures the difference between the rates of the decays $\bar B\to X_s\ga$ and $B\to X_s \ga$ and is defined via the ratio
\begin{align}
   A_\tx{CP}^{b\to s\ga}(\delta) 
   = \frac{\tx{Br}(\bar B\to X_s\ga) - \tx{Br}(B\to X_s\ga)}{\tx{Br}(\bar B\to X_s\ga) - \tx{Br}(B\to X_s\ga)}\bigg|_{E_\ga>(1-\delta)E_\ga^\tx{max}} \,,
\end{align}
with a lower cut on the photon energy, which depends on the experiment. The experimental value $A^{b\to s\ga,\tx{exp}}_\tx{CP}=(1.5\pm 2.0)\%$ \cite{Amhis:2014hma} is compatible with zero. Theoretically, the CP asymmetry is affected by perturbative ``direct photon contributions'', in which the photon couples to a local operator mediating the weak decay in the effective low-energy theory, as well as non-perturbative ``resolved photon contributions'', which account for the hadronic substructure of the photon. Taking both effects into account the SM prediction lies in the region $-0.6\%\leq A_\tx{CP}^{b\to s\ga,\SM}\leq 2.28\%$ \cite{Benzke:2010tq} and is compatible with the experimental result. 

\begin{figure}[t]
\begin{center}
\psfrag{x}[]{\small $\Delta A_\tx{CP}^{b\to s \ga} \, [\%]$}
\psfrag{y}[]{\small $A_\tx{CP}^{b\to s \ga} \, [\%] $}
\psfrag{a}[l]{\scriptsize \parbox{3cm}{$M_{g^{(1)}}=10\,\TeV$}}
\includegraphics[width=0.50\linewidth]{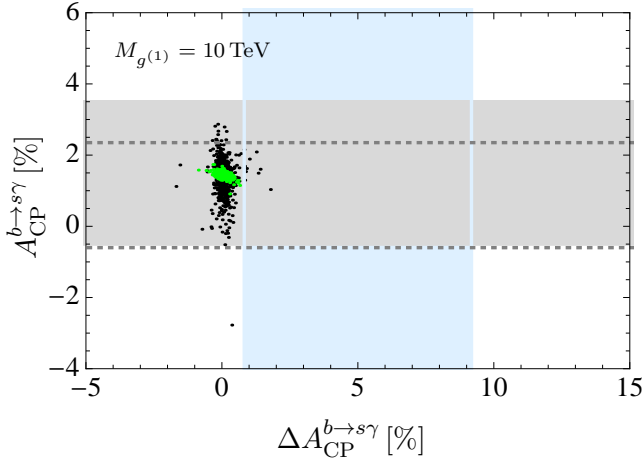} 
\parbox{15.5cm}
{\caption{\label{fig:ACPbsga} 
Shown is the direct CP asymmetry of $\bar B\to X_s\ga$ as a function of the difference of the CP asymmetries of the charged and neutral $B$ mesons.~The light gray and blue bands denote the experimental $1\sigma$ error margins and the area between the two horizontal dashed lines shows the $1\sigma$ error margin of the SM prediction.~The black (green) points represent RS points with $y_\ast=3\,(0.5)$ and $M_{g^{(1)}}=10\,\TeV$.}}
\end{center}\vspace{-2mm}
\end{figure}

Let us now investigate the CP asymmetry in the RS model, where we can have additional contributions from chirality-flipped Wilson coefficients. The perturbative contribution \cite{Kagan:1998bh} 
\begin{align}\label{eqn:ACP}
\begin{split}
   A_\tx{CP}^{b\to s\ga,\tx{dir}}(\delta) 
   = \frac{\al_s(m_b)}{|C_{7\ga}|^2+|\tilde C_{7\ga}|^2}\,\bigg\{& \frac{40}{81}\im[C_2 C_{7\ga}^\ast] 
    - \frac{8z}{9} [v(z)+b(z,\delta)] \im[(1+\e_s)C_2 C_{7\ga}^*] \\ 
   &\mbox{}-\frac{4}{9}\im[C_{8g}C_{7\ga}^* + \tilde C_{8g}\tilde C_{7\ga}^*]
    + \frac{8z}{27}b(z,\delta) \im[(1+\e_s) C_2 C_{8g}^* \\
   &\mbox{}+ \frac{16z}{27} \tilde b(z,\delta) |C_2|^2 \im[\e_s] \bigg\}
\end{split}
\end{align}  
is suppressed at leading order by a factor $\al_{s}(m_b)$ arising from the strong-interaction phases. In the SM $\e_s=\lambda_u/\lambda_t$ is the only source of a CP-violating weak phase, since all Wilson coefficients are real. In addition to this CKM suppression the SM result is further suppressed by the mass ratio $z=m_c^2/m_b^2$ resulting from the GIM mechanism. The functions $v(z),\,b(z,\delta)$ and $\tilde b(z,\delta)$ can be found in \cite{Kagan:1998bh,Benzke:2010tq}. The non-perturbative contribution \cite{Benzke:2010tq}
\begin{align}\label{eqn:ACPresolved}
\begin{split}
   A_\tx{CP}^{b\to s\ga,\tx{res}} = - \frac{\pi}{m_b} \Bigg[
   &\im \left(\e_s \frac{C_2 C_{7\ga}^*}{|C_{7\ga}|^2+|\tilde C_{7\ga}|^2}\right) 
    \tilde \Lambda^u_{27} 
    - \im \left((1+\e_s) \frac{C_2 C_{7\ga}^*}{|C_{7\ga}|^2+|\tilde C_{7\ga}|^2}\right) 
    \tilde\Lambda_{27}^c \\ 
  &\mbox{}- \im \left(\frac{C_{7\ga}^*C_{8g}+\tilde C_{7\ga}^*\tilde C_{8g}}%
   {|C_{7\ga}|^2+|\tilde C_{7\ga}|^2} \right) 4\pi\,\al_s(m_b)\,\tilde\Lambda_{78} \Bigg]
\end{split}
\end{align}
starts at leading order in $\Lambda_\tx{QCD}/m_b$ and involves hadronic parameters with values in the range $-330\,\tx{MeV}<\tilde \Lambda_{27}^u<525\,\tx{MeV}$, $-9\,\tx{MeV}<\tilde \Lambda_{27}^c<11\,\tx{MeV}$ and $-17\,\tx{MeV}<\tilde \Lambda_{78}<190\,\tx{MeV}$. For our analysis we choose the values $\tilde \Lambda_{27}^u=96\,\MeV$, $\tilde \Lambda_{27}^c=1\,\MeV$ and $\tilde \Lambda_{78}=104\,\MeV$.

Adding both contributions \eqref{eqn:ACP} and \eqref{eqn:ACPresolved} we obtain the black (green) RS points shown in the left plot of Figure~\ref{fig:ACPbsga}, for $y_\ast=3\,(0.5)$ and $M_{g^{(1)}}=10\,\TeV$.~We observe that all points lie inside the experimental $2\sigma$ area. The effects decrease for smaller values of $y_\ast$. In general, the corrections to the observable $A_\tx{CP}^{b\to s\ga}$ are too small in order to constrain the RS parameter space.

\subsection{The CP asymmetry difference in $\bar {\bs B}\bs{\to X_s \ga}$}

Another observable is the difference of the CP asymmetries in charged and neutral $B$-meson decays. Its formula is given by \cite{Benzke:2010tq}
\begin{align}
\begin{split}
   \Delta A_\tx{CP}^{b\to s\gamma} 
   &= A_\tx{CP}(\bar B^+\to X_s^+ \gamma)-A_\tx{CP}(\bar B^0\to X_s^0 \gamma) \\
   &= 4\pi^2\al_s(m_b)\,\frac{\tilde\Lambda_{78}}{m_b}\,\im\left( \frac{C_{7\ga}^* C_{8g}
    + \tilde C_{7\ga}^* \tilde C_{8g}}{|C_{7\ga}|^2+|\tilde C_{7\ga}|^2} \right) ,
\end{split}
\end{align}
where the hadronic parameter $\tilde\Lambda_{78}$ is predicted to lie in the range $12\,\tx{MeV}<\tilde\Lambda_{78}<190\,\tx{MeV}$. Recently the BarBar collaboration has published the first experimental result $\Delta A_\tx{CP}^{b\to s\gamma}=(5\pm 3.9_\tx{stat}\pm 1.5_\tx{syst})\%$, which is compatible with a null asymmetry difference at the level of $1\sigma$. For the analysis we take the average value $\tilde \Lambda_{78}=89\,\MeV$.

Figure~\ref{fig:ACPbsga} shows the predictions in the RS model for $y_\ast=3\,(0.5)$. For most of the points the corrections do not exceed the 1\% level. All points are included in the experimental $2\sigma$ error band. The effects are also decreasing for smaller values of $y_\ast$.~Currently, the RS parameter space cannot be constrained by the CP asymmetry difference $ \Delta A_\tx{CP}^{b\to s\gamma}$.

\subsection{Comparison with (almost) model-independent fits}

Several tensions at the 2-3$\sigma$ level in $\bar B\to\bar K^*\mu^+\mu^-$ angular observables have shown up in the collected data of the LHCb run during 2011 and 2012, including an integrated luminosity of 3\,fb$^{-1}$. One result is that the observable $P_5'$ in the invariant lepton mass region $4.0\,\GeV<q^2<8.0\,\GeV$ is only compatible with the SM prediction at the level of $3.7\sigma$ \cite{LHCb:2015dla}. Several model-independent theoretical analyses including also additional observables in $B$ decays have been performed, showing that the deviations can be explained by new physics \cite{Descotes-Genon:2013wba,Altmannshofer:2013foa,Beaujean:2013soa,Hurth:2013ssa,Alonso:2014csa,Hiller:2014yaa,Ghosh:2014awa}. 

The full 4-body angular distribution of the decay $\bar B\to\bar K^*\mu^+\mu^-$ provides a sensitivity to the operators $Q_{7\ga},\,Q_{9},\,Q_{10}$ and to the scalar and pseudo-scalar operators $Q_P,\,Q_S$ and their chirality-flipped counterparts. In the RS model we neglect the corrections to $b\to s \mu^+\mu^-$ arising from the scalar operators $Q_P$ and $Q_S$, since they follow from the tree-level exchange of the Higgs boson and are suppressed by the small lepton masses \cite{Bauer:2009cf}. Here, we will focus on $C_{7\ga,9,10}$ and their chirality-flipped counterparts. Unfortunately, in the literature there is no general fit where new physics is allowed to enter all Wilson coefficients at once and including both real and imaginary parts. Therefore, we have to consider certain scenarios where the underlying assumptions are partly violated in our model. This fact needs to be kept in mind. 

\begin{figure}[t!]
\begin{center}
\begin{minipage}{0.48\linewidth}
\psfrag{x}[]{\small $\re C^\RS_{7\ga}(\mu_b)$}
\psfrag{y}[]{\small $\re C^\RS_{9}(\mu_b)$}
\psfrag{a}[l]{\scriptsize \parbox{3cm}{$M_{g^{(1)}}=10\,\TeV$\\}}
\includegraphics[width=0.85\textwidth]{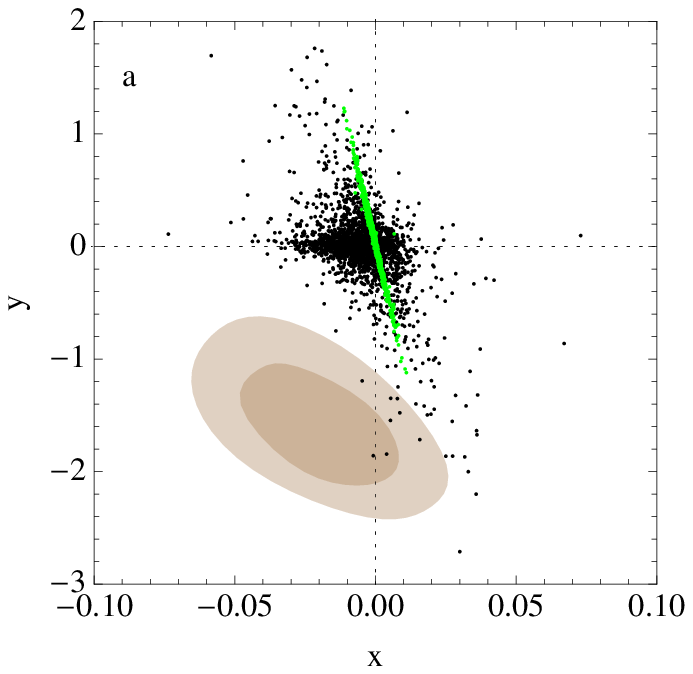} 
\end{minipage} 
\begin{minipage}{0.48\linewidth}
\psfrag{x}[]{\small $\re C^\RS_{9}(\mu_b)$}
\psfrag{y}[]{\small $\re C^\RS_{10}(\mu_b)$}
\psfrag{a}[l]{\scriptsize \parbox{3cm}{$M_{g^{(1)}}=10\,\TeV$\\}}
\includegraphics[width=0.85\textwidth]{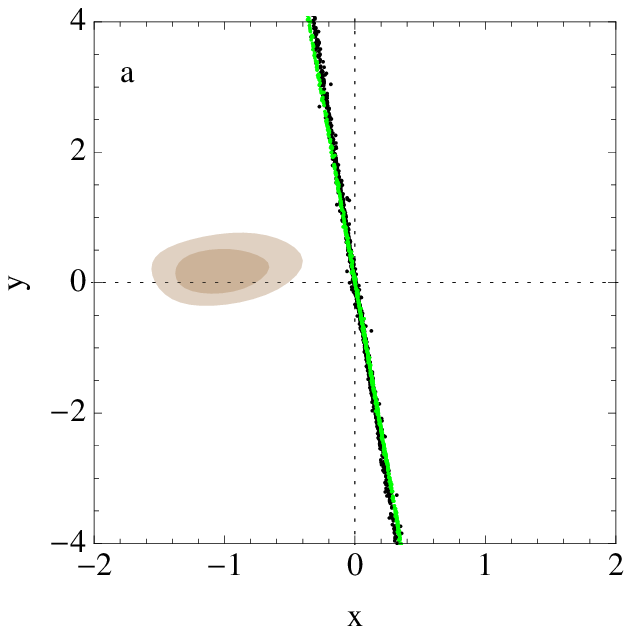} 
\end{minipage}
\parbox{15.5cm}
{\caption{\label{fig:Fitplanes} 
RS predictions for the real parts of the Wilson coefficients $C_{7\ga}^\RS$\,--\,$C_{9}^\RS$ (left) and $C_{9}^\RS$\,--\,$C_{10}^\RS$ (right). Black (green) dots correspond to RS scenarios with $y_\ast=3\,(0.5)$ and $M_{g^{(1)}}=10\,\TeV$. The brown contours correspond to the $1\sigma$ and $2\sigma$ best fit regions obtained in \cite{Descotes-Genon:2013wba} and \cite{Altmannshofer:2013foa,Altmannshofer:2015sma}, respectively. The dotted lines represents the SM values.}} 
\end{center}\vspace{-2mm}
\end{figure}

A general fit where new physics can modify $C_{7\ga,9,10}(\mu_b)$ and $\tilde C_{7\ga,9,10}(\mu_b)$ simultaneously has been performed in \cite{Descotes-Genon:2013wba}. It includes experimental data on the $\bar B\to\bar K^*\mu^+\mu^-$ observables $P_{1,2}$, $P'_{4,5,6,8}$ and $A_{\rm FB}$ in various $q^2$ bins, as well as data on $\tx{Br}(\bar B\to X_s\ga)$, $\tx{Br}(\bar B\to X_s\,\mu^+\mu^-)$, $\tx{Br}(\bar B_s\to\mu^+\mu^-)$, the isospin asymmetry $A_I(B\to K^* \ga)$ and $S_{K^*\ga}$. Deviations of the Wilson coefficients from their SM values are defined by $C^\NP_i(\mu_b) = C_i(\mu_b) - C^\SM_i(\mu_b)$. The best fit regions are given by (at $95\%$ CL)
\begin{equation}\label{eqn:FitC7910}
\begin{aligned}
   C_7^\NP(\mu_b) &\in [-0.06,0.01] \,, \quad & 
    C_9^\NP(\mu_b) &\in [-1.8,-0.6] \,,\quad & 
    C_{10}^\NP(\mu_b) &\in [-1.2,2.0] \,,\\
   \tilde C_7^\NP(\mu_b) &\in [-0.09,0.06] \,, & 
    \tilde C_9^\NP(\mu_b) &\in [-0.8,1.4] \,, & 
    \tilde C_{10}^\NP(\mu_b) &\in [-1.0,0.8] \,,
\end{aligned}
\end{equation}
showing that only $C_9^\NP(\mu_b)$ is inconsistent with zero. The most economical scenario corresponds to a negative new-physics contribution to $C_9(\mu_b)$ with all other Wilson coefficients close to their SM values. Unfortunately, the RS model does not lead to such a large correction to $C_9(\mu_b)$, as is evident from the upper left plot in Figure~\ref{fig:C9C10size}. The corrections to the remaining Wilson coefficients are however compatible at $95\%$ CL with the fit regions given in \eqref{eqn:FitC7910}.

In order to clarify the role played by some of the Wilson coefficients some constrained scenarios have been considered, in which only two Wilson coefficients at a time are assumed to receive contributions from new physics. While this assumption does not hold true for the RS model, we still like to compare our results with the two scenarios where only $C_{7\ga}(\mu_b)$ and $C_{9}(\mu_b)$ or $C_9(\mu_b)$ and $C_{10}(\mu_b)$ are assumed to be modified by (real) new physics contributions. We do not consider modifications of $\tilde C_9(\mu_b)$ and $\tilde C_{10}(\mu_b)$, since the corresponding corrections in the RS model are very small. In the left plot in Figure~\ref{fig:Fitplanes} we consider the $C_{7\ga}^\RS$\,--\,$C_9^\RS$ plane. The brown contours correspond to the $1\sigma$ and $2\sigma$ best fit regions obtained from the analysis in \cite{Descotes-Genon:2013wba}. The best fit points are given by $C_{7\ga}^\NP(\mu_b)\approx -0.02$ and $C_{9}^\NP(\mu_b)\approx -1.5$. We observe that almost none of the RS points touch the best fit region, since $C_9(\mu_b)$ does not receive large enough corrections. As a second scenario, we consider in the right plot of Figure~\ref{fig:Fitplanes} the $C_9^\RS$\,--\,$C_{10}^\RS$ plane, where the best fit regions are obtained from a global fit performed in \cite{Altmannshofer:2013foa,Altmannshofer:2015sma}. The fit includes 88 measurements of 76 different observables, including $\bar B\to\bar K^*\mu^+\mu^-$ angular observables and branching ratios as well as the branching ratios of $\bar B\to\bar K\mu^+\mu^-$, $\bar B\to X_s\,\mu^+\mu^-$, $\bar B_s\to\phi\mu^+\mu^-$, $\bar B\to\bar K^*\ga $, $\bar B\to X_s\ga$ and $\bar B_s\to\mu^+\mu^-$. We observe an anti-correlation between $C_9^\RS(\mu_b)$ and $C_{10}^\RS(\mu_b)$, which is also favored by the best fit regions. While the corrections $C_{10}^\RS(\mu_b)$ can be quite large for a few points in parameter space, the corrections $C_9^\RS(\mu_b)$ are too small to reach the best fit regions. 

\section{Conclusions}

We have investigated the electro- and chromomagnetic (quark) dipole coefficients for $b\to s\ga$ and $b\to s g$ transitions in the minimal Randall-Sundrum model with a brane-localized Higgs sector.~We have derived integral expressions for all contributions arising at one-loop order using 5D fermion and gauge-boson propagators and retaining the full dependence on the Yukawa interactions. The expressions are formally valid to all orders in $v^2/\MKK^2$, in contrast to \cite{Csaki:2010aj}, where the Yukawa interactions were treated as small perturbations. Our final results involve one momentum and two extra-dimensional integrations, and each integrand contains one 5D gauge-boson and one 5D fermion propagator. 

By analyzing the UV behavior of the 5D propagators we have confirmed the finiteness of the penguin loops, as shown first in \cite{Csaki:2010aj}. In addition, we have derived expressions in the KK-decomposed (4D) theory and shown analytically and numerically that the dipole coefficients coincide in both pictures, presenting a highly non-trivial cross-check of our calculations. We have derived approximate formulas for the KK contributions to $C_{7\ga,8g}$ and $\tilde C_{7\ga,8g}$ and have shown that the dominant corrections originate from the $W^\pm$-boson penguin diagrams. More precisely, when working in Feynman-'t Hooft gauge the dominant corrections stem from the parts of the diagrams which involve the scalar component of the 5D gauge-boson field and the charged Goldstone bosons from the Higgs sector. We find that for not too small values of the anarchic (5D) Yukawa matrix entries the latter contributions dominate and the size of the KK corrections to the dipole coefficients increases proportional to $y_\ast^2$. In contrast to \cite{Blanke:2012tv}, we have not found a significant contribution of the triple gluon vertex penguin diagram on $C_{8g}$ and $\tilde C_{8g}$. In agreement with \cite{Blanke:2012tv}, we have observed the general pattern that the chirality-flipped Wilson coefficients $\tilde C_{7\ga,8g}$ receive larger corrections than $C_{7\ga,8g}$, since the left-handed bottom-quark profile is more localized towards the IR brane.  For the Higgs and $Z$-boson penguin diagrams we have obtained results in the brane-localized Higgs and narrow bulk-Higgs scenarios.~For $y_\ast\gtrsim1$ both contributions cancel to good approximation for the case of a brane-localized Higgs, while there remains a non-zero (zero-mode) contribution for the case of a narrow bulk-Higgs. A similar observation was reported in \cite{Beneke:2012ie,Moch:2014ofa,Beneke:2014sta,Beneke:2015lba} for the case of the leptonic dipole coefficients. 

In our phenomenological analysis we have RG evolved the dipole coefficients to the $B$-meson scale $\mu_b$. We have then performed a numerical scan of the RS parameter space with anarchic 5D Yukawa matrices and investigated the branching ratio $\text{Br}(\bar B\to X_s\ga)$, the time-dependent CP asymmetry $S_{K^*\ga}$, the direct CP asymmetry $A_\tx{CP}^{b\to s\ga}$ and the CP asymmetry difference $\Delta A_\tx{CP}^{b\to s\ga}$, all of which are sensitive to corrections to the dipole coefficients. Currently, the observables $\text{Br}(\bar B\to X_s\ga)$ and $S_{K^*\ga}$ can be used to constrain the RS parameter space. Requiring that at least 10\% of the RS model points lie in the $2\sigma$ experimental error margins, we can derive a lower bound on the mass of the first KK gluon resonance of $M_{g^{(1)}}\geq 3.8\,\TeV$ for Yukawa matrix entries bounded from above by $y_\ast=3$. For smaller values of $y_\ast$ the bound gets weaker.~We further discussed the branching ratio $\tx{Br}(\bar B\to X_s l^+l^-)$, which is dominated by the tree-level corrections to $C_{10}(\mu_b)$ in the RS model. For this observable we can derive a lower bound of $M_{g^{(1)}} \geq 3.3\,\TeV\,(6.2\,\TeV)$ for $y_\ast=3\,(0.5)$, showing that the RS corrections increase for smaller values of $y_\ast$. Finally, we have compared the Wilson coefficients $C_{7\ga,9,10}$ and $\tilde C_{7\ga,9,10}$ with the results of (almost) model-independent fits performed in \cite{Descotes-Genon:2013wba,Altmannshofer:2013foa,Altmannshofer:2015sma}. In general, the tree-level corrections to $C_{9}(\mu_b)$ are too small in the RS model in order to cover the best fit regions. 

\vspace{3mm}\noindent
{\em Node added:} While this paper was under review the work \cite{Moch:2015oka} appeared, which contains a detailed analysis of the decay $\bar B\to X_s\ga$ in the minimal and custodial RS model with an IR-localized bulk Higgs.~Their implementation of the scalar sector includes contributions from the Higgs zero mode and its KK excitations.~In contrast to our approach the Yukawa interactions are treated as perturbations and results for the one-loop penguin diagrams to leading order in $v^2/\MKK^2$ are derived.~In addition to our work, the mixing of the dipole operators with four-quark operators obtained by integrating out KK gluons and the Higgses are included.~Comparing our results for the branching ratio $\tx{Br}(\bar B\to X_s\ga)$ in the minimal RS model with the analysis of \cite{Moch:2015oka}, we find that they are of similar size.

\vspace{3mm}
{\em Acknowledgements:\/}
We are grateful to Martin Beneke and Paul Moch for useful discussions, and to Jennifer Mutschall for collaboration at an earlier stage of this project.~M.N.~thanks the Institute for Theoretical Physics in Heidelberg for hospitality and support under a J.~Hans D.~Jensen Professorship.~This research has been supported by the Advanced Grant EFT4LHC of the European Research Council (ERC), the Cluster of Excellence {\em Precision Physics, Fundamental Interactions and Structure of Matter\/} (PRISMA -- EXC 1098), grant 05H12UME of the German Federal Ministry for Education and Research (BMBF), and the DFG Graduate School GRK~1581 {\em Symmetry Breaking in Fundamental Interactions}. 

\begin{appendix}

\newpage

\section{Summary of Feynman Rules}
\label{app:FR}
\renewcommand{\theequation}{A.\arabic{equation}}
\setcounter{equation}{0}

All particle momenta are flowing into the vertex and amplitudes are denoted by $\A$. 

\subsection*{5D Theory}
We begin with the vertices that couple two quarks and one boson. Each vertex is accomplished additionally by an integral $\int_\e^1 dt$ and we obtain for the corresponding amplitudes $(q=u,d)$
\begin{align}\label{eqn:ABmuFR}
\begin{split}
\A^{\{\bar qA_\mu q,\,\bar q_a G_\mu^c q_b,\bar qZ_\mu q,\,\bar qW_\mu^\pm q'\}} & = \Big\{i Q_q \, e_5,\,  i g_{s,5} \, T_{ab}^c ,\,\frac{i g_5 \, g_L^q}{c_w}  ,\,\frac{i g_5}{\sqrt2} \Big\}\,\ga_\mu\, \Pb_B ,\\
\A^{\{\,\bar qA_5 q,\,\bar q_a G_5 q_b,\,\bar qZ_5 q,\,\bar qW_5^\pm q'\}}& = \Big\{Q_q\,e_5,\,g_{s,5} \, T_{ab}^c,\, \frac{g_5 \, g_L^q}{c_w},\, \frac{g_5}{\sqrt2} \Big\} \Big[{\bs V}_{B_5^\pm}(t) P_L + \tilde {\bs V}_{B_5^\pm}(t) P_R \Big] \,,\\
\A^{\bar qhq} & = \frac{-i}{\sqrt2}\,\delta^\eta(t-1) \, \Big[\M_q^Y P_R + \M_q^{Y\da} P_L\Big]\,,
\end{split}
\end{align}
where $\M_q^Y = \Yb_q^C \, \Pb_{12} + \Yb_q^{S\da} \, \Pb_{21}$, $g_L^q = T_3^q - Q_d s_w^2$, $e=e_5 / \sqrt{2\pi r}$ is the 4D electromagnetic and $g_{s}=g_{s,5}/\sqrt{2\pi r}$ is the QCD 4D gauge-coupling. In the first two lines the subscript $B$ of $\Pb_B$, ${\bs V}_{B_5^\pm}(t)$ and $\tilde {\bs V}_{B_5^\pm}(t)$ must be replaced by the corresponding boson label $B=A,G,Z,W$ on the left side\footnote{The superscripts of $B_5^\pm$ are only relevant for $B=W$ and can be ignored otherwise.}. The $t$-dependent functions are given by
\begin{equation}\begin{aligned} \label{eqn:VBs}
\Vb_{A_5,G_5}(t) & = \frac{\e}{t} \, \one\,,& \tilde \Vb_{A_5,G_5}(t)&= - \Vb_{A_5,G_5}(t) \,,\quad \\
\Vb_{W_5^+}(t) & = \frac{\e}{t}\bigg[\Pb_W +  \frac{\rh\MKK^2}{L\tilde m^2_W} \delta^\eta(t-1) \, \M^{Y\da}_{ud} \bigg]\,, & \tilde \Vb_{W_5^+}(t) & = - \Vb_{W_5^+}(t) \Big|_{\M_{ud}^{Y\da}\to-\M_{du}^Y}\,, \\
\Vb_{W_5^-}(t) & = \frac{\e}{t}\bigg[\Pb_W +  \frac{\rh\MKK^2}{L\tilde m^2_W} \delta^\eta(t-1)  \, \M^{Y\da}_{du} \bigg]\,, & \tilde \Vb_{W_5^-}(t) & = - \Vb_{W_5^-}(t)\Big|_{\M_{du}^{Y\da}\to-\M_{ud}^Y} \,, \\
\Vb_{Z_5}(t) &= \frac{\e}{t}\bigg[\Pb_Z +  \bigg(1 - \frac{g_R^q}{g_L^q} \bigg) \frac{\rh\MKK^2}{L\tilde m^2_Z} \, \delta^\eta(t-1) \, \M^{Y\da}_{qq}\bigg]\,,& \tilde \Vb_{Z_5}(t) & = -\Vb_{Z_5}(t) \Big|_{\M_{qq}^{Y\da}\to-\M_{qq}^Y} \,,
\end{aligned}\end{equation}
where $g_R^q= - Q_q s_w^2$ and $\M^Y_{qq'} \equiv \Yb_{q}^C \, \bs P_{12} - \Yb_{q'}^{S\da}\bs P_{21}$. We continue with the Feynman rules for the triple gauge-boson vertices, where we assume that the external photon line is a zero mode.~The corresponding amplitudes read
\begin{align} \notag
\A^{W^\pm_\beta(p) A_\al(q) W^\mp_\ga(k)} & = \mp \frac{i\,e_5\,2\pi}{L\,t} \Big[\eta^{\al\beta} (q-p)^\ga + \eta^{\beta\ga} (p-k)^\al + \eta^{\ga\al} (k-q)^\beta \Big] \,,\\\notag
\A^{W_5^\pm(p) A_\al(q) W^\mp_\beta(k)} & = \pm e_5 \, \eta^{\al\beta} \frac{2\pi \MKK }{L} \, \bigg[\pa_t^{W_5^\pm} + \delta^\eta(t-1)\bigg] \frac{\e}{t^2} \,,\\\notag 
\A^{W_5^\pm(p) A_\al(q) W_5^\mp(k)}  & = \pm (p - k)^\al i \, e_5 \frac{2\pi }{L} \bigg[1 + \frac{\MKK^2}{L \tilde m_W^2} \delta^\eta(t-1) \bigg] \frac{\e^2}{t^3} \,,\\\notag
\A^{G^b_\beta(p) G^a_\alpha(q) G^c_\gamma(k)} & = \frac{g_{s,5} \, f^{abc}\,2\pi}{L} \frac{1}{t} \Big[\eta^{\al\beta} (q-p)^\ga + \eta^{\beta\ga} (p-k)^\al + \eta^{\ga\al} (k-q)^\beta \Big]\,, \\\notag
\A^{G^b_\beta(p) G^a_\alpha(q) G^c_5(k) } & = i\,g_{s,5}\,\eta^{\al\beta}\,f^{abc} \frac{2\pi\MKK}{L} \, \pa^{G_5^c}_t \, \frac{\e}{t^2} \,,\\
\A^{G^b_5(p) G^a_\alpha G^c_5(k)}  & = g_{s,5}\,f^{abc}\,(p- k)^\alpha \frac{2\pi}{L} \frac{\e^2}{t^3}\,,
\end{align}
where the superscript of the $t$-derivatives indicates the field it should act on. 

\subsection*{4D Theory}

In the KK-decomposed (4D) theory the amplitudes for the vertices coupling two quarks and one boson can be summarized by
{\small\begin{align} \label{eqn:4DFRampl}
\begin{split}
\A^{\{\bar q A_\mu q,\,\bar q_a G_\mu^c q_b,\bar qZ_\mu q,\,\bar qW_\mu^\pm q'\}} & = \Big\{i Q_q e_5,\,  i g_{s,5} \, T_{ab}^c ,\,\frac{i g_5 \,g_L^q}{c_w},\,\frac{i g_5}{\sqrt2}\Big\}\,\frac{\ga_\mu}{\sqrt{2\pi r}} \, \left[ V_{nmk}^{B} \, P_L + \Tilde V_{nmk}^{B} \, P_R \right] \,,\\
\A^{\{\,\bar q\varphi_A q,\,\bar q_a \varphi_{G}^c q_b,\bar q \varphi_Z q,\,\bar q\varphi_W^\pm q'\}}& = \Big\{Q_q e_5,\,g_{s,5} \, T_{ab}^c,\, \frac{g_5 \,g_L^q}{c_w},\, \frac{g_5}{\sqrt2} \Big\} \, \frac{1}{\sqrt{2\pi r}} \, \left[V_{nmk}^{\varphi_B} \, P_L + \Tilde V_{nmk}^{\varphi_B} \, P_R \right] \,,\\
\A^{\bar q^{}h^{}q^{}} & = -i \, \Big[(\tilde {\bs g}^q_h)_{nk} \, P_L + ({\bs g}^q_h)_{nk} \, P_R \Big]\,,
\end{split}
\end{align}
}where $n,m,k$ are the mode-numbers of the anti-quarks, bosons and quarks respectively. The labels $B,\varphi_B$ on the right side must be replaced by the corresponding boson (label) on the left side. The vector and scalar overlap integrals for the neutral gauge bosons are given by
\begin{align}\label{eqn:VB4D}
\begin{split}
V_{nmk}^{B} & = \sqrt{2\pi} \int_\e^1 \hspace{0mm} dt \, \chi_m^{B}(t) \, \Q_L^{(n)\da}(t) \, \bs P_B \, \Q_L^{(k)}(t) \,,\quad \tilde V_{nmk}^B=V_{nmk}^B|_{L\leftrightarrow R}\,,\\
V_{nmk}^{\varphi_{B}} & = \sqrt{2\pi} \int_\e^1 \hspace{0mm} dt \, \frac{-k \, t\,\pa_t\chi_m^{B}(t)}{m_{B_m}} \, \Q_R^{(n)\da}(t) \, \Vb_{B_5}(t)\,  \Q_L^{(k)}(t)\,, \quad \tilde V_{nmk}^{\varphi_B} = V_{nmk}^{\varphi_B}|_{L\leftrightarrow R,\,\Vb_{B_5} \rightarrow \tilde\Vb_{B_5}} \,, 
\end{split}
\end{align}
where $B=A,G,Z$. The structures $\Vb_{B_5}(t)$ are defined in \eqref{eqn:VBs}. For the $W^\pm$ boson we find the following overlap integrals
{\small\begin{align}\label{eqn:VW4D}
\begin{split}
V_{nmk}^{W^+} & = \sqrt{2\pi} \int_\e^1 \hspace{0mm} dt \, \chi_m^W(t) \, \U_L^{(n)\da}(t) \, \bs P_W \,  \D_L^{(k)}(t)\,,\quad V_{nmk}^{W^-} = V_{nmk}^{W^+}|_{\U\leftrightarrow \D}\,,\quad \tilde V_{nmk}^{W^\pm} = V_{nmk}^{W^\pm}|_{L\leftrightarrow R}\,,\\
V_{nmk}^{\varphi_W^+}  & = \sqrt{2\pi}  \int_\e^1\hspace{0mm} dt \, \frac{-k \,t\,\pa_t \chi_m^W(t)}{m_{W_m}}  \, \U_{R}^{(n)\da}(t)  \, \Vb_{W_5^+} (t)\,  \D_{L}^{(k)}(t)\,, \qquad \tilde V_{nmk}^{\varphi_W^+}  = V_{nmk}^{\varphi_W^+} \big|_{L\leftrightarrow R,\,\Vb_{W_5^+} \rightarrow \tilde \Vb_{W_5^+}}\,, \\
V_{nmk}^{\varphi_W^-}  & = \sqrt{2\pi}  \int_\e^1 \hspace{0mm}dt \, \frac{- k \, t \, \pa_t \chi_m^W(t)}{m_{W_m}}   \, \D_{R}^{(n)\da}(t) \, \Vb_{W_5^-}(t) \,\U_{L}^{(k)}(t) \,,\qquad \tilde V_{nmk}^{\varphi_W^-} =V_{nmk}^{\varphi_W^-} \big|_{L\leftrightarrow R,\,\Vb_{W_5^-} \rightarrow \tilde \Vb_{W_5^-}}\,.
\end{split}
\end{align}
}The Goldstone-boson contributions from the Higgs sector are contained in the brane-localized terms of $\Vb_{B_5^\pm}(t)$ in \eqref{eqn:VBs}. We can simplify the scalar overlap integrals by performing a partial $t$-integration noting that boundary terms at $t=\e,1$ vanish, since we work with the regularised $\delta$-function \eqref{eqn:deltaReg}. We can apply the equation of motions for the quark profiles such that the terms with the $\delta$-functions cancel. For instance, in case of  $V_{nmk}^{\varphi_W^+}$ we can use that
\begin{align}\label{eqn:DerURDL}
\begin{split}
\pa_t \Big[ \U_R^{(n)\da}(t) \, P_+ \, \D_L^{(k)}(t) \Big] & = \frac{m_{u_n}}{\MKK} \, \U_L^{(n)\da}(t) \, P_+ \, \D_L^{(k)}(t) - \frac{m_{d_k}}{\MKK} \, \U_R^{(n)\da}(t) \, P_+ \, \D_R^{(k)}(t)  \\
&\hspace{5mm} -  \varrho \,\delta^\eta(t-1)\,\U_R^{(n)\da}(t)\, M_{ud}^{Y\da}\,\D_L^{(k)}(t) \,,\\
\end{split}
\end{align}
where the last term in \eqref{eqn:DerURDL} cancels the $\delta$-function appearing in $\Vb_{W_5^+}(t)$. Repeating the steps for the remaining cases we find
\begin{align}\label{eqn:Vnmkscalar}
\begin{split}
V_{nmk}^{\varphi_B} & = \frac{m_{q_n}}{m_{B_m}} V_{nmk}^{B} - \frac{m_{q_k}}{m_{B_m}}  \Tilde V_{nmk}^{B}\,,\quad\Tilde V_{nmk}^{\varphi_B} = \frac{m_{q_n}}{m_{B_m}} \Tilde V_{nmk}^{B} - \frac{m_{q_k}}{m_{B_m}}  V_{nmk}^{B} \,, \quad (B=A,G,Z) \\
V_{nmk}^{\varphi_W^+} &= \frac{m_{u_n}}{m_{W_m}} V_{nmk}^{W^+} - \frac{m_{d_k}}{m_{W_m}} \Tilde V_{nmk}^{W^+}\,,\quad \Tilde V_{nmk}^{\varphi_W^+} = \frac{m_{u_n}}{m_{W_m}} \Tilde V_{nmk}^{W^+} - \frac{m_{d_k}}{m_{W_m}}  V_{nmk}^{W^+}\,, \\
V_{nmk}^{\varphi_W^-}& = \frac{m_{d_n}}{m_{W_m}} V_{nmk}^{W^-} - \frac{m_{u_k}}{m_{W_m}} \Tilde V_{nmk}^{W^-}\,,\quad \Tilde V_{nmk}^{\varphi_W^-} = \frac{m_{d_n}}{m_{W_m}} \Tilde V_{nmk}^{W^-} - \frac{m_{u_k}}{m_{W_m}}  V_{nmk}^{W^-}\,,
\end{split}
\end{align}
where all scalar overlap integrals can be expressed in terms of the vector overlap integrals. Concerning the quark-Higgs-quark couplings in \eqref{eqn:4DFRampl} the overlap integrals are given by $(q=u,d)$
\begin{align}
( {\bs g}_h^q)_{nk} & = \frac{1}{\sqrt2}\int_\e^1 \hspace{0mm} dt \,\delta^\eta(t-1) \, \Q_L^{(n)\da}(t) \, \M_q^{Y} \, \Q_R^{(k)}(t) \,, && (\tilde {\bs g}_h^q)_{nk} = \big[({\bs g}_h^q)_{kn}\big]^\da \,.
\end{align}
In the brane-localized Higgs scenario with $\Yb_q^C=\Yb_q^S$ we can perform the $t$-integration analytically and find
\begin{align}
(\bs g_h^q)_{nk} = \frac{1}{\sqrt2}\,\Q_L^{(n)\da}(1^-) \,\Pb_{12}\, \frac{2\Xb_q}{\sinh 2\Xb_q} \, \tilde \Yb_q \, \Q_R^{(k)}(1^-)\,,
\end{align}
where $\Xb_q = \rh (\Yb_q\Yb_q^\da)^{1/2}$, $\tilde \Yb_q=(\tanh \Xb_q/\Xb_q) \Yb_q$ and $(\tilde {\bs g}_h^q)_{nk} = \big[({\bs g}_h^q)_{kn}\big]^\da$. We continue with the triple gauge-boson vertices where photon or gluon zero modes are attached. We obtain
\begin{align}
\begin{split}
\A^{W^{\pm(n)}_\beta(p) A^{(0)}_\al(q) W^{\mp(k)}_\ga(k)}  &= \mp \frac{i\,e_5}{\sqrt{2\pi r}}  \Big[\eta^{\al\beta} (q-p)^\ga + \eta^{\beta\ga} (p-k)^\al + \eta^{\ga\al} (k-q)^\beta \Big]\delta_{nk}\,, \\
\A^{ \varphi_W^{\pm(n)}(p)  A^{(0)}_\al(q) W_\beta^{\mp(k)}(k)} & = \pm \frac{e_5\,\eta^{\al\beta}}{\sqrt{2\pi r}} \,m_{W_n} \,\delta_{nk}\,, \\
\A^{\varphi_W^{\pm(n)}(p) A^{(0)}_\al \varphi_W^{\mp(k)}(k)}  & = \pm (p - k)^\alpha \frac{i\,e_5}{\sqrt{2\pi r}} \,\delta_{nk}\,, \\
\A^{G_{\beta,b}^{(n)}(p) G^{(0)}_{\alpha,a}(q) G_{\ga,c}^{(k)}(k)} & = \frac{g_{s,5}\,f^{abc}}{\sqrt{2\pi r}} \Big[\eta^{\al\beta} (q-p)^\ga + \eta^{\beta\ga} (p-k)^\al + \eta^{\ga\al} (k-q)^\beta \Big]\delta_{nk}\,, \\
\A^{G_{\beta,b}^{(n)}(p) G^{(0)}_{\alpha,a}(q) \varphi_{G,c}^{(k)}(k)} & =\frac{i\,g_{s,5}\,\eta^{\al\beta} f^{abc}}{\sqrt{2\pi r}} m_{G_n} \delta_{nk}\,,\\
\A^{\varphi_{G,b}^{(n)}(p) G^{(0)}_{\alpha,a}(q) \varphi_{G,c}^{(k)}(k)}  & = (p - k )^\al \, \frac{g_{s,5}\,f^{abc}}{\sqrt{2\pi r}} \delta_{nk}\,.
\end{split}
\end{align}

\section{Solutions for the 5D quark propagator} 
\label{app:5DFermionProp}
\renewcommand{\theequation}{B.\arabic{equation}}
\setcounter{equation}{0}

For details on the procedure of calculating the 5D quark propagator \eqref{eqn:Sprop} in the mixed position-momentum space we refer the reader to \cite{Malm:2013jia}, where the solutions for the propagator functions $\bs\Delta_{LL}^{q}(t,t';k_E^2)$ and $\bs\Delta_{RL}^q(t,t';k_E^2)$ have been derived (with Euclidean momentum $k_E^2=-k^2$). Here we extend their results and also include the solutions for $\bs\Delta_{RR}^{q}(t,t';k_E^2)$. The solution for $\bs\Delta_{LR}^q(t,t';k_E^2)$ can be obtained by complex conjugation $\bs\Delta_{LR}^q(t,t';k_E^2) = \left[\bs\Delta_{RL}^q(t,t';k_E^2)\right]^\da$. We begin with the solutions in the minimal RS model for a brane-localized Higgs sector where the regulator $\eta$ of the regularized $\delta$-function in \eqref{eqn:deltaReg} fulfills the constraint $\eta  \ll y_\ast v / \Lambda_\TeV$, implying that $\eta k_E \ll y_\ast v$. As a consequence the $\eta$ dependence drops out of the solutions for the propagator functions. In this limit $\eta k_E \ll y_\ast v$ we obtain for $q=u,d$: 
\begin{align}
\begin{split} \notag 
 \bs \Delta_{LL}^{q,11} & = \frac{- \sqrt{tt'}}{k_E \MKK}  \bigg[\frac{\bs D_1^Q(\kEh, t)}{\bs D_1^Q(\kEh,1)} \bs R_Q  \frac{1}{1+ \bs Z_q } \frac{\Db_1^Q(\kEh, t')}{\Db_1^Q(\kEh,1)}  - \frac{\bs D_1^Q(\kEh , t_<)}{\bs D_1^Q(\kEh,1)} \bs L_3^Q(\kEh,1,t_>) \bigg]\,, \\
\bs \Delta_{LL}^{q,12} & = \frac{\sqrt{t t'}}{k_E \MKK}\frac{\bs D_1^Q(\kEh, t)}{\bs D_1^Q(\kEh,1)} \bs R_Q   \frac{1}{1+ \bs Z_q } \rh\Tilde \Yb_q  \frac{\Db_2^q(\kEh, t')}{\Db_2^q(\kEh,1)} \,,  \\*
 \bs \Delta_{LL}^{q,21}  & = \frac{\sqrt{t t'}}{k_E \MKK} \frac{\bs D_2^q(\kEh, t)}{\bs D_2^q(\kEh,1)} \rh\Tilde \Yb_q^\da  \bs R_Q  \frac{1}{1+\bs Z_q } \frac{\Db_1^Q(\kEh, t')}{\Db_1^Q(\kEh,1)} \,, \\*
\bs \Delta_{LL}^{q,22}  & = 
\frac{-\sqrt{t t'}}{k_E \MKK} \bigg[\frac{\bs D_2^q(\kEh, t)}{\bs D_2^q(\kEh,1)}\rh\Tilde \Yb_q^\da \bs R_Q \frac{1}{1+\bs Z_q} \rh\Tilde \Yb_q  \frac{\Db_2^q(\kEh, t')}{\Db_2^q(\kEh,1)}  + \frac{\bs D_2^q(\kEh ,t_<)}{\bs D_2^q(\kEh,1)} \bs L_2^q(\kEh,1, t_>) \bigg]\,, 
\end{split} \\ \notag\\
\begin{split} \label{eqn:BranePropSol}
 \bs \Delta_{RL}^{q,11}  & = \frac{-\sqrt{tt'}}{\MKK} \left\{
\begin{array}{ll} 
\frac{\bs D_2^Q(\kEh, t)}{\bs D_2^Q(\kEh,1)} \frac{\bs Z_q}{1 + \bs Z_q} \frac{\Db_1^Q(\kEh, t')}{\Db_1^Q(\kEh,1)} + \frac{ \bs D_2^Q(\kEh, t)}{\bs D_2^Q(\kEh,1)}   \bs L_4^Q(\kEh, t',\e) & \quad , {t< t'} \\
\frac{ \bs D_2^Q(\kEh, t)}{\bs D_2^Q(\kEh,1)}  \frac{\bs Z_q}{1+\bs Z_q} \frac{\Db_1^Q(\kEh, t')}{\Db_1^Q(\kEh,1)} + \frac{\bs D_1^Q(\kEh ,t')}{\bs D_1^Q(\kEh,1)} \bs R_Q \bs L_2^Q(\kEh, 1,t) & \quad,  {t > t'} 
\end{array}\right.\,, \\
\bs \Delta_{RL}^{q,12} & = - \frac{\sqrt{tt'} }{\MKK} \frac{\bs D_2^Q(\kEh, t)}{\bs D_2^Q(\kEh,1)}  \frac{1}{1 + \bs Z_q} \rh\Tilde \Yb_q \frac{\Db_2^q(\kEh, t')}{\Db_2^q(\kEh,1)} \,, \\
 \bs \Delta_{RL}^{q,21} & = - \frac{\sqrt{t t'} }{\MKK} \frac{ \bs D_1^q(\kEh , t)}{\bs D_1^q(\kEh,1)} \frac{1}{\rh\Tilde \Yb_q} \frac{\bs Z_q}{1+\bs Z_q} \frac{\Db_1^Q(\kEh, t')}{\Db_1^Q(\kEh,1)}\,, \\
\bs \Delta_{RL}^{q,22} & = \frac{\sqrt{t t'}}{\MKK} \left\{ 
\begin{array}{ll}
\frac{\bs D_1^q(\kEh, t)}{\bs D_1^q(\kEh,1)} \frac{1}{\Tilde \Yb_q}  \frac{\Zb_q}{1+\bs Z_q}  \Tilde \Yb_q \frac{\Db_2^q(\kEh, t')}{\Db_2^q(\kEh,1)} +  \frac{\bs D_1^q(\kEh, t)}{\bs D_1^q(\kEh,1)}\bs R_q  \bs L_2^q(\kEh,1,t')     & \quad , {t <  t'} \\
\frac{\bs D_1^q(\kEh, t)}{\bs D_1^q(\kEh,1)} \frac{1}{\Tilde \Yb_q}  \frac{\Zb_q}{1+\bs Z_q}  \Tilde \Yb_q  \frac{\Db_2^q(\kEh, t')}{\Db_2^q(\kEh,1)} + \frac{\bs D_2^q(\kEh, t') }{\bs D_2^q(\kEh,1)} \bs L_4^q(\kEh,1,t)    & \quad , {t >  t'}
\end{array}\right.\,,  
\end{split}\\ \notag\\\notag
\bs \Delta_{RR}^{q,11}  & = \frac{- \sqrt{tt'}}{k_E \MKK}  \bigg[\frac{\bs D_2^Q(\kEh, t)}{\bs D_2^Q(\kEh,1)}   \frac{\Zb_q}{1+ \bs Z_q } \frac{1}{\Rb_Q} \frac{\Db_2^Q(\kEh, t')}{\Db_2^Q(\kEh,1)}  + \frac{\bs D_2^Q(\kEh, t_<)}{\bs D_2^Q(\kEh,1)} \bs L_2^Q(\kEh,1,t_>) \bigg]\,, \\\notag
 \bs \Delta_{RR}^{q,12} & = \frac{-\sqrt{t t'}}{k_E \MKK}\frac{\bs D_2^Q(\kEh, t)}{\bs D_2^Q(\kEh,1)}  \frac{\Zb_q}{1+ \bs Z_q} \frac{1}{\Rb_Q} \frac{1}{\rh \Tilde \Yb_q^\da}  \frac{\Db_1^q(\kEh ,t')}{\Db_1^q(\kEh,1)} \,,  \\\notag
 \bs \Delta_{RR}^{q,21} & = \frac{-\sqrt{t t'}}{k_E \MKK} \frac{\bs D_1^q(\kEh, t)}{\bs D_1^q(\kEh,1)} \frac{1}{\rh\Tilde \Yb_q}  \frac{\Zb_q}{1+\Zb_q } \frac{1}{\Rb_Q} \frac{\Db_2^Q(\kEh, t')}{\Db_2^Q(\kEh,1)} \,, \\\notag
\bs \Delta_{RR}^{q,22} & = 
\frac{-\sqrt{t t'}}{k_E \MKK} \bigg[\frac{\bs D_1^q(\kEh, t)}{\bs D_1^q(\kEh,1)} \frac{1}{\rh\Tilde \Yb_q} \frac{\Zb_q}{1+\Zb_q} \frac{1}{\Rb_Q} \frac{1}{\rh\Tilde \Yb_q^\da}  \frac{\Db_1^q(\kEh, t')}{\Db_1^q(\kEh,1)}  - \frac{\bs D_1^q(\kEh, t_<)}{\bs D_1^q(\kEh,1)} \bs L_3^q(\kEh,1, t_>) \bigg]\,.
\end{align}
For the sake of readability we have suppressed the arguments of the propagator functions $\bs\Delta_{AB}^q(t,t';k_E^2)$ for $A,B\in\{L,R\}$ and of $\Zb_q(k_E^2),\,\Rb_{Q,q}(\kEh)$ defined in \eqref{eqn:Zq}, \eqref{eqn:RAdef}. The modified Yukawa matrix is defined by $\tilde\Yb_q=(\tanh\Xb_q/\Xb_q) \Yb_q$ with $\Xb_q=\rh(\Yb_q\Yb_q^\da)^{1/2}$ and $\rh=v/(\sqrt 2 \MKK)$. We have used the abbreviations $\kEh=k_E/\MKK$, $t_> = \tx{Max}(t,t')$ and $t_< = \tx{Min}(t,t')$. In \eqref{eqn:BranePropSol} we used the functions $\bs D_i^{Q,q}( \hat{k}_E,t)$ \cite{Malm:2013jia} which are related to the more general functions $\bs D_i^{Q,q}( \hat{k}_E,t,t')$ defined below via $\bs D_i^{Q,q}( \hat{k}_E,t) \equiv \bs D_i^{Q,q}( \hat{k}_E , t , \epsilon)$. The generalized functions read
\begin{align} 
\begin{split}
\bs D_{1,2}^A ( \kEh,t, t^\prime ) &\equiv I_{-\cA -\frac{1}{2}} ( \kEh t^\prime) \,  I_{\cA\mp\frac{1}{2}} (\kEh t) - I_{\cA+\frac{1}{2}} (\kEh t^\prime) \,  I_{-\cA\pm\frac{1}{2}} (\kEh t)\,,\\
\bs D_{3,4}^A ( \kEh,t, t^\prime ) &\equiv I_{-\cA +\frac{1}{2}} ( \kEh t^\prime ) \,  I_{\cA\mp \frac{1}{2}} (\kEh t) - I_{\cA-\frac{1}{2}} (\kEh t^\prime) \,  I_{-\cA\pm\frac{1}{2}} (\kEh t)\,,
\end{split}
\end{align}
and
\begin{align}\label{eqn:Lstruc}
\R{{A}}{\pE, t}{t^\prime}{i} \equiv \frac{\pi \pE 1_\eta}{2\cos\left(\pi \boldsymbol{c}_{{A}}\right)}\, \bs D_i^{{A}} ( \kEh, t, t^\prime  )\,; \qquad i=1,2,3,4\,,
\end{align}
for $A=Q,q$. We introduced the shorthand notation $1_\eta\equiv 1-\eta$ and the bulk-mass parameters are denoted by $\bs c_{Q,q}$. 

Next we focus on the case of a narrow bulk-Higgs where the regulator $\eta$ takes values in the range $y_\ast v/\Lambda_\TeV\ll \eta \ll y_\ast v/\MKK$, implying that the propagator solutions explicitly depend on the product $\eta \hat k_E$. For the calculation of the scalar contributions in Section~\ref{sec:Wils5Dbsga} we need to evaluate the 5D propagator functions in the region near the IR brane, where the extra-dimensional coordinates take values in the range $t,t'\in [1_\eta,1]$. Focusing on the solutions in this region we obtain the results $(t,t' \in [1_\eta,1])$: 
\begin{align} \notag
\bs \Delta_{LL}^{q,11} & =  
\frac{- 1}{k_E \MKK}  \bigg[ \frac{\Cm(t)}{ \Cm(\et)}\Big( \eta \kEh  \frac{\coth\Sb_\qv}{\Sb_\qv} \bs Z_\At^{\eta,1} +  \bs R_\Av  \Big) \frac{1}{\bs{N}_\At^{\eta,1}} \frac{\Cm(t') }{\Cm(\et)} - \eta \kEh \frac{\Cm(t_>)}{ \Cm(\et)} \frac{\Sm(t_< + \eta)}{\Sb_\qv}  \bigg]\,, \\\notag
\bs \Delta_{LL}^{q,12} & = 
\frac{1}{k_E \MKK} \frac{\Cm(t)}{\Cm(\et)} \bs R_\Av  \frac{1}{\bs{N}_\At^{\eta,2} } \frac{\Sm(t')}{\Sm(\et)} \rh \Tilde \Yb_\av\,, \\\notag
\bs \Delta_{LL}^{q,21} & = 
\frac{1}{k_E \MKK} \rh \Tilde \Yb_\av^\da \frac{\Sm(t)}{\Sm(\et)} \bs R_\Av(\kEh) \frac{1}{\bs{N}_\At^{\eta,1}} \frac{\Cm(t')}{\Cm(\et)}\,, \\\notag
\bs \Delta_{LL}^{q,22}  & = 
 \frac{-1}{k_E \MKK} \rh \Tilde \Yb_\av^\da  \frac{\Sb_\qv^2}{\Xb_\av^2}  \bigg[\frac{ \Sm(t)}{ \Sm(\et)} \Big(\eta \kEh  \frac{\coth\Sb_\qv}{\Sb_\qv} + \bs R_\Av \Big) \frac{1}{\bs{N}_\At^{\eta,2}}\frac{ \Sm(t')}{\Sm(\et)} \\*\notag &\quad \hspace{32mm} - \eta \kEh \frac{ \Sm(t_>)}{ \Sm(\et)}\frac{\Sm(t_<+\eta)}{\Sb_\qv \tanh^2\Sb_\qv} \bigg]  \rh \Tilde \Yb_\av\,, \\\notag\\
\begin{split}\label{eqn:DABbulk}
\bs\Delta_{RR}^{q,11} &= \frac{-1}{k_E\MKK}\bigg[\frac{\Sm(t)}{\Sm(1_\eta)}\Big(\Zb_q^{\eta,1} + \eta\kEh\frac{\tanh\Sb_q}{\Sb_q}\Rb_Q\Big)\frac{1}{\Nb_q^{\eta,1}}\frac{1}{\Rb_Q}\frac{\Sm(t')}{\Sm(1_\eta)} \\
& \quad\hspace{17mm} - \eta\kEh\frac{\Sm(t_>)\Sm(t_<+\eta)}{\Sm(1_\eta)\Sb_q}\bigg]\,,\\
\bs\Delta_{RR}^{q,12} &= \frac{-1}{k_E\MKK}\frac{\Sm(t)}{\Sm(1_\eta)}\frac{1}{\Nb_q^{\eta,2}}\Zb_q^{\eta,2}\frac{1}{\Rb_Q}\frac{\Cm(t')}{\Cm(1_\eta)} \frac{\Xb_q^2}{\Sb_q^2}\frac{1}{\tilde\Yb_q^\da}\,,\\
\bs\Delta_{RR}^{q,21} &= \frac{-1}{k_E\MKK}\frac{1}{\tilde\Yb_q}\frac{\Xb_q^2}{\Sb_q^2}\frac{\Cm(t)}{\Cm(1_\eta)}\Zb_q^{\eta,1}\frac{1}{\Nb_q^{\eta,1}}\frac{1}{\Rb_Q}\frac{\Sm(t')}{\Sm(1_\eta)}\,,\\
\bs\Delta_{RR}^{q,22} &= \frac{-1}{k_E\MKK}\frac{1}{\rh\tilde\Yb_q}\bigg[\frac{\Cm(t)}{\Cm(1_\eta)}\Big(1 + \eta\kEh\frac{\tanh\Sb_q}{\Sb_q}\Rb_Q\Big)\frac{1}{\Nb_q^{\eta,2}}\Zb_q^{\eta,2} \frac{1}{\Rb_Q}\frac{\Cm(t')}{\Cm(1_\eta)} \\
& \quad \hspace{25mm} - \eta\kEh\frac{\Cm(t_>)\Sm(t_<+\eta)}{\Cm(1_\eta)\Sb_q\coth^2\Sb_q}\bigg]\,,
\end{split} \\ \notag \\
\begin{split}\notag
\bs\Delta_{RL}^{q,11} & = \frac{-1}{\MKK} \bigg[\frac{\mathcal{S}(t)}{\mathcal{S}(1_\eta)}\Big(\bs Z_\At^{\eta,1} + \eta \kEh \frac{\tanh\Sb_\av}{\Sb_\av} \bs R_\Av \Big) \frac{1}{\bs{N}_\At^{\eta,1}}   \frac{\mathcal{C}(t')}{\mathcal{C}(1_\eta)}  - \frac{\Cm(t+\eta)\Cm(t')}{\Cm(1_\eta)} \\
&\quad \hspace{13mm}+ \theta(t-t') \Cm(1+t-t')\bigg]\,,\\
\bs\Delta_{RL}^{q,12} & =\frac{-1}{\MKK}\bigg[\frac{\mathcal{S}(t)}{\mathcal{S}(1_\eta)} \frac{1-\bs{N}_\At^{\eta,2}}{\bs{N}_\At^{\eta,2}} \rh \tilde \Yb_{q} \frac{ \bar {\cal S}(t')}{\bar {\cal S}(1_\eta)} + \frac{\Sm(t_>) \Cm(t_<+\eta)}{\Sm(1_\eta)}  \rh \tilde \Yb_{q}\bigg]\,,\\
\bs\Delta_{RL}^{q,21} & =\frac{-1}{\MKK}\bigg[\frac{\bar {\cal C}(t)}{\bar {\cal C}(1_\eta)}  \frac{1}{\rh \tilde \Yb_{q}} \frac{\Xb_{q}^2}{\Sb_{q}^2}  \bs Z_\At^{\eta,1}   \frac{1}{\bs{N}_\At^{\eta,1}} \frac{\mathcal{C}(t')}{\mathcal{C}(1_\eta)}  - \frac{1}{\rh \tilde \Yb_{q}} \frac{\Xb_{q}^2}{\Sb_{q}^2} \frac{\Cm(t_>)\Sm(t_<+\eta)}{ \Cm(1_\eta)\coth\Sb_{q}} \bigg]\,,\\
\bs\Delta_{RL}^{q,22} &=\frac{-1}{\MKK}\bigg[ \frac{\bar {\cal C}(t)}{\bar {\cal C}(1_\eta)} \frac{1}{\tilde \Yb_{q}} \Big[ 1 - \Nb_{q}^{\eta,2} + \eta\kEh \frac{\tanh\Sb_\av}{\Sb_\av} \bs R_\Av \Big]  \frac{1}{\bs{N}_\At^{\eta,2}} \tilde \Yb_{q}  \frac{ \bar {\cal S}(t')}{\bar {\cal S}(1_\eta)} + \frac{\bar \Sm(t+\eta) \bar \Sm(t')}{\bar \Cm(1_\eta)}  \\
&\quad \hspace{13mm} + \theta(t-t') \Cm(1+t-t') \bigg]\,.
\end{split}
\end{align}
Again we have suppressed the arguments of the functions for the sake of readability. The step functions is denoted by $\theta(t-t')$. The $t$-dependent functions ${\cal S}(t),\,{\cal C}(t)$ are defined in \eqref{eqn:SqtCqt} where we have to replace $\Xb_q$ by $\Sb_q$ as defined in \eqref{eqn:Sq}, analogously for $\bar {\cal S}(t)$ and $\bar{\cal C}(t)$. In \eqref{eqn:DABbulk} the modified Yukawa matrix is defined by $\tilde\Yb_q=(\tanh \Sb_q/\Sb_q) \Yb_q$. We also introduced new structures ($q=u,d$) \cite{Malm:2013jia}
\begin{align}
\begin{split}
\bm{N}_{q}^{\eta,1}(k_E^2) &\equiv 1 +\bm{Z}_{q}^{\eta,1}(k_E^2)   + \eta \hat k_E \bigg[\bs R^{-1}_{Q}(\hat k_E) \frac{\coth \bs S_{q}}{\bs S_{q}}  \bm{Z}_{q}^{\eta,1}(k_E^2) + \frac{\tanh \bar\Sb_{q}}{\bar\Sb_{q}} \bs R_{Q}(\hat k_E) \bigg]\,,  \\
\bm{N}_{q}^{\eta,2}(k_E^2)
&\equiv 1 + \bm{Z}_{q}^{\eta,2}(k_E^2)
+ \eta\hat k_E\,
\left[\bs Z_{q}^{\eta,2}(k_E^2)\bs R^{-1}_{Q}(\hat k_E) \frac{\coth\Sb_{q}}{\Sb_{q}} + \frac{\tanh\bm{S}_{q}}{\bm{S}_{q}} \bm{R}_{{Q}}( \hat k_E)\right]\,,
\end{split}
\end{align}
with
\begin{align}
\Zm_{q}^{\eta,1}(k^2_E) &=\varrho^2 \frac{\bs S_{q}^{2}}{\Xm_{q}^{2}}\, \tilde{\Y}_{q}\, \Rmq{{q}}{ \hat k_E} \,\tilde{\Y}^{\dagger}_{q} \Rmq{{{Q}}}{ \hat k_E} \,, &
\Zm_{q}^{\eta,2}(k^2_E) & = \varrho^2 
\tilde{\Y}_{q}\,\Rmq{{q}}{ \hat k_E}\, \tilde{\Y}^{\dagger}_{q}  \, \frac{\bs S_{q}^{2}}{\Xm_{q}^{2}}\, \Rmq{{{Q}}}{ \hat k_E} \;.
\end{align}

\section{Ultra-violet behavior of 5D propagators}
\label{sec:UV5Dprop}
\renewcommand{\theequation}{C.\arabic{equation}}
\setcounter{equation}{0}

This section discusses the behavior of the 5D propagator functions in the brane-localized Higgs scenario for large momenta $k_E \gg \MKK / t$ exceeding the effective Planck scale at each point in the extra dimension. The results are used to show the finiteness of the penguin diagrams and to calculate the boundary terms in Section~\ref{sec:Wils5Dbsga}.

\subsubsection*{Gauge-boson propagator functions}

Expanding the scalar and vector parts of the gauge-boson propagator functions in Euclidean momentum space we find to leading order ($\kEh \gg 1/t,1/t'$)
\begin{align}\label{eqn:BBexpansion}\notag
B_B^\tx{scalar}(t,t';k_E^2) & \approx \frac{L (t t')^\frac{3}{2}}{2\pi \e^2} \frac{e^{- \hat k_E (t_> -t_<)}}{2 k_E \MKK} \bigg[1 - e^{2\hat k_E(t_>-1)} + \frac{L \tilde m^2_B \big(1 + e^{2\hat k_E(t_>-1)}\big) }{k_E \MKK} \bigg] \bigg[ 1 - e^{2 \hat k_E(\e-t_<)}\bigg]\,,\\
B_B(t,t';k_E^2) &\approx \frac{L \sqrt{t t'}}{2\pi} \frac{e^{- \hat k_E (t_> -t_<)}}{2 k_E \MKK} \Big[1 + e^{2\hat k_E(t_>-1)} \Big] \Big[ 1 + e^{2 \hat k_E(\e-t_<)}\Big]\,, 
\end{align}
for (subscript) $B=A,G,W,Z$. In case of the massive gauge bosons $B=W,Z$ the scalar propagator function includes the contributions from the fifth component of the 5D gauge-boson fields and from the corresponding Goldstone-bosons in the Higgs sector, which gives rise to the term proportional to $L \tilde m_B^2/k_E \MKK$. This term is absent in case of the photon or gluon scalar propagator $(B=A,G)$. Integrating \eqref{eqn:BBexpansion} with a (well-behaved) function $f(t,t')$ along both extra dimensional coordinates we can show that
\begin{align}\label{eqn:BBf}
\begin{split}
\int_\e^1 dt  dt' \, B_B(t,t';k_E^2) \, f(t,t') & \approx \frac{L}{4\pi k_E^2} \int_\e^1 dt \, f(t,t)  \,,    \\
\int_\e^1 dt  dt' \, B_B^\tx{scalar}(t,t';k_E^2) \, f(t,t')& \approx \frac{L}{8\pi\e^2 k_E^2} \int_\e^1 dt \, f(t,t)   \,.
\end{split}
\qquad(\kEh \gg 1/\e)
\end{align}
We cannot prove \eqref{eqn:BBf} in general but we checked analytically that the relations are valid for the functions relevant in the calculations of this paper. Relations \eqref{eqn:BBf} imply that for large Euclidean momenta the propagator functions can be effectively replaced by the $\delta$-function $\delta(t-t')$ apart from a constant factor. 

\subsubsection*{Quark propagator functions}

The results for the 5D quark propagator functions in the brane-localized Higgs scenario are listed in \eqref{eqn:BranePropSol}. The solutions contain several functions that have a simple form when expanded for large Euclidean momenta. To leading order we find the expressions ($\kEh \gg 1/t$)
\begin{align} \label{eqn:DLUVexp} \notag
\frac{ \Db^{A}_{1,4}(\kEh, t) }{ \Db^{A}_{1,4}(\kEh,1) }  & \approx \frac{e^{\kEh(t-1)}}{\sqrt t} {\left[1 + e^{2\kEh(\e-t)} \right]}\,, \hspace{2mm}  \frac{ \Db^{ A}_{2,3}(\kEh,t) }{ \Db^{A}_{2,3}(\kEh,1) }  \approx \frac{e^{\kEh(t-1)}}{\sqrt t} {\left[1 - e^{2\kEh(\e-t)} \right]}\,,\hspace{2mm} \Rb_{A}(\kEh) \approx 1\,,\\
\Lb^A_{1,2}(\kEh,1, t)  & = - \Lb^A_{4,3}(\kEh,1, t) \approx  \frac{e^{\kEh(1-t)}}{2\sqrt t}  \left[1 \pm e^{2\kEh(t-1)} \right]\,,\hspace{2mm} \Zb_q(k_E^2) \approx \rh^2\tilde\Yb_q\tilde\Yb_q^\da\,,
\end{align}
for $A=Q,q$ and $q=u,d$. The expansions are independent of the bulk-mass parameters. Using the above expressions in case of the propagator function $\bs\Delta_{LL}^q(t,t';k_E^2)$ we find ($\kEh \gg 1/t,\,1/t'$)  
\begin{align}\label{eqn:DeltaLLUVexp}
\begin{split}
\bs \Delta_{LL}^{q,11}(t,t';k_E^2) & \approx - \frac{e^{- \hat k_E(t_> - t_<)}}{2  k_E \MKK} \bigg[ 1 + e^{2 \hat k_E(\e-t_<)}\bigg] \bigg[1+\frac{1-\rh^2\tilde\Yb_q\tilde\Yb_q^\da}{1+\rh^2\tilde\Yb_q\tilde\Yb_q^\da} e^{2\hat k_E(t_>-1)}\bigg]  \,,\\
\bs \Delta_{LL}^{q,12}(t,t';k_E^2) & \approx  \frac{e^{- \hat k_E(2-t-t')}}{k_E \MKK} \frac{1}{1+\rh^2\tilde\Yb_q\tilde\Yb_q^\da} \varrho \tilde{\bs Y}_q\Big[1 + e^{2 \hat k_E(\e-t)}\Big] \Big[1 - e^{2 \hat k_E(\e-t')}\Big]  \,, \\
\bs \Delta_{LL}^{q,21}(t,t';k_E^2) & \approx  \frac{e^{- \hat k_E(2-t-t')}}{k_E \MKK} \varrho \tilde{\bs Y}^\da_q \frac{1}{1+\rh^2\tilde\Yb_q\tilde\Yb_q^\da} \Big[1 - e^{2 \hat k_E(\e-t)}\Big] \Big[1 + e^{2 \hat k_E(\e-t')}\Big] \,,\\
\bs \Delta_{LL}^{q,22}(t,t';k_E^2) & \approx -  \frac{e^{- \hat k_E(t_> - t_<)}}{2 k_E \MKK} \bigg[1 - e^{2 \hat k_E(\e-t_<)}\bigg] \bigg[1 - \frac{1-\rh^2\tilde\Yb_q^\da\tilde\Yb_q}{1+\rh^2\tilde\Yb_q^\da\tilde\Yb_q} e^{2\hat k_E(t_>-1)}\bigg]  \,,
\end{split}
\end{align}
at leading order in $\kEh^{-1}$. For a (well-behaved) function $f(t,t')$ we can show that 
\begin{align}\label{eqn:DeltaLLpEhlim}
\int_\e^1 dt  dt'  \, \bs\Delta^{q}_{LL}(t,t';k_E^2) \, f(t,t')& \approx - \frac{1}{k_E^2} \int_\e^1 dt \, f(t,t) \,,\qquad (\kEh \gg 1/\e)
\end{align}
implying that the propagator function behaves like the $\delta$-function $\delta(t-t')$ for large momenta. For functions that are localized near the IR brane the equation is already a good approximation for $\kEh\gg1$. Equation \eqref{eqn:DeltaLLpEhlim} is also valid in case of the propagator function $\bs \Delta_{RR}^q(t,t';k_E^2)$. We continue with the chirality-changing propagator function $\bs\Delta_{LR}^q(t,t';k_E^2)$.~Using the coupled differential equation \cite{Malm:2013jia}
\begin{align}
\bs\Delta_{LR}^q(t,t';k_E^2)= \MKK(-\pa_{t'} + \M_q(t')) \bs\Delta_{LL}^q(t,t';k_E^2)\,,
\end{align}
we can show for a (well-behaved) function $f(t,t')$ that
\begin{align}\label{eqn:DeltaLRexp1}\notag
\int_\e^1 dt dt' \, \bs \Delta^{q}_{LR}(t,t';k_E^2)\,f(t,t') &=  \MKK  \int_\e^1 dt \, \Big[ - \bs \Delta^{q}_{LL}(t,1;k_E^2) f(t,1) +  \bs \Delta^{q}_{LL}(t,\e;k_E^2) f(t,\e) \Big]  \\
& \hspace{5mm} +\MKK  \int_\e^1 dtdt' \bs \Delta^{q}_{LL}(t,t';k_E^2) \Big[\pa_{t'} \, + \M_q(t')\Big] f(t,t')  \,,
\end{align}
where we have performed a partial integration in $t'$. In the large Euclidean momentum region we can show that $(\kEh\gg1/\e)$
\begin{align}\label{eqn:DeltaLLft1exp}
\begin{split}
\int_\e^1 dt \, \bs \Delta_{LL}^q(t,1;k_E^2)\, f(t,1) & \approx \frac{-1}{k_E^2} \bigg[\frac{ \bs P_{+}}{1 + \rh^2\tilde\Yb_q\tilde\Yb_q^\da} + \frac{\bs P_{-} \,\rh^2\tilde\Yb_q^\da\tilde\Yb_q}{1 + \rh^2\tilde\Yb_q^\da\tilde\Yb_q}  -\frac{\bs P_{12}}{1 + \rh^2\tilde\Yb_q\tilde\Yb_q^\da}  \varrho \tilde{\bs Y}_q \\
& \hspace{5mm} - \varrho \tilde{\bs Y}_q^\da   \frac{\bs P_{21}}{1 + \rh^2\tilde\Yb_q\tilde\Yb_q^\da}  \bigg] f(1,1)\,,\\
\int_\e^1 dt \, \bs \Delta_{LL}^q(t,\e;k_E^2) \, f(t,\e)& \approx - \frac{1}{k_E^2} \, \bs P_+ \,f(\e,\e)\,,
\end{split}
\end{align}
where higher order terms are suppressed at least by $k_E^{-3}$. The first relation is approximately valid already for $\kEh\gg1$ if $f(t,1)$ has most of its support near the IR brane. Using \eqref{eqn:DeltaLLft1exp} in \eqref{eqn:DeltaLRexp1} we finally find for a (well-behaved) function $f(t,t')$ ($\kEh\gg1/\e$)
\begin{align}\label{eqn:DeltaLRlim}
\begin{split}
\int_\e^1 dt dt' \, \bs\Delta^{q}_{LR}(t,t';k_E^2) \, f(t,t')& \approx   \frac{\MKK}{k_E^2} \bigg\{ \bigg[\frac{ \bs P_{+}}{1 + \rh^2\tilde\Yb_q\tilde\Yb_q^\da} + \frac{\bs P_{-} \,\rh^2\tilde\Yb_q^\da\tilde\Yb_q}{1 + \rh^2\tilde\Yb_q^\da\tilde\Yb_q}  -\frac{\bs P_{12}}{1 + \rh^2\tilde\Yb_q\tilde\Yb_q^\da}  \varrho \tilde{\bs Y}_q \\
& \hspace{-45mm} - \varrho \tilde{\bs Y}_q^\da   \frac{\bs P_{21}}{1 + \rh^2\tilde\Yb_q\tilde\Yb_q^\da}  \bigg] f(1,1) - \bs P_+ \,f(\e,\e)  - \int_\e^1 dt \, \Big[\frac{f'(t,t^+) + f'(t,t^-)}{2} + \M_q(t) \, f(t,t)  \Big] \bigg\}\,,
\end{split}
\end{align}
where $f'(t,t^\pm) = \lim_{s\rightarrow t \pm 0}  \pa_s \, f(t,s)$ is understood as a limiting procedure. An analogous equation can be derived for the propagator function $\bs\Delta_{RL}^q(t,t';k_E^2)$. 

\section{Loop functions} 
\label{sec:LoopFunctions}
\renewcommand{\theequation}{D.\arabic{equation}}
\setcounter{equation}{0}

In the KK-decomposed 4D theory the Wilson coefficients in \eqref{eqn:C78KK} involve the loop functions $I_{3,4}(x)$ and $I_{6-11}(x)$. They are defined by the integral representations
\begin{align}\label{eqn:LoopFuncsInt} \notag
I_3(x^n_m) & = \frac{m_m^2}{m_n^2}\bigg(\frac{1}{2} + m_m^2 \int_0^\infty dk_E \, \frac{1}{k_E^2+m_n^2} \bigg[\frac{3k_E^2}{4} \pa_{k_E} + \frac{k_E^3}{4}\pa_{k_E}^2\bigg]\, \frac{1}{k_E^2+m_m^2}\bigg) \,,\\\notag
I_4(x^n_m) & =\frac{m_m^2}{m_n^2}\bigg( \frac{1}{12} + m_m^2 \int_0^\infty dk_E \, \frac{1}{k_E^2+m_n^2} \bigg[\frac{k_E^2}{16}\pa_{k_E} - \frac{k_E^3}{16}\pa_{k_E}^2 - \frac{k_E^4}{48}\pa_{k_E}^3\bigg]\, \frac{1}{k_E^2+m_m^2}\bigg)\,, \\\notag
I_6(x^n_m) & = -\frac{1}{2} + m_m^2 \int_0^\infty dk_E \frac{1}{k_E^2+m_n^2} \bigg[\frac{9k_E^2}{4}\pa_{k_E}+\frac{3k_E^3}{4}\pa_{k_E}^2\bigg] \frac{1}{k_E^2+m_m^2}\,,\\\notag
I_7(x^n_m) & = \frac{5}{12} + m_m^2 \int_0^\infty dk_E \frac{1}{k_E^2+m_n^2} \bigg[-\frac{3k_E^2}{16}\pa_{k_E}+\frac{3k_E^3}{16}\pa_{k_E}^2+\frac{k_E^4}{16}\pa_{k_E}^3\bigg] \frac{1}{k_E^2+m_m^2}\,,\\\notag
I_8(x^m_n) & = -\frac{1}{4} + m_m^2 \int_0^\infty dk_E \frac{1}{k_E^2+m_n^2} \bigg[\frac{3k_E^2}{8}\pa_{k_E}-\frac{3k_E^3}{8}\pa_{k_E}^2\bigg] \frac{1}{k_E^2+m_m^2}\,,\\\notag
I_9(x^m_n) & = \frac{1}{6} + m_m^2 \int_0^\infty dk_E \frac{1}{k_E^2+m_n^2} \bigg[-\frac{3k_E^2}{32}\pa_{k_E}+\frac{3k_E^3}{32}\pa_{k_E}^2 - \frac{k_E^4}{32}\pa_{k_E}^3 \bigg] \frac{1}{k_E^2+m_m^2}\,,\\\notag
I_{10}(x^m_n) & = -\frac{1}{4} + m_m^2 \int_0^\infty dk_E \frac{1}{k_E^2+m_n^2} \bigg[-\frac{3k_E^2}{8}\pa_{k_E}+\frac{3k_E^3}{8}\pa_{k_E}^2\bigg] \frac{1}{k_E^2+m_m^2}\,,\\
I_{11}(x^m_n) & = \frac{1}{6} + m_m^2 \int_0^\infty dk_E \frac{1}{k_E^2+m_n^2} \bigg[\frac{5k_E^2}{32}\pa_{k_E}-\frac{5k_E^3}{32}\pa_{k_E}^2 + \frac{k_E^4}{96}\pa_{k_E}^3 \bigg] \frac{1}{k_E^2+m_m^2}\,,
\end{align}
where $x_b^a=m_a^2/m_b^2$. Note the different arguments of the loop functions $I_{3,4,6,7}(x)$ and $I_{8-11}(x)$. Performing the momentum integrals the loop functions explicitly read
\begin{align}\label{eqn:LoopFuncsExpl}
\begin{split}
I_3(x) & = \frac{3-4x+x^2+2\ln x}{2(x-1)^3}\,,\\
I_4(x) & = \frac{2+3x-6x^2+x^3+6x\ln x}{12(x-1)^4}\,,\\
I_6(x) & = \frac{4 - 3 x - x^3 + 6 x \ln x}{2(x-1)^3}\,, \\
I_7(x) & =\frac{8 - 38 x + 39 x^2 - 14 x^3 + 5 x^4 - 18 x^2 \ln x}{12 (x-1)^4}\,,\\
I_8(x) & = \frac{1 - 12 x + 15 x^2 - 4 x^3 - 6 x \ln x}{4 (x-1)^3}\,,\\
I_9(x) & = \frac{4 - 49 x + 78 x^2 - 43 x^3 + 10 x^4 - 18 x \ln x}{24 (x-1)^4} \,,\\
I_{10}(x) & =\frac{1 + 6 x - 9 x^2 + 2 x^3 + 6 x \ln x}{4 (x-1)^3} \,,\\
I_{11}(x) &= \frac{4 + 13 x - 36 x^2 + 23 x^3 - 4 x^4 - 6 x (2x - 3) \ln x}{24 (x-1)^4}\,.
\end{split}
\end{align}

\end{appendix}


\newpage
\begin{thebibliography}{99}

\bibitem{ATLAS:2012gk} 
  G.~Aad {\it et al.}  [ATLAS Collaboration],
  %``Observation of a new particle in the search for the Standard Model Higgs boson with the ATLAS detector at the LHC,''
  Phys.\ Lett.\ B {\bf 716}, 1 (2012)
  [arXiv:1207.7214 [hep-ex]].
  %%CITATION = ARXIV:1207.7214;%%
  
\bibitem{CMS:2012gu} 
  S.~Chatrchyan {\it et al.}  [CMS Collaboration],
  %``Observation of a new boson at a mass of 125 GeV with the CMS experiment at the LHC,''
  Phys.\ Lett.\ B {\bf 716}, 30 (2012)
  [arXiv:1207.7235 [hep-ex]].
  %%CITATION = ARXIV:1207.7235;%%  

\bibitem{Randall:1999ee}  
  L.~Randall and R.~Sundrum,  
  %``A large mass hierarchy from a small extra dimension,''  
  Phys.\ Rev.\ Lett.\  {\bf 83}, 3370 (1999)  
  [hep-ph/9905221].  
  %%CITATION = HEP-PH 9905221;%%

\bibitem{Grossman:1999ra}
  Y.~Grossman and M.~Neubert,
  %``Neutrino masses and mixings in non-factorizable geometry,''
  Phys.\ Lett.\ B {\bf 474}, 361 (2000)
  [hep-ph/9912408].
  %%CITATION = HEP-PH 9912408;%%

\bibitem{Gherghetta:2000qt}
  T.~Gherghetta and A.~Pomarol,
  %``Bulk fields and supersymmetry in a slice of AdS,''
  Nucl.\ Phys.\ B {\bf 586}, 141 (2000)
  [hep-ph/0003129].
  %%CITATION = HEP-PH 0003129;%%

\bibitem{Huber:2000ie}
  S.~J.~Huber and Q.~Shafi,
  %``Fermion masses, mixings and proton decay in a Randall-Sundrum model,''
  Phys.\ Lett.\ B {\bf 498}, 256 (2001)
  [hep-ph/0010195].
  %%CITATION = HEP-PH 0010195;%%

\bibitem{Agashe:2004ay} 
  K.~Agashe, G.~Perez and A.~Soni,
  %``B-factory signals for a warped extra dimension,''
  Phys.\ Rev.\ Lett.\  {\bf 93}, 201804 (2004)
  [hep-ph/0406101].
  %%CITATION = HEP-PH/0406101;%%

\bibitem{Agashe:2004cp} 
  K.~Agashe, G.~Perez and A.~Soni,
  %``Flavor structure of warped extra dimension models,''
  Phys.\ Rev.\ D {\bf 71}, 016002 (2005)
  [hep-ph/0408134].
  %%CITATION = HEP-PH/0408134;%%

\bibitem{Csaki:2008zd}
  C.~Csaki, A.~Falkowski and A.~Weiler,
  %``The Flavor of the Composite Pseudo-Goldstone Higgs,''
  JHEP {\bf 0809}, 008 (2008)
  [arXiv:0804.1954 [hep-ph]].
  %%CITATION = JHEPA,0809,008;%%  
   
\bibitem{Casagrande:2008hr}
  S.~Casagrande, F.~Goertz, U.~Haisch, M.~Neubert and T.~Pfoh,
  %``Flavor Physics in the Randall-Sundrum Model: I. Theoretical Setup and
  %Electroweak Precision Tests,''
  JHEP {\bf 0810}, 094 (2008)
  [arXiv:0807.4937 [hep-ph]].
  %%CITATION = JHEPA,0810,094;%%  

\bibitem{Blanke:2008zb} 
  M.~Blanke, A.~J.~Buras, B.~Duling, S.~Gori and A.~Weiler,
  %``$\Delta$ F=2 Observables and Fine-Tuning in a Warped Extra Dimension with Custodial Protection,''
  JHEP {\bf 0903}, 001 (2009)
  [arXiv:0809.1073 [hep-ph]].
  %%CITATION = ARXIV:0809.1073;%%

\bibitem{Blanke:2008yr} 
  M.~Blanke, A.~J.~Buras, B.~Duling, K.~Gemmler and S.~Gori,
  %``Rare K and B Decays in a Warped Extra Dimension with Custodial Protection,''
  JHEP {\bf 0903}, 108 (2009)
  [arXiv:0812.3803 [hep-ph]].
  %%CITATION = ARXIV:0812.3803;%%

\bibitem{Bauer:2009cf} 
  M.~Bauer, S.~Casagrande, U.~Haisch and M.~Neubert,
  %``Flavor Physics in the Randall-Sundrum Model: II. Tree-Level Weak-Interaction Processes,''
  JHEP {\bf 1009}, 017 (2010)
  [arXiv:0912.1625 [hep-ph]].
  %%CITATION = ARXIV:0912.1625;%%

\bibitem{Haisch:2008ar} 
  U.~Haisch,
  %``$\bar{B} \to X_{s} \gamma$: Standard Model and Beyond,''
  arXiv:0805.2141 [hep-ph].
  %%CITATION = ARXIV:0805.2141;%%
  
\bibitem{Agashe:2006iy} 
  K.~Agashe, A.~E.~Blechman and F.~Petriello,
  %``Probing the Randall-Sundrum geometric origin of flavor with lepton flavor violation,''
  Phys.\ Rev.\ D {\bf 74}, 053011 (2006)
  [hep-ph/0606021].
  %%CITATION = HEP-PH/0606021;%%

\bibitem{Csaki:2010aj} 
  C.~Csaki, Y.~Grossman, P.~Tanedo and Y.~Tsai,
  %``Warped penguin diagrams,''
  Phys.\ Rev.\ D {\bf 83}, 073002 (2011)
  [arXiv:1004.2037 [hep-ph]].
  %%CITATION = ARXIV:1004.2037;%%

\bibitem{Blanke:2012tv} 
  M.~Blanke, B.~Shakya, P.~Tanedo and Y.~Tsai,
  %``The Birds and the Bs in RS: The $b to s \gamma$ penguin in a warped extra dimension,''
  JHEP {\bf 1208}, 038 (2012)
  [arXiv:1203.6650 [hep-ph]].
  %%CITATION = ARXIV:1203.6650;%%
  
\bibitem{Schmell:2014lka} 
  C.~Schmell,
  ``Hunting for warped extra-dimensions using loop-induced Processes'',
  Doctoral Thesis, Johannes Gutenberg University Mainz (2015).
  %%CITATION = INSPIRE-1358355;%%
  
\bibitem{Biancofiore:2014wpa} 
  P.~Biancofiore, P.~Colangelo and F.~De Fazio,
  %``Rare semileptonic $B\to K^* \ell^+ \ell^- $ decays in RS$_c$ model,''
  Phys.\ Rev.\ D {\bf 89}, no. 9, 095018 (2014)
  [arXiv:1403.2944 [hep-ph]].
  %%CITATION = ARXIV:1403.2944;%%
  
  \bibitem{Beneke:2015lba} 
  M.~Beneke, P.~Moch and J.~Rohrwild,
  %``Lepton flavour violation in RS models with a brane- or nearly brane-localized Higgs,''
  arXiv:1508.01705 [hep-ph].
  %%CITATION = ARXIV:1508.01705;%%
  
\bibitem{Carena:2004zn} 
  M.~S.~Carena, A.~Delgado, E.~Ponton, T.~M.~P.~Tait and C.~E.~M.~Wagner,
  %``Warped fermions and precision tests,''
  Phys.\ Rev.\ D {\bf 71}, 015010 (2005)
  [hep-ph/0410344].
  %%CITATION = HEP-PH/0410344;%%

\bibitem{Baak:2014ora} 
  M.~Baak {\it et al.} [Gfitter Group Collaboration],
  %``The global electroweak fit at NNLO and prospects for the LHC and ILC,''
  Eur.\ Phys.\ J.\ C {\bf 74}, 3046 (2014)
  [arXiv:1407.3792 [hep-ph]].
  %%CITATION = ARXIV:1407.3792;%%
  
\bibitem{Agashe:2003zs}
  K.~Agashe, A.~Delgado, M.~J.~May and R.~Sundrum,
  %``RS1, custodial isospin and precision tests,''
  JHEP {\bf 0308}, 050 (2003)
  [hep-ph/0308036].
  %%CITATION = HEP-PH 0308036;%%

\bibitem{Csaki:2003zu} 
  C.~Csaki, C.~Grojean, L.~Pilo and J.~Terning,
  %``Towards a realistic model of Higgsless electroweak symmetry breaking,''
  Phys.\ Rev.\ Lett.\  {\bf 92}, 101802 (2004)
  [hep-ph/0308038].
  %%CITATION = HEP-PH/0308038;%%
  
\bibitem{Agashe:2006at}
  K.~Agashe, R.~Contino, L.~Da Rold and A.~Pomarol,
  %``A custodial symmetry for Z b anti-b,''
  Phys.\ Lett.\ B {\bf 641}, 62 (2006)
  [hep-ph/0605341].
  %%CITATION = HEP-PH 0605341;%%
 
\bibitem{Bauer:2011ah} 
  M.~Bauer, R.~Malm and M.~Neubert,
  %``A Solution to the Flavor Problem of Warped Extra-Dimension Models,''
  Phys.\ Rev.\ Lett.\  {\bf 108}, 081603 (2012)
  [arXiv:1110.0471 [hep-ph]].
  %%CITATION = ARXIV:1110.0471;%%
  
\bibitem{Malm:2014gha} 
  R.~Malm, M.~Neubert and C.~Schmell,
  %``Higgs Couplings and Phenomenology in a Warped Extra Dimension,''
  JHEP {\bf 1502}, 008 (2015)
  [arXiv:1408.4456 [hep-ph]].
  %%CITATION = ARXIV:1408.4456;%%

\bibitem{Hahn:2013nza} 
  J.~Hahn, C.~Hoerner, R.~Malm, M.~Neubert, K.~Novotny and C.~Schmell,
  %``Higgs Decay into Two Photons at the Boundary of a Warped Extra Dimension,''
  Eur.\ Phys.\ J.\ C {\bf 74}, no. 5, 2857 (2014)
  [arXiv:1312.5731 [hep-ph]].
  %%CITATION = ARXIV:1312.5731;%%
  
\bibitem{Malm:2013jia} 
  R.~Malm, M.~Neubert, K.~Novotny and C.~Schmell,
  %``5D perspective on Higgs production at the boundary of a warped extra dimension,''
  JHEP {\bf 1401}, 173 (2014)
  [arXiv:1303.5702 [hep-ph]].
  %%CITATION = ARXIV:1303.5702;%%
  
\bibitem{Agashe:2014jca} 
  K.~Agashe, A.~Azatov, Y.~Cui, L.~Randall and M.~Son,
  %``Warped Dipole Completed, with a Tower of Higgs Bosons,''
  JHEP {\bf 1506}, 196 (2015)
  [arXiv:1412.6468 [hep-ph]].
  %%CITATION = ARXIV:1412.6468;%%
    
\bibitem{Delaunay:2012cz} 
  C.~Delaunay, J.~F.~Kamenik, G.~Perez and L.~Randall,
  %``Charming CP Violation and Dipole Operators from RS Flavor Anarchy,''
  JHEP {\bf 1301}, 027 (2013)
  [arXiv:1207.0474 [hep-ph]].
  %%CITATION = ARXIV:1207.0474;%%
  
\bibitem{Randall:2001gb} 
  L.~Randall and M.~D.~Schwartz,
  %``Quantum field theory and unification in AdS5,''
  JHEP {\bf 0111}, 003 (2001)
  [hep-th/0108114].
  %%CITATION = HEP-TH/0108114;%%
  
\bibitem{Puchwein:2003jq} 
  M.~Puchwein and Z.~Kunszt,
  %``Radiative corrections with 5-D mixed position / momentum space propagators,''
  Annals Phys.\  {\bf 311}, 288 (2004)
  [hep-th/0309069].
  %%CITATION = HEP-TH/0309069;%%

\bibitem{Contino:2004vy} 
  R.~Contino and A.~Pomarol,
  %``Holography for fermions,''
  JHEP {\bf 0411}, 058 (2004)
  [hep-th/0406257].
  %%CITATION = HEP-TH/0406257;%%

\bibitem{Casagrande:2010si}
  S.~Casagrande, F.~Goertz, U.~Haisch, M.~Neubert and T.~Pfoh,
  %``The Custodial Randall-Sundrum Model: From Precision Tests to Higgs
  %Physics,''
  JHEP {\bf 1009}, 014 (2010)
  [arXiv:1005.4315 [hep-ph]].
  %%CITATION = JHEPA,1009,014;%%

\bibitem{Djouadi:2007fm} 
  A.~Djouadi and G.~Moreau,
  %``Higgs production at the LHC in warped extra-dimensional models,''
  Phys.\ Lett.\ B {\bf 660}, 67 (2008)
  [arXiv:0707.3800 [hep-ph]].
  %%CITATION = ARXIV:0707.3800;%%

\bibitem{Falkowski:2007hz}
  A.~Falkowski,
  %``Pseudo-Goldstone Higgs Production via Gluon Fusion,''
  Phys.\ Rev.\  D {\bf 77}, 055018 (2008)
  [arXiv:0711.0828 [hep-ph]].
  %%CITATION = PHRVA,D77,055018;%%

\bibitem{Cacciapaglia:2009ky} 
  G.~Cacciapaglia, A.~Deandrea and J.~Llodra-Perez,
  %``Higgs ---> \Gamma \Gamma beyond the Standard Model,''
  JHEP {\bf 0906}, 054 (2009)
  [arXiv:0901.0927 [hep-ph]].
  %%CITATION = ARXIV:0901.0927;%%}

\bibitem{Bhattacharyya:2009nb} 
  G.~Bhattacharyya and T.~S.~Ray,
  %``Probing warped extra dimension via gg ---> h and h ---> gamma gamma at LHC,''
  Phys.\ Lett.\ B {\bf 675}, 222 (2009)
  [arXiv:0902.1893 [hep-ph]].
  %%CITATION = ARXIV:0902.1893;%%

\bibitem{Bouchart:2009vq} 
  C.~Bouchart and G.~Moreau,
  %``Higgs boson phenomenology and VEV shift in the RS scenario,''
  Phys.\ Rev.\ D {\bf 80}, 095022 (2009)
  [arXiv:0909.4812 [hep-ph]].
  %%CITATION = ARXIV:0909.4812;%%
    
\bibitem{Azatov:2010pf}
  A.~Azatov, M.~Toharia and L.~Zhu,
  %``Higgs Production from Gluon Fusion in Warped Extra Dimensions,''
  Phys.\ Rev.\  D {\bf 82}, 056004 (2010)
  [arXiv:1006.5939 [hep-ph]].
  %%CITATION = PHRVA,D82,056004;%%
  
\bibitem{Azatov:2011qy} 
  A.~Azatov and J.~Galloway,
  %``Light Custodians and Higgs Physics in Composite Models,''
  Phys.\ Rev.\ D {\bf 85}, 055013 (2012)
  [arXiv:1110.5646 [hep-ph]].
  %%CITATION = ARXIV:1110.5646;%%

\bibitem{Goertz:2011hj} 
  F.~Goertz, U.~Haisch and M.~Neubert,
  %``Bounds on Warped Extra Dimensions from a Standard Model-like Higgs Boson,''
  Phys.\ Lett.\ B {\bf 713}, 23 (2012)
  [arXiv:1112.5099 [hep-ph]].
  %%CITATION = ARXIV:1112.5099;%%  
 
\bibitem{Carena:2012fk} 
  M.~Carena, S.~Casagrande, F.~Goertz, U.~Haisch and M.~Neubert,
  %``Higgs Production in a Warped Extra Dimension,''
  JHEP {\bf 1208}, 156 (2012)
  [arXiv:1204.0008 [hep-ph]].
  %%CITATION = ARXIV:1204.0008;%%
    
\bibitem{Archer:2014jca} 
  P.~R.~Archer, M.~Carena, A.~Carmona and M.~Neubert,
  %``Higgs Production and Decay in Models of a Warped Extra Dimension with a Bulk Higgs,''
  JHEP {\bf 1501}, 060 (2015)
  [arXiv:1408.5406 [hep-ph]].
  %%CITATION = ARXIV:1408.5406;%%

\bibitem{Azatov:2009na} 
  A.~Azatov, M.~Toharia and L.~Zhu,
  %``Higgs Mediated FCNC's in Warped Extra Dimensions,''
  Phys.\ Rev.\ D\ {\bf 80}, 035016  (2009)
  [arXiv:0906.1990 [hep-ph]].
  %%CITATION = PHRVA,D80,035016;%%  

\bibitem{Agashe:2014kda} 
  K.~A.~Olive {\it et al.} [Particle Data Group Collaboration],
  %``Review of Particle Physics,''
  Chin.\ Phys.\ C {\bf 38}, 090001 (2014).
  %%CITATION = CHPHD,C38,090001;%%
  %1615 citations counted in INSPIRE as of 07 Aug 2015

\bibitem{Beneke:2012ie} 
  M.~Beneke, P.~Dey and J.~Rohrwild,
  %``The muon anomalous magnetic moment in the Randall-Sundrum model,''
  JHEP {\bf 1308}, 010 (2013)
  [arXiv:1209.5897 [hep-ph]].
  %%CITATION = ARXIV:1209.5897;%%

\bibitem{Moch:2014ofa} 
  P.~Moch and J.~Rohrwild,
  %``g-2 In the custodially protected RS model,''
  J.\ Phys.\ G {\bf 41}, 105005 (2014)
  [arXiv:1405.5385 [hep-ph]].
  %%CITATION = ARXIV:1405.5385;%%

\bibitem{Beneke:2014sta}
  M.~Beneke, P.~Moch and J.~Rohrwild,
  %``Muon anomalous magnetic moment and penguin loops in warped extra dimensions,''
  Int.\ J.\ Mod.\ Phys.\ A {\bf 29} (2014) 1444011
  [arXiv:1404.7157 [hep-ph]].
  %%CITATION = ARXIV:1404.7157;%%
      
\bibitem{Buras:2011zb} 
  A.~J.~Buras, L.~Merlo and E.~Stamou,
  %``The Impact of Flavour Changing Neutral Gauge Bosons on $\bar{B} -> X_s \gamma$,''
  JHEP {\bf 1108}, 124 (2011)
  [arXiv:1105.5146 [hep-ph]].
  %%CITATION = ARXIV:1105.5146;%%

\bibitem{Gambino:2001ew} 
  P.~Gambino and M.~Misiak,
  %``Quark mass effects in anti-B ---> X(s gamma),''
  Nucl.\ Phys.\ B {\bf 611}, 338 (2001)
  [hep-ph/0104034].
  %%CITATION = HEP-PH/0104034;%%

\bibitem{Misiak:2006ab} 
  M.~Misiak and M.~Steinhauser,
  %``NNLO QCD corrections to the anti-B ---> X(s) gamma matrix elements using interpolation in m(c),''
  Nucl.\ Phys.\ B {\bf 764}, 62 (2007)
  [hep-ph/0609241].
  %%CITATION = HEP-PH/0609241;%%
  
\bibitem{Amhis:2014hma} 
  Y.~Amhis {\it et al.}  [Heavy Flavor Averaging Group (HFAG) Collaboration],
  %``Averages of $b$-hadron, $c$-hadron, and $\tau$-lepton properties as of summer 2014,''
  arXiv:1412.7515 [hep-ex].
  %%CITATION = ARXIV:1412.7515;%%

\bibitem{Misiak:2015xwa} 
  M.~Misiak {\it et al.},
  %``Updated NNLO QCD predictions for the weak radiative B-meson decays,''
  Phys.\ Rev.\ Lett.\  {\bf 114}, no. 22, 221801 (2015)
  [arXiv:1503.01789 [hep-ph]].
  %%CITATION = ARXIV:1503.01789;%%

\bibitem{Kagan:1998bh} 
  A.~L.~Kagan and M.~Neubert,
  %``Direct CP violation in B ---> X(s) gamma decays as a signature of new physics,''
  Phys.\ Rev.\ D {\bf 58}, 094012 (1998)
  [hep-ph/9803368].
  %%CITATION = HEP-PH/9803368;%%

\bibitem{Benzke:2010js} 
  M.~Benzke, S.~J.~Lee, M.~Neubert and G.~Paz,
  %``Factorization at Subleading Power and Irreducible Uncertainties in $\bar B\to X_s\gamma$ Decay,''
  JHEP {\bf 1008}, 099 (2010)
  [arXiv:1003.5012 [hep-ph]].
  %%CITATION = ARXIV:1003.5012;%%

\bibitem{TheATLAScollaboration:2013kha} 
  The ATLAS collaboration,
  %``A search for $t\bar{t}$ resonances in the lepton plus jets final state with ATLAS using 14 fb$^{?1}$ of pp collisions at $\sqrt{s}=8$ TeV,''
  ATLAS-CONF-2013-052.
  %%CITATION = ATLAS-CONF-2013-052, ATLAS-COM-CONF-2013-052;%%
  %61 citations counted in INSPIRE as of 22 août 2015
  
  \bibitem{CMS:lhr} 
  The CMS Collaboration,
  %``Search for ttbar resonances in semileptonic final state,''
  CMS-PAS-B2G-12-006.
  %%CITATION = CMS-PAS-B2G-12-006;%%
  %14 citations counted in INSPIRE as of 22 août 2015

\bibitem{Iwasaki:2005sy} 
  M.~Iwasaki {\it et al.} [Belle Collaboration],
  %``Improved measurement of the electroweak penguin process B ---> X(s) l+ l-,''
  Phys.\ Rev.\ D {\bf 72}, 092005 (2005)
  [hep-ex/0503044].
  %%CITATION = HEP-EX/0503044;%%
  %180 citations counted in INSPIRE as of 04 août 2015
  
\bibitem{Lees:2013nxa} 
  J.~P.~Lees {\it et al.} [BaBar Collaboration],
  %``Measurement of the B->Xsl+l- branching fraction and search for direct CP violation from a sum of exclusive final states,''
  Phys.\ Rev.\ Lett.\  {\bf 112}, 211802 (2014)
  [arXiv:1312.5364 [hep-ex]].
  %%CITATION = ARXIV:1312.5364;%%
  %25 citations counted in INSPIRE as of 04 Aug 2015
  
\bibitem{Huber:2015sra} 
  T.~Huber, T.~Hurth and E.~Lunghi,
  %``Inclusive $ \overline{B}\to {X}_s{\ell}^{+}{\ell}^{-} $ : complete angular analysis and a thorough study of collinear photons,''
  JHEP {\bf 1506}, 176 (2015)
  [arXiv:1503.04849 [hep-ph]].
  %%CITATION = ARXIV:1503.04849;%%
  
\bibitem{Guetta:1997fw} 
  D.~Guetta and E.~Nardi,
  %``Searching for new physics in rare B ---> tau decays,''
  Phys.\ Rev.\ D {\bf 58}, 012001 (1998)
  [hep-ph/9707371].
  %%CITATION = HEP-PH/9707371;%%
  %67 citations counted in INSPIRE as of 11 Aug 2015
    
\bibitem{Buras:1998raa} 
  A.~J.~Buras,
%  ``Weak Hamiltonian, CP violation and rare decays'',
  hep-ph/9806471.
  %%CITATION = HEP-PH/9806471;%%

\bibitem{DescotesGenon:2011yn} 
  S.~Descotes-Genon, D.~Ghosh, J.~Matias and M.~Ramon,
  %``Exploring New Physics in the C7-C7' plane,''
  JHEP {\bf 1106}, 099 (2011)
  [arXiv:1104.3342 [hep-ph]].
  %%CITATION = ARXIV:1104.3342;%%

\bibitem{Ball:2006cva} 
  P.~Ball and R.~Zwicky,
  %``Time-dependent CP Asymmetry in B ---> K* gamma as a (Quasi) Null Test of the Standard Model,''
  Phys.\ Lett.\ B {\bf 642}, 478 (2006)
  [hep-ph/0609037].
  %%CITATION = HEP-PH/0609037;%%

\bibitem{Altmannshofer:2011gn} 
  W.~Altmannshofer, P.~Paradisi and D.~M.~Straub,
  %``Model-Independent Constraints on New Physics in b --> s Transitions,''
  JHEP {\bf 1204}, 008 (2012)
  [arXiv:1111.1257 [hep-ph]].
  %%CITATION = ARXIV:1111.1257;%%

\bibitem{Amhis:2012bh} 
  Y.~Amhis {\it et al.}  [Heavy Flavor Averaging Group Collaboration],
  %``Averages of B-Hadron, C-Hadron, and tau-lepton properties as of early 2012,''
  arXiv:1207.1158 [hep-ex].
  %%CITATION = ARXIV:1207.1158;%%

\bibitem{Benzke:2010tq} 
  M.~Benzke, S.~J.~Lee, M.~Neubert and G.~Paz,
  %``Long-Distance Dominance of the CP Asymmetry in B->X_{s,d}+gamma Decays,''
  Phys.\ Rev.\ Lett.\  {\bf 106}, 141801 (2011)
  [arXiv:1012.3167 [hep-ph]].
  %%CITATION = ARXIV:1012.3167;%%

\bibitem{LHCb:2015dla} 
  The LHCb Collaboration [LHCb Collaboration],
  %``Angular analysis of the $B^{0} \rightarrow K^{*0} \mu^{+} \mu^{-}$ decay,''
  LHCb-CONF-2015-002, CERN-LHCb-CONF-2015-002.
  %%CITATION = LHCB-CONF-2015-002, CERN-LHCB-CONF-2015-002;%%

\bibitem{Descotes-Genon:2013wba} 
  S.~Descotes-Genon, J.~Matias and J.~Virto,
  %``Understanding the B->K*mu+mu- Anomaly,''
  Phys.\ Rev.\ D {\bf 88}, 074002 (2013)
  [arXiv:1307.5683 [hep-ph]].
  %%CITATION = ARXIV:1307.5683;%%
   
\bibitem{Altmannshofer:2013foa} 
  W.~Altmannshofer and D.~M.~Straub,
  %``New physics in $B \to K^*\mu\mu$?,''
  Eur.\ Phys.\ J.\ C {\bf 73}, 2646 (2013)
  [arXiv:1308.1501 [hep-ph]].
  %%CITATION = ARXIV:1308.1501;%%
  
\bibitem{Beaujean:2013soa} 
  F.~Beaujean, C.~Bobeth and D.~van Dyk,
  %``Comprehensive Bayesian analysis of rare (semi)leptonic and radiative $B$ decays,''
  Eur.\ Phys.\ J.\ C {\bf 74}, 2897 (2014)
  [Eur.\ Phys.\ J.\ C {\bf 74}, 3179 (2014)]
  [arXiv:1310.2478 [hep-ph]].
  %%CITATION = ARXIV:1310.2478;%%
  
\bibitem{Hurth:2013ssa} 
  T.~Hurth and F.~Mahmoudi,
  %``On the LHCb anomaly in B $\to K^*\ell^+\ell^-$,''
  JHEP {\bf 1404}, 097 (2014)
  [arXiv:1312.5267 [hep-ph]].
  %%CITATION = ARXIV:1312.5267;%%
  
\bibitem{Alonso:2014csa} 
  R.~Alonso, B.~Grinstein and J.~Martin Camalich,
  %``$SU(2)\times U(1)$ gauge invariance and the shape of new physics in rare $B$ decays,''
  Phys.\ Rev.\ Lett.\  {\bf 113}, 241802 (2014)
  [arXiv:1407.7044 [hep-ph]].
  %%CITATION = ARXIV:1407.7044;%%
    
\bibitem{Hiller:2014yaa} 
  G.~Hiller and M.~Schmaltz,
  %``$R_K$ and future $b \to s \ell \ell$ physics beyond the standard model opportunities,''
  Phys.\ Rev.\ D {\bf 90}, 054014 (2014)
  [arXiv:1408.1627 [hep-ph]].
  %%CITATION = ARXIV:1408.1627;%%
  
\bibitem{Ghosh:2014awa} 
  D.~Ghosh, M.~Nardecchia and S.~A.~Renner,
  %``Hint of Lepton Flavour Non-Universality in $B$ Meson Decays,''
  JHEP {\bf 1412}, 131 (2014)
  [arXiv:1408.4097 [hep-ph]].
  %%CITATION = ARXIV:1408.4097;%%

\bibitem{Altmannshofer:2015sma} 
  W.~Altmannshofer and D.~M.~Straub,
  %``Implications of $b\to s$ measurements,''
  arXiv:1503.06199 [hep-ph].
  %%CITATION = ARXIV:1503.06199;%%
  
\bibitem{Moch:2015oka} 
  P.~Moch and J.~Rohrwild,
  %``$\bar B\to X_s \gamma$ with a warped bulk Higgs,''
  Nucl.\ Phys.\ B {\bf 902}, 142 (2016)
  [arXiv:1509.04643 [hep-ph]].
  
\end{thebibliography}
\end{document}